%% file: exampleMasterFile.tex
\documentclass[editMode]{ufdissertation}
\usepackage[T1]{fontenc} 
\usepackage{breqn}
\usepackage{url}
\usepackage{derivative}
\usepackage{placeins}
\usepackage{amsmath,mathrsfs}
\usepackage{float}
\usepackage[colorlinks=true]{hyperref}
\usepackage{subcaption}
\usepackage{epigraph}
\usepackage[]{cleveref}

\usepackage{tikz}
\usepackage{pgfplots}
\usepackage{algpseudocode}

\usepackage{scrextend}
\deffootnote{1.5em}{0em}{\thefootnotemark\quad}

\usepackage{graphicx}
\usepackage{soul}
\usepackage{titlesec}

\usepackage{xcolor} 
\definecolor{c1}{HTML}{802410} 
\definecolor{c2}{HTML}{003262} 
\usepackage{tkz-euclide}
\usetikzlibrary{decorations.pathmorphing}	
\tikzset{
    v/.style={decorate, decoration={snake, segment length=3mm, amplitude=0.75mm}, draw},
    f/.style={draw,decoration={markings,mark=at position #1 with {\arrow[very thick]{latex}}},postaction={decorate},node contents=#1},
    f/.default=.6,
    fb/.style={draw,decoration={markings,mark=at position #1 with {\arrowreversed[very thick]{latex}}},postaction={decorate},node contents=#1},
    fb/.default=.4,
    fnar/.style={draw},
    g/.style={decorate, draw,  decoration={coil,amplitude=3pt, segment length=3.5pt}},
    s/.style={dashed,draw, postaction={decorate},
        decoration={markings,mark=at position .55 with {\arrow[very thick]{latex}}}},
    sb/.style={dashed,draw, postaction={decorate},
        decoration={markings,mark=at position .55 with {\arrowreversed[draw=black,very thick]{latex}}}},
    snar/.style={dashed,draw,line width =1.25pt},
}
\usetikzlibrary{shapes}	
\tikzset{every picture/.style={line width=1}}

\newcommand{\Eq}[1]{Eq.~\hyperref[#1]{(\ref*{#1})}}
\newcommand{\Fig}[1]{Fig.~\hyperref[#1]{(\ref*{#1})}}

\newcommand{\sinc}{\operatorname{sinc}}
\usepackage{epstopdf}

\haveFigurestrue

\title{Topics in the phenomenology of axions: The cases of cosmic strings and superradiance}

\degreeType{Doctor of Philosophy}
\major{Physics}
\author{Antonios Kyriazis}
\thesisType{Dissertation}
\degreeYear{2026}
\degreeMonth{August}
\chair[]{Pierre Sikivie}

\setDedicationFile{dedicationFile}
\setAcknowledgementsFile{acknowledgementsFile}
\setAbstractFile{abstractFile}
\setReferenceFile{referenceFile}{amsplain}

\setBiographicalFile{biographyFile}

\setAbbreviationsFile{abbreviations}
\setAppendixFile{appendix}
\multipleAppendixtrue

\begin{document}


\include{chapter1}
\include{chapter2}
\include{chapter3}
\include{chapter4}


\end{document}

%% file: chapter1.tex
\chapter{Introduction to the axion}
\label{section:intro}

The nature of dark matter is one of the most interesting puzzles of modern physics. Its treatment as a pressureless fluid in the $\Lambda$CDM  model has explained successfully a wealth of observational evidence, ranging from the rotation curves of galaxies \cite{Rubin1,Rubin2,Bosma}, to the CMB \cite{Planck}, to the Bullet Cluster \cite{Clowe:2006eq}. In the context of this model, dark matter  comprises 26\% of the energy density of our universe, it is cold, meaning its velocity is much smaller than the speed of light, and it is collisionless, meaning that it interacts with ordinary matter only through gravity. Despite the successes of this model, very little is actually known about the particle nature of dark matter. 

Many particle candidates have been proposed in the literature that span a wide range of masses  (Reference \cite{DM_review} is an extensive review of all these candidates). One candidate that has received significant attention in recent years is the axion. Initially proposed as a solution to the strong CP problem \cite{Peccei, Weinberg}, it was realized quickly afterwards that it can also play the role of dark matter \cite{Preskill,Willy, Sikivie}
\par
The strong CP problem stems from the following term in the QCD Lagrangian:
\begin{equation}
    \label{eqn:qcd}
    \delta \mathcal{L} = \theta_{\rm QCD} \frac{g_{s}^{2}}{32 \pi^{2}} G_{\mu \nu}^{a} \tilde{G}^{a\mu \nu},
\end{equation}
where $\theta_{\rm QCD}$ is an angle that can have any value in the domain $[0,2 \pi]$, $g_{s}$ is the coupling constant of QCD, $G^{a}_{\mu \nu}$ is the field strength tensor of QCD and $\tilde{G}^{a}_{\mu \nu}=\frac{1}{2} \epsilon_{\mu \nu \lambda \rho} G^{a \lambda \rho}$. This term contributes to the electric dipole moment of the neutron \cite{neeutron_edm_th}:
\begin{equation}
    d_{n} \approx 10^{-16} \theta_{\rm QCD} \hspace{0.1 cm} e \cdot cm .
\end{equation}
The measured value of this dipole moment is $d_{n} < 2.9 \times 10^{-26} e \cdot cm $ \cite{Pendlebury:2015lrz}. This implies that $\theta_{\rm QCD} \leq 10^{-10}$ and this poses a fine tuning problem, because there is no a priori reason for $\theta_{\rm QCD}$ to be this small. 

\par
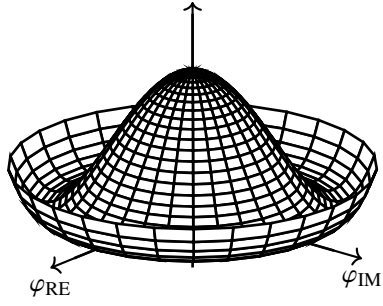
\begin{figure}
\centering
\begin{tikzpicture}
    \begin{axis}[
        axis lines=center,
        axis line style={-},
        axis on top=false,
        view={130}{20},
        axis equal,
        colormap={blackwhite}{gray(0cm)=(1); gray(1cm)=(0)},
        samples=30,
        domain=0:360,
        y domain=0:1.25,
        ticks=none,
        zmin=0,
        zmax=1,
        yticklabels={,,},
        xticklabels={,,},
        zticklabels={,,}
    ]
    \addplot3[surf, shader=flat, draw=black, fill=white, z buffer=sort]
        ({sin(x)*y}, {cos(x)*y}, {(y^2-1)^2});

    \draw[black,thick,->] (axis cs:1,0,0) -- (axis cs:1.5,0,0)
        node[below,font=\footnotesize]{$\varphi_{\text{RE}}$};

    \draw[black,thick,->] (axis cs:0,1,0) -- (axis cs:0,1.5,0)
        node[below,font=\footnotesize]{$\varphi_{\text{IM}}$};

    \draw[black,thick,->] (axis cs:0,0,1) -- (axis cs:0,0,1.5);
    \end{axis}
\end{tikzpicture}
\caption[The Mexican hat potential]{The Mexican hat potential of the scalar field $\varphi$ that realizes spontaneous symmetry breaking}
\label{fig:mexican}
\end{figure}

 The axion is a proposal to solve this problem dynamically. It is the Nambu-Goldstone boson associated with the spontaneous breaking of a new U(1) symmetry called the PQ symmetry \cite{Peccei}. This symmetry breaks spontaneously at some scale $f_{a}$, which can be higher or lower than the inflationary energy scale. To realize this breaking, a scalar field $\varphi$ is introduced:

\begin{equation}
    \label{eqn:pq field}
    \varphi(x) = \frac{\rho(x)}{\sqrt{2}} e^{i \phi/f_{a}},
\end{equation}
where $\rho(x)$ is the radial degree of freedom that becomes massive after the spontaneous breaking of the symmetry and $\phi$ is the angular degree of freedom that will be the axion. The potential that determines the breaking of the PQ symmetry is given by the usual Mexican hat configuration, shown in \Fig{fig:mexican}.

Chiral rotations of quarks that are charged under this new U(1) symmetry are anomalous and generate a term similar to \Eq{eqn:qcd}:

\begin{equation}
    \label{eqn: axion_term}
    \delta \mathcal{L} = \frac{g^{2}_{s}}{32 \pi^{2}} \frac{\phi}{f_{a}} G \tilde{G}.
\end{equation}

Only derivative couplings of $\phi$ appear in the Lagrangian, and therefore a shift $\phi \rightarrow \phi + \kappa$ can be performed and $\kappa$ can be chosen in such a way that \Eq{eqn:qcd} is eliminated. The remaining piece of the puzzle is to show that the potential of the axion enjoys a minimum at $\phi=0$ and the strong CP problem is solved.
 
Indeed, at temperatures above $\Lambda_{\rm QCD} \approx 160 \textrm{MeV}$, non-perturbative effects known as instantons, will induce a potential for the axion:
\begin{equation}
    \label{eqn: potential}
    V(\phi)=m^{2}_{a}(T) f_{a}^{2}\left( 1-\cos 
    \left(\frac{\phi}{f_{a}} + \theta_{\rm QCD} \right)\right),
\end{equation}
where $m_{a}(T)$ is the mass of the axion that depends on temperature. In \cite{Willy,Sikivie,Preskill}, it was found that $m_{a}(T) \propto T^{-4}$. This potential is minimized at $\frac{\phi}{f_{a}}=-\theta_{\rm QCD}$ and the strong CP problem is solved. 

At temperatures below $\Lambda_{\rm QCD}$, the same conclusion holds and chiral perturbation theory gives us the zero temperature limit of the axion mass $m_{a}= 5.7 \textrm{eV} \hspace{0.05cm} \left(10^{12} \textrm{GeV}/f_{a} \right) $ \cite{Weinberg,Wilczek1}. The particle with this mass is called the $\rm QCD$ axion. 
\par

More generally, the spontaneous symmetry breaking of a global $U(1)$ symmetry, similar to the PQ symmetry discussed above, produces pseudo Nambu-Goldstone bosons, which we shall refer to as ALPs, since they are neutral pseudoscalars. In these general constructions, the symmetry breaking scale $f_{a}$ and the mass $m_{a}$ of the boson are independent of each other. These axion-like particles also appear in certain realizations of string theory, with masses $10^{-33} \textrm{eV} \leq m_{a} \leq 10^{-10} \textrm{eV}$\cite{Arvanitaki:2009fg}. Reference \cite{AxionLimits} displays various constraints on $m_{a}$ and $f_{a}$, for both ALPs and $\rm QCD$ axions.

\section{Cosmological Evolution}
For small values of the axion's amplitude, we can approximate the potential in \Eq{eqn: potential} as $V(\phi) \approx \frac{1}{2} m_{a}^{2} \phi^{2}$. The equation of motion for the zero mode of the scalar field is:
\begin{equation}
    \label{eqn: zero_mode}
    \ddot{\theta}_{0}(t)+3H\dot{\theta}_{0}(t)+m_{a}^{2}(t)\theta_{0}(t)=0,
\end{equation}
where we defined $\theta \equiv \phi/f_{a}$, and $H$ as the Hubble parameter, given by $H = \dot{a}/a$, with $a(t)$ the scale factor and with dot denoting a time derivative. 

When $H \gg m_{a}$, the solution to \Eq{eqn: zero_mode} is \cite{Sikivie_2008}:
\begin{equation}
    \theta_{0}(t)=c_{1}+c_{2} t^{-1/2},
\end{equation}
where $c_{1}$ and $c_{2}$ are constants. In this regime, where Hubble friction dominates, the axion is essentially frozen. The axion mass "turns on" at time $t=t_{m}$, when $m_{a}(t_{m})= H(t_{m})$. The number density at that time is:
\begin{equation}
    n_{a}(t_{m})=\frac{1}{2} m_{a}(t_{m}) \phi^{2}(t_{m}) \approx \frac{1}{2 t_{m}} f^{2}_{a} \theta^{2}_{1},
\end{equation}
where $\theta_{1}$ is the initial misalignment angle. 
\par
The expression for $m_{a}(t)$ at times $t>t_{m}$  varies slowly, i.e. $\frac{d \log(m_{a})}{dt} \ll m_{a}$. Thus, we can use the WKB approximation to find the solution to \Eq{eqn: zero_mode} for times $t>t_{m}$:
\begin{equation}
    \theta_{0}(t)=C \frac{1}{t^{3/4} \sqrt{m_{a}(t)}} \textrm{cos} \left(\int \omega(t')dt' +\delta  \right), \omega^{2}(t)=m^{2}_{a}(t)+\frac{3}{16 t^{2}},
\end{equation}
where C is a constant and $\delta$ is a phase. The number density of these oscillations is given by:
\begin{equation}
    \label{eqn: number density}
    n_{a}(t)=\frac{1}{2 m_{a}(t)} \dot{\phi}^{2}_{0}(t)+\frac{1}{2}m_{a}(t)\phi^{2}_{0}(t) = 4 \cdot 10^{47} \textrm{cm}^{-3}\left( \frac{f_{a}}{5 \times 10^{12} \textrm{GeV}} \right)^{3} \left( \frac{a(t_{m})}{a(t)} \right)^{3},
\end{equation}
where we used $a(t) \propto t^{1/2}$, which holds in a radiation dominated universe \cite{Sikivie_2008}. Therefore, these oscillations of the zero mode behave as ordinary matter and this is partly what makes the axion a dark matter candidate.

\section{The Contribution from Cosmic Strings}

The above discussion has ignored the contribution to the dark matter density from cosmic strings. These are topological defects that arise from the Kibble mechanism during the spontaneous symmetry breaking of a global symmetry \cite{1976JPhA....9.1387K}. In brief, if along a closed loop in physical space, the phase of $\varphi$ changes by $2 \pi$, that implies that a cosmic string passes through that loop. This takes place because in causally disconnected regions of space, the phase of the scalar field obtains a random value. 

More formally, if the original symmetry is called $G$ and the subgroup that remains unbroken after the spontaneous symmetry breaking is called $H$, the manifold that spans the vacuua that minimize the potential of the scalar field is the left coset $M = G/H$. If the mapping of a circle $S^{1}$ to this manifold is non-trivial (in the language of homotopy, this mapping is known as the first homotopic group $\pi_{1}(M)$ \cite{coleman1985aspects}) then cosmic strings form.

The scalar field is in its symmetric phase inside the string core and in its broken phase outside the core. At time $t$, the typical curvature of a string is of order $t$. An important process that governs the evolution of the string network is the intercommuting of strings, that breaks up two intersecting strings into smaller loops \cite{Vilenkin:1984ib}. In cases where the original symmetry is local, the loops primarily decay due to emission of gravitational waves, while when the symmetry is global, they decay through the emission of the Nambu-Goldstone particle \cite{Battye:1993jv}. Due to the emission of these particles, the number of strings per Hubble volume reaches asymptotically a value of $\mathcal{O}(1)$, irregardless of the initial conditions, and the network is said to be in the ``scaling regime''. 

The string network will eventually collapse at $t_{m}$, when the strings become the boundaries of domain walls and they subsequently decay into mildly relativistic axions \cite{Sikivie_2008,Chang:1998tb}. A fraction of the Nambu-Goldstone particles ,that are emitted by all these components, become non-relativistic at time $t_{m}$ or shortly afterwards, just like the zero mode of the previous section.    

The gravitational coupling of these axion-like particles will contribute to the density perturbations in our universe, and one could search for its particular shape, once it is derived, in cosmological observables. To be more specific, the gravitational influence of axion-like dark matter can be understood from the statistical properties of fluctuations in its stress--energy tensor~\cite{Hu:2000ke,Khmelnitsky:2013lxt,Hui:2016ltb}. Given a phase-space distribution, the induced metric perturbations can be described statistically~\cite{Kim:2023kyy,Kim:2024,Boddy:2025oxn,Dror:2025nvg}. Since the stress--energy tensor scales quadratically with the field, its temporal Fourier spectrum contains two characteristic classes of modes: (i) fast modes, with frequencies near twice the particle mass, corresponding to sums of single-particle energies; and (ii) slow modes, with frequencies set by the average kinetic energy, determined by the product of the mass and the velocity dispersion squared. Both classes fluctuate on the same characteristic spatial scale—the de~Broglie wavelength—but the slow modes dominate in amplitude and will be the focus of this study.

For a Maxwell--Boltzmann phase-space distribution, the statistical properties of these metric perturbations have been extensively studied. The associated density perturbations exhibit stochastic fluctuations on spatial separations larger than the coherence length, set by the inverse of the product of the particle mass and its velocity dispersion. This scale governs the structure of the matter power spectrum: for wavenumbers below the inverse coherence scale, the spectrum approaches a white-noise plateau, corresponding to a scale-independent (constant) isocurvature power spectrum. Together with the associated free-streaming length, this feature has motivated a rapidly developing program of cosmological searches for ultralight dark matter using probes of the matter power spectrum, including the $\textrm{Lyman-}\alpha$ forest, UV luminosity functions, large-scale structure surveys, and the CMB~\cite{Feix:2019lpo,Feix:2020txt,Irsic:2017yje,Kobayashi:2017jcf,Irsic:2019iff,Gorghetto:2021fsn,Gorghetto:2025uls,Chathirathas:2025aan,Gorghetto:2022ikz} (see also \cite{amin2024lowerbounddarkmatter,Harigaya:2025pox,Long:2024cak}). While these searches have rapidly become leading probes of ultralight dark matter from topological defect decay, they typically assume that the dominant constraints arise entirely from the white-noise plateau region of the matter power spectrum.

In \Cref{sec:strings}, we develop a formalism to calculate the isocurvature power spectrum of ultralight dark matter for an arbitrary phase-space distribution, extending existing approaches that model the galactic dark matter field as a superposition of plane waves~\cite{Foster:2017hbq,Boddy:2025oxn} to a cosmological framework that consistently accounts for the expansion of the Universe. Using this formalism, we derive the matter power spectrum for ultralight dark matter produced via cosmic string decay and characterize its behavior across all wavenumbers. We then estimate the sensitivity of existing and future cosmological probes. Furthermore, previous searches assumed a time-independent mass, such that the time of network collapse coincides with when the Hubble rate becomes comparable to its present-day mass; we go beyond this assumption by considering a temperature-dependent boson mass and studying its consequences.

The key result is the isocurvature power spectrum of post-inflationary axion-like particles today $P_{\rm iso}(k)$, denoting the bosonic field as $\phi$ and assuming it forms a sub-component of dark matter:
\begin{equation}
    P_{\rm iso}(k) = D^{2}_{\rm iso}(k,a_{0})
    \left(\frac{\bar{\rho}_{\phi}(a_{0})}{\bar{\rho}_{\rm DM}(a_{0})}\right)^{2}
    k_\star^{-3}\,\mathcal{T}(k/k_\star).
    \label{eqn:general power spectrum}
\end{equation}
The power spectrum is proportional to the ratio of the relic density $\bar{\rho}_{\phi}$ to the total dark matter density $\bar{\rho}_{\rm DM}$, since fluctuations are defined relative to the total dark matter density. The growth factor $D_{\rm iso}(k,a_{0})$ accounts for the evolution of perturbations after matter--radiation equality. The transfer function $\mathcal{T}(x)$ tends to unity for small $x$, with a characteristic scale $k_\star$ above which it begins to fall. For cosmic string emission, the scale $k_\star$ is set by the infrared cutoff of the boson spectrum at the time the string network collapses, reflecting the fact that the emitted boson number density is dominated by particles with momenta near this cutoff scale. As a result, the features at $k>k_\star$ encode the information needed to distinguish between different ultralight dark matter production mechanisms. We calculate $\mathcal{T}(x)$ and determine $k_\star$ from first principles, comparing the resulting isocurvature spectrum with that from adiabatic inflationary fluctuations.

\section{The Superradiant Mechanism}
If a light particle such as the axion exists, it may form a bound cloud around a spinning black hole, a process known as superradiance. This process can extract mass and angular momentum from the black hole. It is triggered when a wave that scatters by the black hole satisfies the condition \cite{Brito_2020}: 

\begin{equation} \label{eqn:superradiance condition}
    \omega < m \Omega_{H},
\end{equation}
where $\omega$ is the frequency of the wave, $\Omega_{H}$ is the angular velocity of the black hole and $m$ is the azimuthal number with respect to the rotation axis. One simple argument to understand the mechanism was presented by Zel'dovich in \cite{1972JETP...35.1085Z}. Consider an axisymmetric cylinder that has reached equilibrium with well-defined temperature $T$ and entropy $S$, rotating with angular velocity $\Omega_{H}$. Assuming that radiation with power $P_{m}(\omega) d\omega$ in the frequency range $(\omega,\omega + d\omega)$ and azimuthal number $m$  is incident on the cylinder, the energy transferred to the cylinder per unit time is:

\begin{equation}
    \frac{dE}{dt} = Z_{m} P_{m}(\omega) d\omega
\end{equation}

where $Z_{m}$ is the fraction absorbed and is negative for the case of superradiance. The amount of angular momentum transferred $\delta J$, given an infinitesimal amount of energy $\delta E$, is given by \cite{Brito_2020}:

\begin{equation}
    \delta J = \frac{m}{\omega} \delta E
\end{equation}

and therefore the rate that angular momentum is absorbed by the cylinder is:

\begin{equation}
    \frac{dJ}{dt} = Z_{m} \frac{m}{\omega} P_{m}(\omega) d \omega
\end{equation}

In the frame co-rotating with the cylinder, the energy change is:

\begin{equation}
    dE_{0} = dE - \Omega dJ = dE \left(1 - \frac{m \Omega_{H}}{\omega} \right) 
\end{equation}

and therefore the entropy change per unit time is given by:

\begin{equation}
    \frac{dS}{dt} = \frac{dE_{0}/dt}{T} = \frac{dE}{dt} \frac{\left( \omega - m \Omega_{H} \right)}{\omega} = Z_{m} P_{m}(\omega) d\omega \frac{\omega - m \Omega_{H}}{\omega T}
\end{equation}
The second law of thermodynamics dictates that $\frac{dS}{dt} > 0$ and hence:

\begin{equation}
    Z_{m} (\omega - m \Omega_{H})>0
\end{equation}
and therefore for superradiance, $Z_{m}<0$, we obtain \Eq{eqn:superradiance condition}.  
When this mechanism is applied to black holes, as it was first done in \cite{1969NCimR...1..252P}, it is discovered that the incident radiation can extract mass and angular momentum from the black hole.  In addition, if the bosonic particles that make up the wave have a small mass, such as axions and axion-like particles, they can form hydrogen-like bound states around the black hole, hence the term ``gravitational atom'' \cite{Detweiler:1980uk,spectra} Using the measurements of black holes' spins in X-ray binaries, constraints have been placed on these light particles \cite{Arvanitaki_2011,Arvanitaki_2015,_nal_2021,witte2025steppingsuperradianceconstraintsaxions,hoof2024gettingblackholesuperradiance,mehta2021superradianceexclusionslandscapetype,self-interactions,superradiance_string_theory}. \par 
Searches for exotic bosons by gravitational wave emissions from the GA in an isolated black hole system have been done in two main channels. These are the annihilations of the GA into gravitons to produce GWs with frequency $2 \mu$, where $\mu$ is the boson's mass, and the spontaneous transition between two superradiant states, producing GWs with frequency equal to the energy difference between the states. These types of signals can be searched for in LIGO and LISA \cite{Arvanitaki_2011,Gravitaitonal_wave_searches,Yang_2023,LIGOScientific:2021rnv}. The inclusion of self-interactions of bosons induces mixing between superradiant states that leads to a GW signal in the deci-Hz frequency range and can also be searched for in ground-based interferometers \cite{self-interactions,deci_Hz,Collaviti_2024,DellaMonica:2025zby}. 

\par
The GA presents rich and intriguing phenomena when perturbed via a companion compact object \cite{Baumann_2019}.
The tidal field of the companion induces resonant transitions between the growing and decaying states of the cloud, causing its demise. The question is: are there any observable signals from these transitions? By using conservation of angular momentum, it has been shown that a transition can back-react to the orbit, causing the orbital frequency to either ``float'' or ``sink'', depending on the type of transition. This can leave distinct imprints on the binary's waveform, a smoking gun signature for the presence of a GA \cite{legacy,Baumann_2020,resonant_history,axion_cloud_backreaction,Ionization,sharp_signals,self_interaction_binary,Guo:2024iye,extreme_mass_ratio,Zhang:2018kib}. Another distinct effect of the back-reaction is the increase of the orbit's eccentricity, while the orbit is within the resonance band. This will drastically alter the distribution of black hole masses and eccentricities in black hole binary systems that are expected to be observed by LISA \cite{Bo_kovi__2024}. Off-resonant mixing between growing and decaying states may also prevent superradiance altogether or cause the decay of the GA, if it has grown, while the back-reaction can, also in this case, leave observable imprints on the inspiral's waveform \cite{Tong_2022}.
\par
Whereas most of the literature on this topic has focused on the signatures that a GA imprints on the inspiral GW signal of the binary, we point out in \Cref{sec:ga} a {\it novel} GW signal that comes from the tidally perturbed GA itself. The GWs are generated by the time-varying quadrupole moment of the GA due to the interference of two states during level transitions and we classify them as a monochromatic signal.\par      
We study the waveform and spectrum features of this GW signal in a binary system, including duration, strength and peak frequency. We focus on hyperfine and fine transitions, which occur at small orbital frequencies, when the companion is at a much larger distance than the size of the GA and analytical results for the GW waveform and spectrum can be obtained. \par
The frequency of these types of events can be in DECIGO's and LISA's frequency band, from milli-Hz to deci-Hz, and so we scan the parameter space of black hole masses, mass ratios of companions, and boson masses to determine which type of systems produce the most promising SNR. For these computations, we assess the validity of the non-relativistic approximation that we employ throughout for the cloud's wavefunction and the superradiant rates and use the relativistically computed quantities in the parameter space where that approximation fails. 

%% file: chapter2.tex
\chapter{Echoes of Global Cosmic Strings}\label{sec:strings}

In this Chapter, we derive the density power spectrum of axion-like particles emitted by cosmic strings and impose constraints on the parameter space spanned by $(m_{a},f_{a})$ using cosmological observables. This Chapter is based on the paper \cite{2c5c-vz3h}.

In \Cref{sec:field correlations}, we construct the scalar field describing the axions and derive its two-point correlation function. In \Cref{sec:scaling regime}, we discuss the scaling regime and the energy spectrum of the axions as they are emitted by the cosmic strings. In \Cref{sec:density correlations}, we compute the density correlations of the slow modes in terms of the energy spectrum of particles emitted by cosmic strings, and derive the corresponding density power spectrum and transfer function. Finally, in \Cref{sec:constraints}, we impose constraints from large-scale structure data.

\section{Field Correlations}
\label{sec:field correlations}
The evolution of a given mode of the scalar field $\phi$ in an expanding Universe is governed by the Klein--Gordon equation:~\footnote{We assume a quadratic potential, so the scalar field evolves linearly. Previous works have considered a cosine potential, which introduces non-linearities and leads to non-conservation of particle number after the collapse of the string network \cite{Gorghetto:2020qws,Chathirathas:2025aan,Gorghetto:2025uls,Gorghetto:2021fsn}.}
\begin{equation}
\ddot{\phi}_{\bf k}+ 3 H \dot{\phi}_{\bf k} + \phi_{\bf k}\left( m^{2} + \frac{|{\bf k}|^{2}}{a^{2}} \right) = 0,
\label{eqn:real k-g}
\end{equation}
where ${\bf k}$ is the comoving wavenumber, $a(t)$ is the scale factor, $H$ is the Hubble parameter, and overdots denote derivatives with respect to physical time.

The mass parameter may depend on temperature, as in the case of the QCD axion \cite{Sikivie_2008}. For temperatures $T > T_{\rm c} \equiv \sqrt{m_{a} f_{a}}$, the temperature dependence can be approximated as
\begin{equation}
    m(T) = m_{a} \left( \frac{T_{\rm c}}{T} \right)^{n},
\label{eqn:mass temperature}
\end{equation}
while for $T < T_{\rm c}$ one has $m(T) = m_{a}$, the zero-temperature boson mass~\cite{2018PhRvD..97h3502F,Maseizik:2024qly}. The temperature at which the field begins to oscillate, $T_m$, is determined numerically from $H(T_{m}) = m(T_{m})$, and the corresponding scale factor follows from entropy conservation. In this work, we consider both temperature-independent and temperature-dependent masses.

We model the ultralight dark matter field as a superposition of classical particles, treating each mode as a plane wave satisfying \Eq{eqn:real k-g}. Applying the WKB approximation, we obtain
\begin{align}
\phi_{\bf k}(x)
    &= \frac{\phi_{{\bf k},0}}{a^{3/2}}
       \sum_{j=1}^{N_{\bf k}}
       \cos\!\left[
           \int^{t}\!\omega_{\bf k}(t')\,dt'
           - {\bf k}\!\cdot\!{\bf x}
           + \varphi_{{\bf k},j}
       \right],
\label{eqn:label j}
\end{align}
where $x\equiv(t,{\bf x})$, $\omega_{\bf k}=\sqrt{m^{2}+{\bf k}^{2}/a^{2}}$, and $N_{\bf k}$ is the number of particles in a fixed phase-space volume (time-invariant since we work with comoving momentum). The sum over $j$ can be performed analytically (see, e.g., Refs.~\cite{Foster:2017hbq,Dror_2021,Boddy:2025oxn}), yielding
\begin{equation}
\phi_{\bf k}(x)
    = \sqrt{\frac{N_{\bf k}}{2a^{3}(t)}}\,\phi_{{\bf k},0}\,
      \alpha_{\bf k}
      \cos\!\left[
          \int^{t}\!\omega_{\bf k}(t')\,dt'
          - {\bf k}\!\cdot\!{\bf x}
          + \varphi_{\bf k}
      \right],
\label{eqn:phi with N}
\end{equation}
where $\alpha_{\bf k}$ and $\varphi_{\bf k}$ are random variables drawn from a Rayleigh distribution with unit scale parameter and a uniform distribution on $[0,2\pi)$, respectively~\cite{Foster:2017hbq}.

To fix the amplitude $\phi_{{\bf k},0}$, we compute the ensemble-averaged energy density of $\phi_{\bf k}$:
\begin{align}
\langle \rho_{\phi,{\bf k}}\rangle
    &= \left\langle
        \frac12\!\left(
            \dot{\phi}_{\bf k}^{2}
            + m^{2}\phi_{\bf k}^{2}
            + \frac{(\nabla\phi_{\bf k})^{2}}{a^{2}}
        \right)
      \right\rangle \notag\\
    &= \frac{N_{\bf k}}{2a^{3}}\,\phi_{{\bf k},0}^{2}\,\omega_{\bf k}^{2}.
\label{eqn:rho}
\end{align}
Equating this to the energy density of $N_{\bf k}$ particles, $\omega_{\bf k} N_{\bf k}/a^{3}$ per comoving volume element $V$, gives
\begin{equation}
\phi_{{\bf k},0}
    = \sqrt{\frac{2}{V\,\omega_{\bf k}(t)}}.
\label{eqn:phi0}
\end{equation}

The number of particles can be related to the phase-space density via $N_{\bf k}=f({\bf k})\,V\,d^{3}k$. Because $f({\bf k})$ depends only on comoving momentum, it evolves self-similarly. Substituting \Eq{eqn:phi0} into \Eq{eqn:phi with N}, we obtain
\begin{align}
\phi_{\bf k}(x)
    &= \sqrt{
        \frac{f({\bf k})\,d^{3}k}
             {(2\pi)^{3}a^{3}\omega_{\bf k}}
       }\,
       \alpha_{\bf k}
       \cos\!\left[
           \int^{t}\!\omega_{\bf k}(t')\,dt'
           - {\bf k}\!\cdot\!{\bf x}
           + \varphi_{\bf k}
       \right].
\end{align}

\begin{figure}[t]
      \centering
    \includegraphics[width=1\columnwidth]{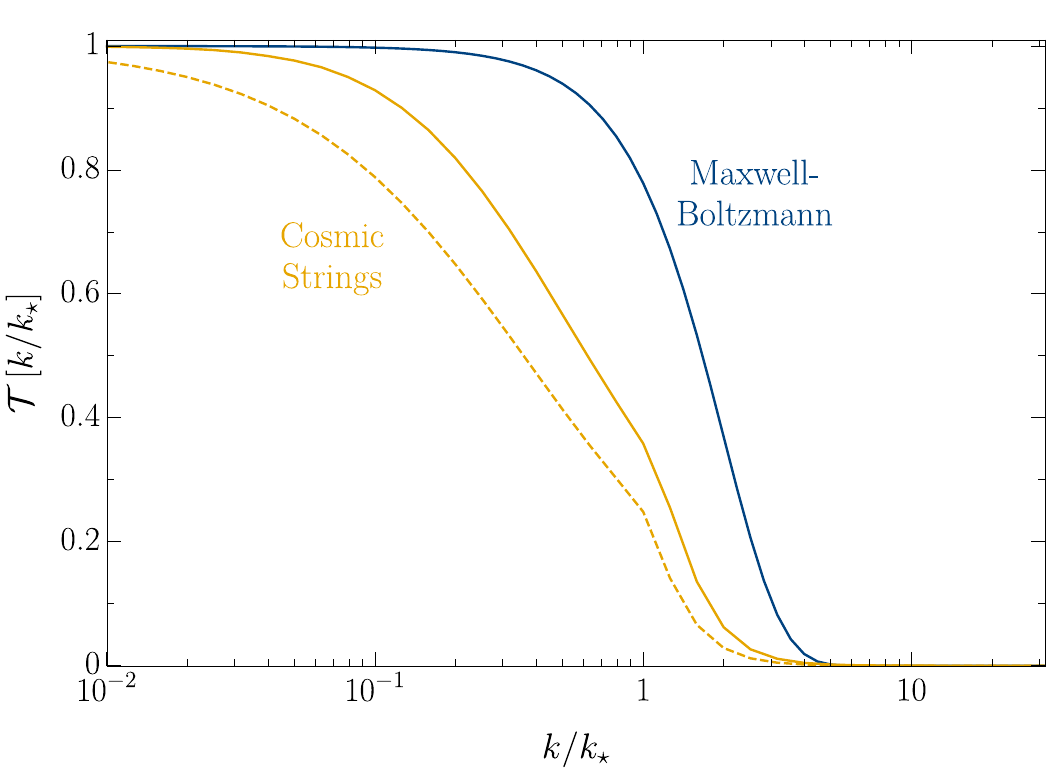}
    \caption[The Maxwell-Boltzmann and the cosmic strings transfer functions.]{The Maxwell-Boltzmann (blue) and the cosmic strings (yellow) transfer functions.  The solid line for cosmic strings corresponds to a numerical evaluation, while the dashed line corresponds to the analytic approximation presented in \Eq{eqn:transfer small k} and \Eq{eqn:transfer large k}, respectively. For $k \gg k_\star$, the cosmic string transfer functions drops $ \propto k^{-4}$, while the Maxwell-Boltzmann transfer function drops exponentially. We have used $k_\star = 2.1 k_{\rm IR}$ to plot the cosmic string curves (see text).} 
    \label{fig:transfer functions}
\end{figure}

Using this expression, the full scalar field is constructed as a sum over plane-wave modes,
\begin{equation}
    \phi(x)=\sum_{\bf k}\phi_{\bf k}(x).
\end{equation}

To compute the two-point correlation function, we note that the ensemble average over the random phase $\varphi_{\bf k}$ vanishes for a single cosine, so only terms with ${\bf k}={\bf k}'$ contribute. We are primarily interested in the equal-time two-point function given by:
\begin{equation}
\langle \phi({\bf x},t)\phi({\bf x}',t)\rangle
    = \int\frac{d^{3}k}{(2\pi)^{3}}
      \frac{f({\bf k})}{a^{3}(t)\,\omega_{\bf k}(t)}
      \cos\!\big[{\bf k}\cdot\Delta{\bf x}\big],
\label{eqn:phi correl}
\end{equation}
where $\Delta{\bf x}={\bf x}-{\bf x}'$.

\section{The Scaling Regime}
\label{sec:scaling regime}
As discussed in \Cref{section:intro}, the number of cosmic strings per Hubble volume reaches an asymptotic value after sufficient time, irregardless of the initial conditions of the network \cite{Gorghetto_2018}. In this section we will study this regime quantitatively and define the energy spectrum of the emitted pseudo Nambu-Goldstone modes. 

In general, the energy density of strings is given by \cite{Gorghetto_2018,Sikivie_2008}:
\begin{equation}\label{eqn: 5.1}
    \rho_{s} = \frac{\xi(t) \mu_{\textrm{eff}}}{t^{2}} 
\end{equation}
where $\xi(t)$ is the average number of strings per Hubble volume and $\mu_{\rm eff}$ is the effective string tension which we will write shortly. In the scaling regime, $\xi$ tends to:
\begin{equation}\label{eqn: 5.2}
    \xi(t) = \alpha \text{log} \left( \frac{m_{\textrm{r}}}{H} \right) + \beta
\end{equation}
where $\alpha$ and $\beta$ are constants and $m_{\textrm{r}}$ is the inverse core size of the string, which we will assume to be $m_{r} \sim f_{a}$. The effective tension is given by:
\begin{equation}\label{eqn: 5.3}
    \mu_{\textrm{eff}} (t) = \pi f^{2}_{a} \log \left( \frac{m_{r} \gamma}{H \sqrt{\xi}} \right)
\end{equation}
where $\gamma$ is of order unity. The rate of emission of axions from strings in the large log limit is given by \cite{Gorghetto_2018,Dror_2021}:
\begin{equation}\label{eqn: 5.4}
    \Gamma_{\phi} \rightarrow 2 H \frac{\rho_{s}}{\rho_{\textrm{SM}}} \rho_{\textrm{SM}}
\end{equation}
where $\rho_{\rm SM} = \frac{\pi^{2}}{30} g_{\ast}(T) T^{4}$ is the energy density in the Standard Model and $g_{\ast}(T)$ are the relativistic degrees of freedom. Using \Eq{eqn: 5.1} and the relation $t = \frac{1}{2H}$, we find for the ratio of the string density to the standard model density: 
\begin{equation}\label{eqn: 5.5}
    \frac{\rho_{s}}{\rho_{\rm SM}} = \frac{4 \xi \mu_{\rm eff}}{3 M^{2}_{\rm pl}}
\end{equation}
We can write the density and the spectrum of these axions as $\rho_{\phi} = \int dk  \frac{\partial \rho_{\phi}}{\partial k}$. The spectrum can also be written as:
\begin{equation}\label{eqn: 5.6}
    \frac{\partial \rho_{\phi}}{\partial k} = \int^{t_{m}}_{t_{\rm min}} dt' \frac{\Gamma_{\phi}(t')}{H(t')} \frac{a^{3}(t')}{a^{4}(t)}  F\left[ \frac{k'}{H'} , \frac{m_{r}}{H'} \right]
\end{equation}
where $k' =  \frac{k}{a(t')}$ and $t_{\rm min}$ is the start of the scaling regime. \footnote{Note that in our notation $k$ is a co-moving momentum,  while in \cite{Gorghetto_2018} it denotes the physical momentum, and hence the absence of the scale factor $a(t)$ in the our definition of $k'$.}. Defining $x=\frac{k}{a H}$ and $y = \frac{m_{r}}{H}$, we can parameterize the spectrum as:
\begin{align}\label{eqn: 5.7}
    \begin{split}
    F[x,y] & = \frac{\mathcal{N}}{x^{q}}, \hspace{0.3cm} x_{_{\rm IR}} < x < y \\ &
    = 0, \hspace{0.3cm} \text{otherwise}.
    \end{split}
\end{align}
$x_{_{\rm IR}}$ is an infrared cut-off of the order $x_{\rm IR} \simeq 10$, $y = \frac{m_{r}}{H}$ is the ultra-violet cut-off and $\mathcal{N}$ is a constant that ensures the spectrum is normalized according to $\int dx F[x,y] = 1$. When $q>1$, the spectrum is IR dominated while when $q<1$ it is UV dominated. The form of the spectral index that we use is \cite{Gorghetto:2020qws}:

\begin{equation}
\label{eqn:q}
    q(a) = 0.51 + 0.053 \log \left( \frac{f_{a}}{H(a)} \right).
\end{equation}

It is convenient to express $\partial \rho_{\phi}/\partial k$ in terms of the critical energy density per unit log frequency: 
\begin{equation}
\label{eqn:Omega def}
    \Omega_\phi (k,a) \equiv \frac{1}{\rho_{c}(a)} \frac{d \rho_\phi}{d \log k},
\end{equation}
where the critical energy density $\rho_{c}(a) \equiv 3 M^{2}_{\rm pl} H^{2}(a)$ is defined as a function of scale factor and $M_{\textrm{pl}} \simeq 2.4 \times 10^{18}~ \textrm{GeV}$ is the reduced Planck mass.

Using the definition \Eq{eqn:Omega def}, \Eq{eqn: 5.4}, \Eq{eqn: 5.5} and the definition of the Hubble constant to express the integral in terms of the scale factor, we get:
\begin{equation}
    \Omega_{\phi} (k,a) = \frac{8 k}{3 a^{4} M^{2}_{\rm pl} \rho_{c}} \int^{a}_{a_{\rm min}} da' \frac{(a')^{2} \xi' \mu'_{\rm eff} \rho'_{\rm SM}}{H'} F\left[ \frac{k'}{H'}, \frac{f_{a}}{H'} \right].
\label{eqn:Omega}\end{equation}
Integrating over the wavenumbers $k$, we obtain the relativistic energy density at time $a$:

\begin{equation}
    \rho_{\phi}(a) = \frac{8}{3 M^{2}_{\rm pl} a^{4}} \int^{a}_{a_{\rm min}} da' (a')^{3} \xi' \mu'_{\rm eff} \rho'_{\rm SM}.
\label{eqn:relativistic energy density}\end{equation}

\section{The Matter-Power Spectrum}

In this section, we build on the formalism of \Cref{sec:scaling regime,sec:field correlations}  to construct the density power spectrum of the non-relativistic Nambu-Goldstone bosons that contribute to the dark matter density at late times. 

\label{sec:density correlations}
\subsection{Density-Density Correlations}
The density is quadratic in $\phi$, which we treat as a Gaussian random field. Consequently, its two-point correlation function, $\langle \rho(x)\rho(x')\rangle$, can be reduced using Wick’s theorem to a sum of products of scalar two-point functions~\cite{Boddy:2025oxn}. This decomposition naturally separates contributions into ``fast'' modes, oscillating at frequencies $\sim 2m$, and ``slow'' modes, with frequencies set by the particle kinetic energies. In the non-relativistic limit, the fast-mode contribution is suppressed by $v^{2}$ relative to the slow component. Since we are interested in time-averaged observables, we retain only the slow-mode contribution in what follows.

A key timescale which sets the physics of ultralight dark matter density perturbations is when the bulk of the boson population begins to evolve non-relativistically, $a_{\rm NR}$. As explained shortly, the dark matter evolution after this point is well approximated by that of CDM and can be captured by the standard CDM growth factor $D(k,a)$. If the mass of the boson does not reach its constant value before radiation-matter equality, the growth factor will differ from that of CDM. We restrict our attention to the parameter space where this does not occur. As such, we now focus on evaluating the spectrum when the dominant component of the pseudo Nambu Goldstone bosons becomes non-relativistic, a condition given by the equation:
\begin{equation}
\label{eqn:tnr}
    m(a_{\rm NR}) a_{\rm NR} = k_\star,
\end{equation}

\begin{figure}[t!]
    \centering
    \begin{tikzpicture}
    \draw [-latex] (0,0)  coordinate (O) --++(8,0) coordinate (X) ;
    \foreach \x/\y/\z in {
    0.1/$a_m$/A,
    0.35/$a_{\rm NR}$/B,
    0.65/$a_{\rm eq}$/C,
    0.9/$a_0$/F
    }
    {\draw ($(O)!\x!(X)+(0.,-0.1)$) node[below] {\large \y} coordinate(\z) --++ (0,0.2);}
    \draw[c1,thick, decorate,decoration={brace,amplitude=5}] ($(A)+(0,0.3) $) --($(B)+(0,0.3) $) node[midway,yshift=0.4cm]{\normalsize relativistic}; 
    \draw[c2,thick, decorate,decoration={brace,amplitude=5}] ($(B)+(0,0.3) $) --($(O)!0.9!(X)+(0.,0.2) $) node[midway,yshift=0.4cm]{\normalsize non-relativistic}; 
    \end{tikzpicture}
    \caption[Timeline of the evolution of particles emitted by comsic strings]{A sketch of the important instances in the problem. From $a_{m}$, the collapse of the string network, up to time $a_{\rm NR}$, the particles evolve relativistically and afterwards, they contribute to the dark matter density. These events take place before $a_{\rm eq}$, the time of radiation-matter equality.}
    \label{fig:timeline}
\end{figure}
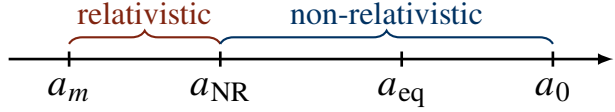

In the case of a temperature-independent mass $m (a_{\rm NR}) = m_a$. This relation is fixed by demanding the characteristic physical momentum of the particles at time $a_{\rm NR}$ is equal to the particles' mass $m (a_{\rm NR})$. For later times, the physical momentum is always smaller than the mass and the particles evolve non-relativistically. The key time scales are summarized in \Fig{fig:timeline}. From $a_{m}$, the collapse of the string network, up to time $a_{\rm NR}$, the particles evolve relativistically. Afterwards, they contribute to the dark matter density. These events take place before $a_{\rm eq}$, the time of radiation-matter equality.

We now evaluate the power spectrum at $a_{\rm NR}$. In the non-relativistic limit, we may set in the $\omega_{\bf k} \simeq m(a_{\rm NR}) \equiv m _{\rm NR} $ and obtain: 
\begin{align}
\begin{split}
      \langle \rho_{\phi}(\textbf{x}) \rho_{\phi}(\textbf{x}') \rangle =  \frac{m^{2}_{\rm NR}}{a^{6}_{\rm NR} } \int \frac{d^{3}k d^{3}k'}{(2 \pi)^{6}}   f(\textbf{k}) f(\textbf{k}')  \cos \left( \Delta \textbf{x} \cdot \Delta \textbf{k}   \right),
\end{split}
\label{eqn: rhorho master}
\end{align}
where $\Delta {\bf k} \equiv {\bf k}' - {\bf k }$. A generalized expression of the two-point function, valid outside of the highly non-relativistic limit, is provided in Appendix \ref{app:density correl}. 

The angular integrals can be performed explicitly, assuming that the phase space density is isotropic, and the result it:
\begin{equation}
    \langle \rho_{\phi}(\textbf{x}) \rho_{\phi}(\textbf{x}') \rangle = \frac{m^{2}_{\rm NR}}{a^{6}_{\rm NR} \pi^{4}} \left[ \int_0^\infty d k k^{2} f(k) \sinc(k |\Delta \textbf{x}|) \right]^{2},
\label{eqn: rhorho strings}\end{equation}
where $\textrm{sinc}(x)  \equiv \textrm{sin}(x)/x$.

Through the evolution of the relativistic density $\rho_{\phi}$ for scale factors $a_{m} < a < a_{\rm NR}$:
\begin{equation}
\label{eqn:density rela}
    \rho_{\phi}(a) = \frac{1}{a^{4}(t)} \int \frac{dk k^{3}}{2 \pi^{2}} f(k),
\end{equation}
Using \Eq{eqn:Omega def}, we can relate the phase space density $f(k)$ and the energy density per unit log momentum $\Omega_{\phi}(k)$. At the point of string network collapse (labeled by scale factor $a_m$):
\begin{align}
\begin{split}
    f(k) & =  \frac{2 \pi^{2}}{k^{4}} \rho_{c}(a_{m}) a^{4}_{m} \Omega_{\phi}(k,a_m)\,.
    \label{eqn:f and Omega}
\end{split}
\end{align}
Plugging \Eq{eqn:f and Omega} into \Eq{eqn: rhorho strings} and using that $\rho_c\propto a^{-4}$, we obtain: 
\begin{align}
\begin{split}
     &\langle \rho_{\phi}(\textbf{x}) \rho_{\phi}(\textbf{x}') \rangle \\ &\hspace{0.5cm}=  4 m^{2}_{\rm NR}a^{2}_{\rm NR} \rho^{2}_{c} (a_{\rm NR}) \bigg[\int \frac{d k}{k^{2}} \Omega_{\phi} (k) \sinc (k |\Delta \textbf{x}|) \bigg]^{2}.
\label{eqn:rho correl cosmic strings}
\end{split}
\end{align} 


\begin{figure*}[]
\includegraphics[width=14cm]{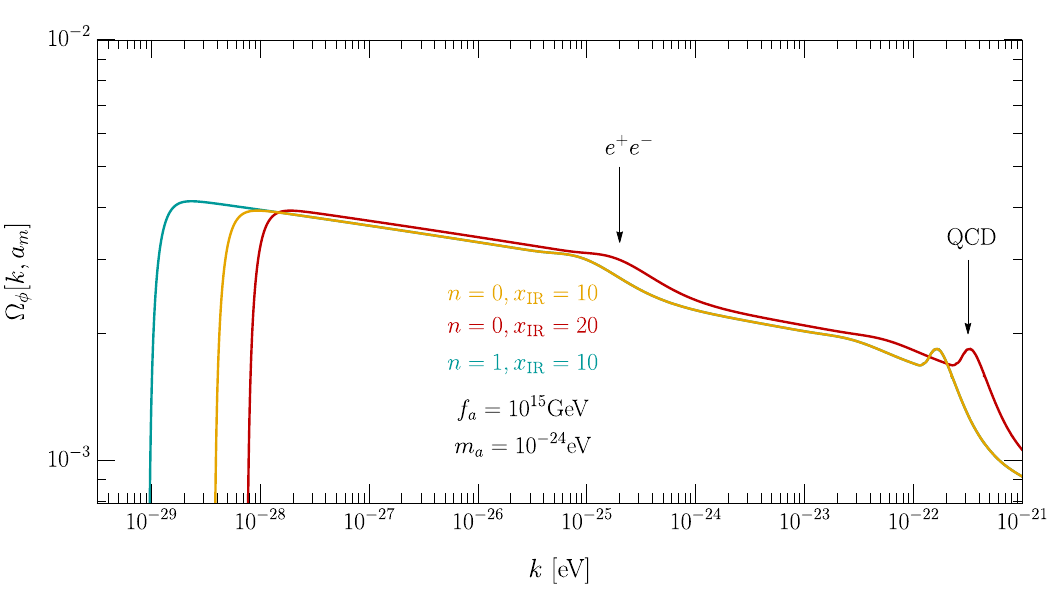}
    \caption[The spectrum $\Omega_{\phi}$ versus the frequency $k$.]{The spectrum $\Omega_{\phi}$ versus the frequency $k$, evaluated at $a_{m}$, for $m_{a}=10^{-24} \rm eV$, $f_{a} = 10^{15} \rm GeV$ and characteristic choices of $x_{\rm IR}$ and $n=0$. The yellow curve corresponds to $n=0, x_{\rm IR}=10$, the red curve to $n=0, x_{\rm IR} = 20$ and the cyan curve to $n=1, x_{\rm IR} = 10$. The $e^{+} e^{-}$ and $\rm QCD$ bumps are due to the changing relativistic degrees of freedom $g_{\ast}$ during the electron positron annihilation and the $\rm QCD$ phase transition respectively.}
    \label{fig:Omegak}
\end{figure*}

The spectrum of particles produced by cosmic string decay can be written as an integral over the string emission spectrum weighted by the string energy density. Because strings continuously radiate into Goldstone modes, their energy density approximately tracks the radiation energy density, leading to the so-called ``scaling solution.'' The emission spectrum is typically modeled as a power law, with a spectral index inferred from simulations (see Refs.~\cite{Sikivie_2008,Gorghetto_2018,Battye:2026whd,Dine:2020pds,Saikawa:2024bta} for reviews of the ongoing debate regarding this exponent). Assuming the string network follows a scaling solution, up to logarithmic corrections, allows the resulting particle spectrum to be evaluated numerically~\cite{Gorghetto_2018,Gorghetto:2020qws,Gorghetto:2022ikz,Gorghetto:2025uls,Benabou:2024msj,Battye:2026whd,Kaltschmidt:2025nkz,Saikawa:2024bta,Benabou:2023ghl}

An important characteristic of the spectrum is its infrared cut-off. This cut-off reflects the physical fact that cosmic strings longer than the horizon scale cannot efficiently radiate; instead, they reconnect and fragment into smaller loops. Consequently, the emitted radiation must be suppressed below a scale set by the inverse horizon size at a given time. Strings shorter than the horizon, on the other hand, decay by radiating Nambu--Goldstone bosons, producing a spectrum that peaks at
\begin{equation}
\label{eqn:kir}
    k_{\rm IR} = a_m H(a_m)\, x_{\rm IR},
\end{equation}
with $x_{\rm IR} \sim \mathcal{O}(10)$ \cite{Gorghetto_2018}. 

For temperature-independent masses, this cut-off is independent of $f_a$, while for $n\neq 0$ it shifts to smaller energies, as discussed in Appendix \ref{section:Cut-off}. The shape of the spectrum at larger energies depends on assumptions about the string decay mechanism. In what follows, we adopt the spectral index reported in \cite{Gorghetto_2018} and given in \Eq{eqn:q}, which yields a spectrum peaked at $k_{\rm IR}$ at late times.

We show the spectrum $\Omega_{\phi}$ in \Fig{fig:Omegak}, evaluated at $a_m$ for $m_a = 10^{-24}~\mathrm{eV}$ and $f_a = 10^{15}~\mathrm{GeV}$. For a temperature-independent mass, we display results for $x_{\rm IR} = 10$ (yellow) and $x_{\rm IR} = 20$ (red). We also show the case of a temperature-dependent mass with $n = 1$ and $x_{\rm IR} = 20$ (blue). In all cases, the infrared cut-off $k_{\rm IR}$ corresponds to the minimum value of $k$ for which $\Omega_\phi$ has support. The characteristic scale $k_\star$ denotes the peak of the spectrum, which appears at energies slightly above $k_{\rm IR}$ and is determined numerically in the following section. 

For fixed $x_{\rm IR}$, the spectra corresponding to different values of $n$ coincide at energies $k \gg k_{\rm IR}$. This behavior reflects the fact that high-energy modes are insensitive to the infrared cut-off, and in turn the temperature dependence scaling of the mass. 
Increasing $x_{\rm IR}$ shifts the entire spectrum to larger energies; spectra with different $x_{\rm IR}$ are related by the rescaling $k \rightarrow x_{\rm IR} k$.

The features (``bumps'') visible in \Fig{fig:Omegak} arise from changes in the relativistic degrees of freedom $g_\ast$ during the expansion of the Universe. To identify the epoch corresponding to a given feature, we solve $k/(aH(a)) = x_{\rm IR}$ for $a$ at fixed $k$. Assuming radiation domination, this yields $a \propto T_{\rm eq}^2 a_{\rm eq}^2 x_{\rm IR}/(k M_{\rm pl})$. For $k \sim 10^{-22}~\mathrm{eV}$ and $k \sim10^{-25}~\mathrm{eV}$, as seen in \Fig{fig:Omegak}, we find that particles with these characteristic energies were first emitted around the time of the QCD phase transition and the epoch of $e^+ e^-$ annihilation, respectively.

\subsection{The Power Spectrum}
With the formalism to calculate $\Omega_{\phi}(k)$ established, we now compute the density correlations in \Eq{eqn:rho correl cosmic strings} and the associated power spectrum, defined through the Fourier transform:
\begin{equation}
    P_{\phi}(k,a_{\rm NR}) = \int d^{3}x \, e^{-i \mathbf{k} \cdot \mathbf{x}}
    \frac{\langle \rho_{\phi}(\mathbf{x}) \rho_{\phi}(0) \rangle}
    {\bar{\rho}_{\rm DM}(a_{\rm NR})^{2}} .
\end{equation}
This quantity is evaluated at the time when the particles become non-relativistic and begin contributing to the dark matter density. Performing the angular integral yields:
\begin{align}
\begin{split}
    & P_{\phi}(k,a_{\rm NR}) =  16 \pi m^{2}_{\rm NR} a^{2}_{\rm NR} \left( \frac{ a^{3}_{\rm NR} \rho_{c}(a_{\rm NR})}{\bar{\rho}_{\rm DM}(a_{0})} \right)^{2}  \\ &  \times \int^{\infty}_{0} dx x^{2} \sinc(k x) \left[ \int \frac{d k} {k^{2}} \Omega_{\phi} (k)\sinc \left( k x \right)  \right]^{2}. 
\end{split}
\label{eqn:power spec strings}\end{align}
Using the identity
\begin{align*}
   & \int^{\infty}_{0} d x x^{2} \sinc(k x) \sinc (\ell x) \sinc(p x)  \\ & \hspace{0.5cm}=\frac{\pi}{4 k p \ell}  
   \theta(p - |k - \ell|) \theta(k + \ell-p).
\end{align*}
we may perform the remaining integral over the spatial separation $x$ leading to the power spectrum
\begin{align}
\begin{split}
     P_{\phi}(k,a_{\rm NR}) = 8 \pi^{2} \left( \frac{\bar{\rho}_{\phi}(a_{0})}{\bar{\rho}_{\rm DM}(a_{0})} \right)^{2} k_\star^{-3} \mathcal{T}\left( k \right),
\label{eqn: power specrum numerical}\end{split}
\end{align}
where the transfer function $\mathcal{T}(k)$ is
\begin{align}
\begin{split}
    \mathcal{T}\left( k \right) = & \frac{1}{2 k \mathcal{N}} \int^{\infty}_{k_{\rm IR}} \frac{dp}{p^{3}} \Omega_{\phi}(p)  \int^{k + p}_{|k - p|} \frac{d \ell}{\ell^{3}} \Omega_{\phi}(\ell)\,.
\end{split}
\label{eqn:transfer numerical}
\end{align}

Note that technically, we should not integrate over momenta for which the pseudo Nambu Goldstone bosons are relativistic. This imposes a condition on the integration limits set by the particle mass. Since the integrals are dominated by modes near $k_{\rm IR}$, and the relativistic cutoff lies parametrically above this scale for the masses considered here, we extend the upper limit to infinity with negligible error. The normalization of the transfer function $\mathcal{N}$ is chosen such that $\mathcal{T}(0) = 1$, fixing:
\begin{equation} 
    \mathcal{N} = \int^{\infty}_{k_{\rm IR}} \frac{dp}{p^{6}} \Omega_{\phi} (p)^{2},
\label{eqn:normalization}
\end{equation}
The relic energy density $\bar{\rho}_{\phi}$ is also determined as an integral over the spectrum:
\begin{align}
    \bar{\rho}_{\phi}(a_{0}) & = \int^{\infty}_{k_{\rm IR}} \frac{dk}{2 \pi^{2}} k^{2} m_{\rm NR} f\left(k\right)  \\ &  =  a^{4}_{\rm NR} \rho_{c}(a_{\rm NR}) m_{\rm NR}\int^{\infty}_{k_{\rm IR}} \frac{dk}{k^{2}} \Omega_{\phi}(k),
\label{eqn:relic density}
\end{align}
where we employ the convention that $a_{0}=1$. The characteristic momentum $k_\star$ is determined by the ratio between $\mathcal{N}$ and $(\bar{\rho}_{\phi}(a_{0})/a^{4}_{\rm NR} \rho_{c}(a_{\rm NR}) m_{\rm NR})^{2}$:
\begin{equation}
    k^{-3}_\star \equiv \left[ \int^{\infty}_{k_{\rm IR}} \frac{d \ell}{\ell^{2}} \Omega_{\phi}(\ell) \right]^{-2}\displaystyle \int^{\infty}_{k_{\rm IR}} \frac{dp}{p^{6}} \Omega^{2}_{\phi}(p)\,.
    \label{eq:kstar}
\end{equation} 
The characteristic momentum is largely set by the IR cutoff. In particular, we found numerically, that $ k_\star \simeq 2.1 k_{\rm IR}$ up to $1 \%$ accuracy for $10^{-27} ~\textrm{eV} \leq m_{a} \leq 10^{-18} ~\textrm{eV}$, $10^{14}~ \textrm{GeV} \leq f_{a} \leq 10^{15} ~\textrm{GeV}$ and $n=0,1$.

With these definitions, \Eq{eqn: power specrum numerical} is of the same form as \Eq{eqn:general power spectrum}, up to accounting for evolution between $a_{\rm NR}$ and $a_0$. The structure of the integrals in \Eq{eqn:transfer numerical} also matches the result of Ref.~\cite{Chathirathas:2025aan}.

In the literature, typically the dimensionless power spectrum is displayed $\Delta^{2}_{\phi}(k) = k^{3} P_{\phi}(k)/2 \pi^{2}$. Note that for $k \rightarrow 0$, we obtain from \Cref{eqn: power specrum numerical}:
\begin{equation}
    \Delta^{2}_{\phi}(k) = 4 \left(\frac{\bar{\rho}_{\phi}(a_{0})}{\bar{\rho}_{\rm DM}} \right)^{2} \left( \frac{k}{k_\star}\right)^{3}.
\label{eqn:dimensionless}\end{equation}
This matches exactly with the results in Refs.~\cite{Gorghetto:2021fsn,Gorghetto:2025uls,Chathirathas:2025aan}.

To obtain the present-day spectrum we must also include the growth of perturbations after radiation–matter equality. This is encoded in the growth factor $D(k,a)$~\cite{Amin:2025ayf,Amin:2025sla,Gorghetto:2025uls}:~\footnote{Note that Ref.~\cite{Gorghetto:2025uls} calls this a transfer function, a name we have reserved in this dissertation for $\mathcal{T}(k)$.}
\begin{equation}
\label{eqn:growth factor}
    D(k,a) \simeq \begin{cases}
        \displaystyle 1+\frac{a}{a_{\rm eq}} & k < k_{\rm J}(a_{\rm eq}) \\ \displaystyle \sqrt{ 1+ \frac{k_{\rm J}(a_{\rm eq})}{k} \left( \frac{a}{a_{\rm eq}} \right)^{2} } & k> k_{\rm J} (a_{\rm eq})
    \end{cases}
\end{equation}
The Jeans wavenumber is 
\begin{equation}
    \label{eqn:Jeans wavenumber}
    k_{\rm J}(a_{\rm eq}) = \sqrt{ \frac{a_{\rm eq}\rho_{\rm DM}(a_{0})}{2}} \frac{m_{a}}{M_{\rm pl}} \frac{1}{k_\star}\,.
\end{equation}
Comparing to $k_\star \simeq 2 k_{\rm IR}$,
\begin{align}
    \frac{k_{\rm J}(a_{\rm eq})}{k_\star} & = \left(\frac{1}{x_{\rm IR}} \right)^{2} \sqrt{\frac{\rho_{\rm DM}(a_{0})}{2 T^{4}_{\rm eq} a^{3}_{\rm eq}}} \left( \frac{M_{\rm pl}}{f_{a}} \right)^{\frac{n}{n+2}}\,, \\ 
    & \simeq 3 \times 10^{-3} \left( \frac{10}{x_{\rm IR}} \right)^{2} \left( \frac{M_{\rm pl}}{f_{a}} \right)^{\frac{n}{n+2}} \,.
\end{align}
Thus the Jeans scale is typically smaller than $k_\star$, mildly suppressing power for $k\lesssim k_\star$. A temperature-dependent mass brings the two scales closer, reducing this suppression.

\begin{figure*}[]
    \centering
    \includegraphics[width=14cm]{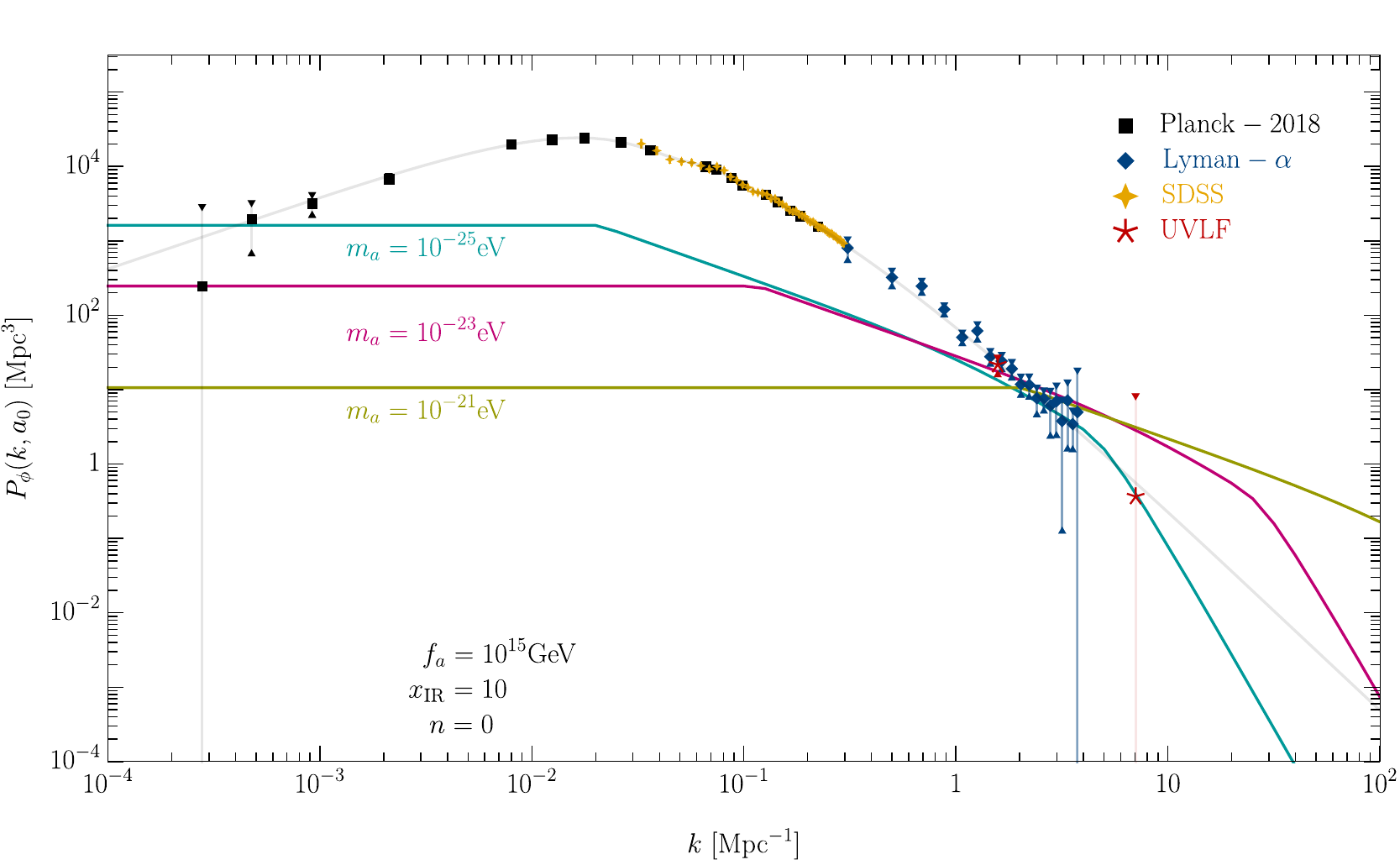}
    \caption[The density power spectrum]{The power spectrum of pseudo-Goldstone bosons emitted by cosmic strings for $f_{a} = 10^{15} \rm GeV$, $n=0$ and $x_{\rm IR} = 10$. The three different curves correspond to $m_{a} = 10^{-25} \rm eV$ (cyan), $m_{a} = 10^{-23} \rm eV$ (purple) and $m_{a} = 10^{-21} \rm eV$ (olive green). The data of the observables were taken from $\rm Planck-2018$ (black) \cite{Planck:2018vyg}, $\textrm{Lyman} - \alpha$ (blue) \cite{eBOSS:2018qyj},  UVLF (red) \cite{Sabti_2022}  and SDSS R7-(yellow) \cite{2010MNRAS.404...60R}. The triangles represent the 68\% confidence intervals for each measurement.}
    \label{fig:power spectrum strings}
\end{figure*}
    
The present-day spectrum is
\begin{equation}
\label{eqn:power spectrum today}
    P_{\phi}(k,a_{0}) = D(k,a_{0})^{2} P_{\phi}(k,a_{\rm NR})
\end{equation}
In \Fig{fig:power spectrum strings} we show this spectrum for $n=0$ and $x_{\rm IR}=10$ for three benchmark masses within observational reach. 
The data of the observables were taken from $\rm Planck-2018$ (black) \cite{Planck:2018vyg}, $\textrm{Lyman} - \alpha$ (blue) \cite{eBOSS:2018qyj},  UVLF (red) \cite{Sabti_2022}  and SDSS R7-(yellow) \cite{2010MNRAS.404...60R}.
The discontinuity in the slope at small $k$ arises from $k_{\rm J}(a_{\rm eq})$, while a slope change at higher $k$ appears near $k\simeq k_\star$, as explained analytically in \Cref{sec:analytics}. For $n\neq0$, the main difference is an overall enhancement of the spectrum at fixed $(m_a,f_a)$, since $k_\star$ decreases when $n\neq0$ (see Appendix \ref{section:Cut-off}). Previous work in this direction cut off the power spectrum at $k_\star$~\cite{Gorghetto:2021fsn,Gorghetto:2025uls}. The utility of our approach is to utilize the $k \geq k_\star$ components of the spectrum. 

\subsection{Analytic Approximation}
\label{sec:analytics}
As is evident from \Fig{fig:Omegak}, $\Omega_{\phi}(k)$ varies slowly with $k$, which reflects the assumption that the string network reaches a near-scaling solution. In this sub-section, we obtain analytic expressions for the power spectrum by approximating $\Omega_\phi$ as independent of $k$ for $k>k_{\rm IR}$ and zero for $k<k_{\rm IR}$.
The integrals in \Eq{eqn:normalization}, \Eq{eq:kstar}, and \Eq{eqn:relic density} can now be easily performed, leading to $\mathcal{N} \simeq \Omega^{2}_{\phi}/5$, $k_\star \simeq 5^{1/3} k_{\rm IR}$, and~\footnote{For $n=0$, this relic-density scaling agrees with the results of Refs.~\cite{Gorghetto:2021fsn,Gorghetto:2025uls} after using
$a^{4}_{\rm NR}\rho_{c}(a_{\rm NR})
= a^{4}_{m}\rho_{c}(a_{m})$
and the parametric relations
$\Omega_{\phi}\propto f_a^{2}$,
$a_m\propto m^{-1/2}$,
$k_\star\propto \sqrt{m}\,x_{\rm IR}$,
and
$\rho_c(a_m)\propto m^{2}$.}
\begin{align}
    \bar{\rho}_{\phi}(a_{0}) & \simeq
    \frac{a^{4}_{\rm NR} m_{\rm NR} \rho_{c}(a_{\rm NR}) \Omega_{\phi}}{k_\star},
\end{align}

Evaluating the transfer function integral in \Eq{eqn:transfer numerical} requires keeping track of the integration limits but is otherwise straightforward. For $k/k_{\rm IR} \leq 2 $, we can approximate its form as:
\begin{align}
\label{eqn:transfer small k}
       \mathcal{T} (y) & \simeq 
       \frac{5}{8y^{5} (1+y)}
      \bigg[ y ^5 + y^4 - 2y^3+6y^2 \\ &+12 y (1-\log (1+y))- 12  \log(1+y) \bigg] \notag
\end{align}
where $y = k/k_{\rm IR}$. When $k/k_{\rm IR} > 2 $,
\begin{equation}
       \mathcal{T} (y)  \simeq 
  \frac{-5}{4y^{5}} \left[ \frac{12y \hspace{-0.05cm}-\hspace{-0.05cm} 8y^{3}}{y^{2}-1} \hspace{-0.075cm}+\hspace{-0.075cm} 3 \log \left[ 
\frac{y+1}{y-1}  \frac{2y+1}{2y-1} \frac{2y^2 + y - 1}{2y^2 - y - 1} 
\right]\right] 
\label{eqn:transfer large k}
\end{equation}
The change in functional form at $y=2$ arises due to the absolute value at the lower limit of the inner integral in \Eq{eqn:transfer numerical} and requires treating the two regimes independently. The transfer function itself is valid for any value of $n$; the dependence on $n$ enters implicitly through the cutoff scale $k_{\rm IR}$. 

In \Fig{fig:transfer functions}, we compare this transfer function with the numerical result of \Eq{eqn:transfer numerical}.The two results diverge by at most $30 \%$ for $10^{-2} \lesssim k/k_\star \lesssim 1$. 

For $n=0$, an estimate for the order of magnitude of the power spectrum in the limit $k \rightarrow 0$ is then given by:
\begin{equation}
    \label{eqn:estimate P}
    P_{\phi} \simeq 800~\textrm{Mpc}^{3} \left( \frac{f_{a}}{10^{15} \textrm{GeV}} \right)^{4} \left( \frac{10^{-25} \textrm{eV}}{m_{a}} \right)^{1/2} \left( \frac{10}{x_{\rm IR}} \right)^{5}
\end{equation}
This matches the numerical results displayed in \Fig{fig:power spectrum strings} up to an $\mathcal{O}(1)$ factor. 

\section{Cosmic String Sensitivity}
\label{sec:constraints}
To search for a cosmic string signal, we construct a Gaussian likelihood from the measured power spectrum values, comparing the background-only likelihood $\mathcal{L}_{\rm B}$ to the signal-plus-background likelihood $\mathcal{L}_{{\rm S}+{\rm B}}$, which includes a contribution from the isocurvature power spectrum from \Eq{eqn: power specrum numerical}, $P_{\phi}(k)$. Explicitly, we have: 
\begin{equation}
    \mathcal{L}_{{\rm S}+{\rm B}} \propto \textrm{exp}\left[- \sum_{i} \frac{(P_{\rm data}(k_{i}) - P_{\rm CDM}(k_{i}) - P_{\phi}(k_{i}))^{2}}{2 \sigma(k_{i})^{2}} \right],
\end{equation}
In this expression, $P_{\rm data}(k_i)$ denotes the observed matter power spectrum and $P_{\rm CDM}(k_i)$ the CDM prediction. The quantities $\sigma(k_i)$ are the experimental uncertainties. The background likelihood $\mathcal{L}_{\rm B}$ is obtained from $ {\cal L} _{ {\rm S} + {\rm B} } $ by setting $P_\phi(k)=0$.

For the CDM spectrum we adopt~\cite{Maggiore:2018sht}
\begin{equation}
    P_{\rm CDM}(k)
    = \frac{18\pi^2}{25}\,A_\Phi
    \left(\frac{a_{\rm eq}}{H_0^2\Omega_M}\right)^2
    k\,\mathcal{T}_{\rm CDM}^2(k)
    \left(\frac{k}{k_\ast}\right)^{n_s-1},
\label{eqn:CDM power spectrum}
\end{equation}
where $\mathcal{T}_{\rm CDM}(k)$ is the Bardeen--Bond--Kaiser--Szalay (BBKS) transfer function~\cite{Bardeen:1985tr}, $n_s=0.967$, $A_\Phi=0.95\times10^{-9}$, and $k_\ast=0.05~\mathrm{Mpc}^{-1}$~\cite{Planck:2018vyg}. In a more complete analysis, the CDM parameters would be varied jointly with $(m_a,f_a)$; however, previous studies indicate that the amplitude of the isocurvature spectrum is only weakly degenerate with these parameters~\cite{Gorghetto:2025uls}. Later, we estimate up to $\sim 30\%$ uncertainty in our limit on $ f _a $ when fixing them to their fiducial values.

Searching the matter power spectrum data for a signal, we do not find any significant evidence for a signal across any mass point. As such, we move to set constraints in the $(m_a,f_a)$ plane.  
The background consists of the adiabatic  CDM perturbations, while the signal corresponds to the ultralight dark-matter--induced isocurvature perturbations derived in this work. We exclude a given parameter point $\{m_a,f_a\}$ when~\cite{2020NatRP...2..245A}
\begin{equation}
   2\log \left( \frac{\mathcal{L}_{{\rm S}+{\rm B}}}{\mathcal{L}_{\rm B}}\right) <-2.71,
\end{equation}
which corresponds to a one-sided $95\%$ confidence-level exclusion.

\begin{figure*}[]
    \centering
    \includegraphics[width=15cm]{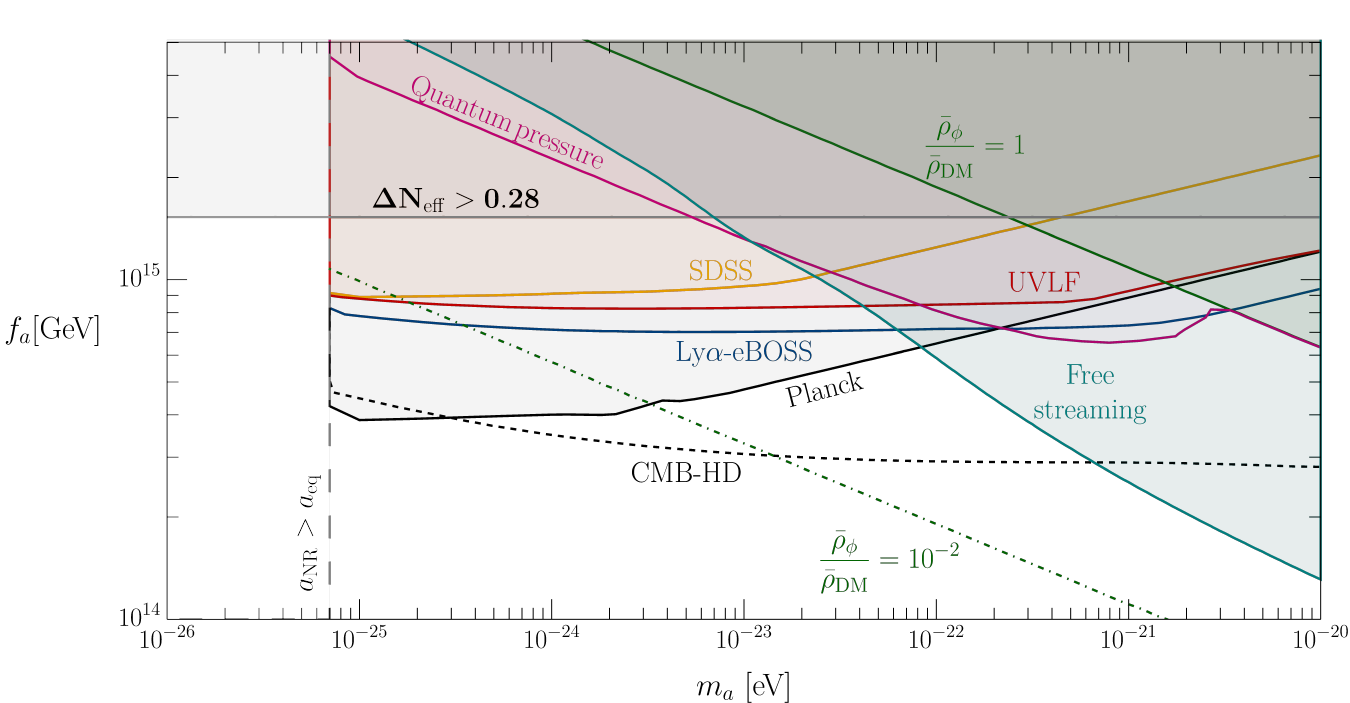}
    \caption[Constraints for $n=0$]{Constraints for cosmic strings in the parameter space of $m - f_{a}$ for $n=0$. The data of the observables were taken from $\rm Planck-2018$ (black) \cite{Planck:2018vyg}, $\textrm{Lyman} - \alpha$ (blue) \cite{eBOSS:2018qyj},  UVLF (red) \cite{Sabti_2022}  and SDSS R7-(yellow) \cite{2010MNRAS.404...60R}. We also include the projections for a future $\rm CMB-HD$ mission \cite{macinnis2025cmbhdprobedarkmatter}. In pink, we depict the constraints from the quantum pressure \cite{Kobayashi:2017jcf} and in cyan the constraints from free-streaming, both affecting the adiabatic component of the power spectrum. The green region is where dark matter is overproduced. The $\Delta N_{\rm eff}$ bound is due to excessive radiation density during BBN \cite{Consiglio_2018,Kirilova:2023rnl}. Finally, for $a_{\rm NR} > a_{\rm eq}$, the bosons are relativistic after equality and our analysis does not apply.}
    \label{fig:constraints cosmic strings}
\end{figure*}

In \Fig{fig:constraints cosmic strings} we present the resulting constraints for $n=0$, using four matter power spectrum data sets: the Planck 2018 results~\cite{Planck:2018vyg}, Lyman--$\alpha$ forest measurements~\cite{eBOSS:2018qyj}, UV galaxy luminosity function data~\cite{Sabti_2022}, and sloan digital sky survey (SDSS) DR7 measurements~\cite{2010MNRAS.404...60R}. 
We also show the projected sensitivity of a future CMB-HD lensing survey~\cite{macinnis2025cmbhdprobedarkmatter}. 
All data sets were compiled following Ref.~\cite{macinnis2025cmbhdprobedarkmatter}.

The constraints from the matter-power spectrum span a wider range of masses than those obtained in Refs.~\cite{Gorghetto:2021fsn,Gorghetto:2025uls} since these prior works focused on the influence of the white noise plateau on the matter power spectrum and assumed a zero value for smaller scales (larger wavenumbers). In particular, our constraints from the Planck data appear to be consistent with those presented in Fig. 4 of \cite{Buckley:2025zgh}. There, the isocurvature power spectrum was parameterized by an amplitude $A_{\rm iso}$ and a characteristic scale $k_\star$. Both were considered free parameters. The functional form of the power spectrum was considered as a broken power law, with the white noise plateau for $k < k_\star$ and a scaling $k^{-3}$ for $k>k_\star$. The constraints were depicted in the $\{ k_\star, A_{\rm iso} \}$ plane. They are particularly strong where $k_\star \leq 0.1 \textrm{Mpc}^{-1}$, as the CMB data are concentrated in this regime, while they scale as $k_\star^{-3}$ for larger wave-numbers. With the identifications $A_{\rm iso} \simeq \left( \bar{\rho}_{\phi} / \bar{\rho}_{\rm DM} \right)^{2}$ and $k_\star \simeq 3 \textrm{Mpc}^{-1} \left( m_{a} / 10^{-25} \textrm{eV} \right)^{1/2}$, we indeed find good agreement between the results presented in our \Fig{fig:constraints cosmic strings} and the results in \cite{Buckley:2025zgh}.
The flat regions of the constraints that are observed for the SDSS, Lyman\text{-}$\alpha$ and UVLF data occur because for these masses, the scales probed by these observables are within the white noise plateau of the isocurvature power spectrum. As we move to lower $m_{a}$, the characteristic scale is also reduced and it becomes increasingly important to consider the tail of the power spectrum for $k > k_\star$ in order to impose constraints from these observables. 
In \cite{Garcia-Gallego:2026phh,Kobayashi:2017jcf,Gorghetto:2025uls}, the Lyman$\text{-}$$\alpha$ constraints appear to be more stringent than the ones we present in this paper.That is because the underlying dataset that is used to impose these constraints originates from the MIKE/HIRES spectrometers \cite{Viel:2013fqw}, which yield higher resolution of individual sources and therefore a larger signal-to-noise ratio\cite{eBOSS:2018qyj}.These constraints are not relevant in the context of the present study, given that the white noise plateau was primarily used in those works, whereas we have attempted to include the high wavenumber tail in this analysis. Including the derivation of these constraints is distinct from the primary objectives of this paper and we defer this analysis to future work.

For large $m_{a}$, the power spectrum is reduced primarily because of the scaling $P_{\phi} \propto k^{-3}_\star \propto m^{-3/2}_{a}$ and the constraints are weakened. The kink that appears at $m_{a} = 4 \times 10^{-24} \rm eV$ in the curve of the Planck data is due to the discontinuity of our power spectrum at $k_{\rm J}(a_{\rm eq})$. This particular mass corresponds to $k_{\rm J}(a_{\rm eq}) \simeq 0.1~ \textrm{Mpc}^{-1}$, close to the maximum of the adiabatic power spectrum.

We also highlight that the increased sensitivity of a future $\textrm{CMB-HD}$ experiment can probe $f_{a}$ down to $f_{a} \simeq 4 \times10^{14} \textrm{GeV}$ for masses $ 3 \times 10^{-25} \textrm{eV} \lesssim m_{a} \lesssim 6 \times 10^{-22} \textrm{eV}$, well below the sensitivity of existing constraints.

The vertical dashed line at $m_{a} = 7 \times 10^{-26} \rm eV$ depicts the boundary beyond which the non-relativistic approximation that we have employed is not valid. Studying the bounds below this mass point requires treating the pseudo-Goldstone boson population relativistically as is beyond the scope of our analysis.

The upper portion of the plot is constrained by the contribution of relativistic Nambu--Goldstone bosons to the radiation density during Big Bang Nucleosynthesis, parameterized by $\Delta N_{\rm eff}$. Across the parameter space considered here, these particles remain relativistic at temperatures $T \sim \mathrm{MeV}$, so this bound is effectively independent of the mass~\cite{Consiglio_2018,Gorghetto:2021fsn}. The abundance of these relativistic particles depends only weakly on the onset of the scaling regime; in drawing this bound, we assume the scaling regime begins at a temperature $T \simeq f_a$.

For densities $\bar{\rho}_{\phi}(a_{0})/\bar{\rho}_{\rm DM}(a_{0}) \gtrsim
0.1$ the leading constraints come from quantum pressure and free-streaming. The wavenumber associated with quantum pressure is given by~\cite{Kobayashi:2017jcf,Hlozek:2014lca}:
\begin{equation}
    k_{\rm Q}(a_{\rm eq}) \simeq 7~\textrm{Mpc}^{-1} \left( \frac{m_{a}}{10^{-22} \textrm{eV}} \right)^{1/2}.
\end{equation}
The adiabatic spectrum is suppressed for wavenumbers larger than this one and this scale is relevant even if no velocity dispersion is assumed, in contrast to $k_{\rm J}$.  
The data to draw this bound were taken from \cite{Kobayashi:2017jcf}. 

The methodology to draw the free-streaming bound follows closely the work of Refs.~\cite{amin2024lowerbounddarkmatter,Long:2024cak,Liu:2024pjg} and is presented in Appendix \ref{sec:free streaming}. Since particles with finite velocity dispersion are unable to cluster, they suppress the matter power spectrum below the co-moving free-streaming scale \cite{Liu:2024pjg,amin2024lowerbounddarkmatter}: 
\begin{equation}
    \lambda_{\rm fs}(k,a) =  \int^{a}_{0} \frac{da'}{(a')^{2} H(a')} \frac{k}{\sqrt{k^{2} + (a')^{2} m^{2}(a')}}
\label{eqn:free streaming scale}
\end{equation}
The free-streaming suppression is encoded in a transfer function that multiplies the adiabatic spectrum and is proportional to the dark matter density fraction of the sub-component \cite{amin2024lowerbounddarkmatter,Gorghetto:2025uls}. In the case where the energy spectrum $\Omega_{\phi}(k)$ is peaked at $k_\star$, an approximate form of this transfer function is:

\begin{equation}
\label{eqn:transfer free approx}
    \mathcal{T}_{\rm fs}(k,a_{0}) \simeq \frac{\bar{\rho}_{\phi}(a_{0})}{\bar{\rho}_{\rm DM}(a_{0})} \sinc \left( k \lambda_{\rm fs}(k_\star,a_{0}) \right).
\end{equation}

To draw the bound, we used the Planck-2018 data. In our approach, we used the numerical results for the spectrum $\Omega_{\phi}(k)$ and the full expression for the free-streaming transfer function provided in Appendix \ref{sec:free streaming} and confirmed that \Eq{eqn:transfer free approx} produces the same constraints.

\begin{figure*}[]
    \centering
    \includegraphics[width=16cm]{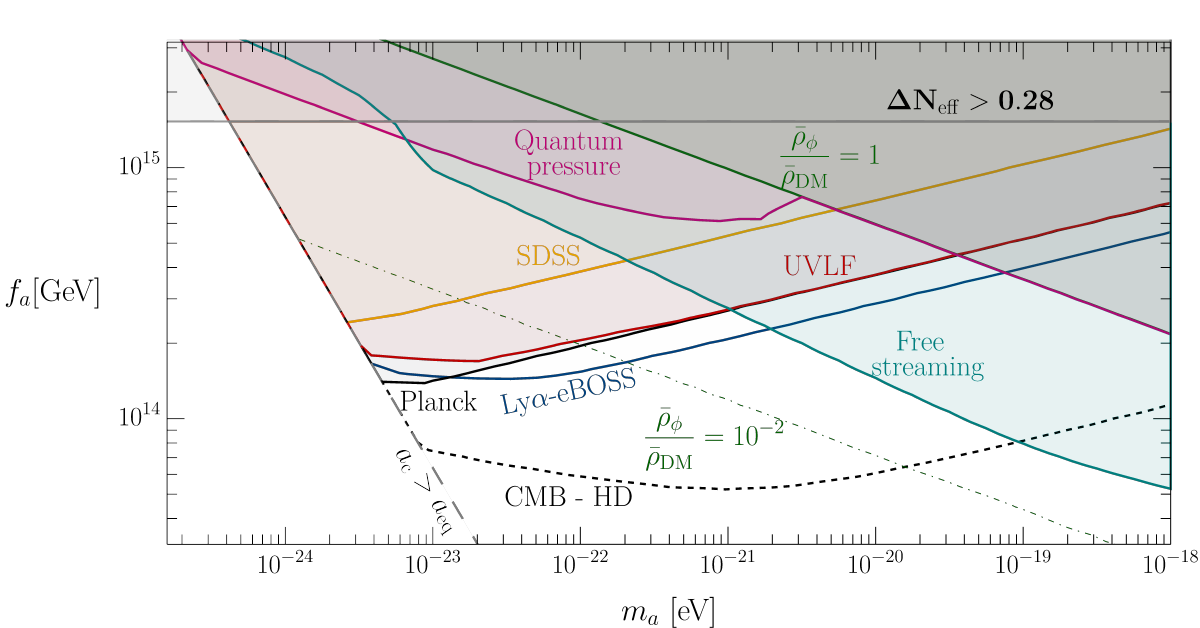}
    \caption[Constraints for $n=1$]{Constraints for cosmic strings in the parameter space of $m_{a} - f_{a}$ for $n=1$. The data of the observables were taken from $\rm Planck$ (black) \cite{Planck:2018vyg}, $\textrm{Lyman} \text{-} \alpha$ (blue) \cite{eBOSS:2018qyj},  UVLF (red) \cite{Sabti_2022}  and SDSS R7-(yellow) \cite{2010MNRAS.404...60R}. We also include the projections for a future $\rm CMB \text{-} HD$ mission \cite{macinnis2025cmbhdprobedarkmatter}.In pink, we depict the constraints from the quantum pressure \cite{Kobayashi:2017jcf} and in cyan the constraints from free-streaming, both affecting the adiabatic component of the power spectrum. The green region is where dark matter is overproduced. The $\Delta N_{\rm eff}$ bound is due to excessive radiation density during BBN \cite{Consiglio_2018,Kirilova:2023rnl}. For $  a_{\rm c} \equiv a (T_{\rm c})>a_{\rm eq}$ the mass obtains its zero temperature value after equality, where our analysis does not apply.}
    \label{fig:constraints cosmic strings n=1}
\end{figure*}

In \Fig{fig:constraints cosmic strings n=1}, we show the constraints discussed above for $n=1$. For $m_{a}$ and $f_{a}$ to the left of the line indicated by the label $a_{\rm c} > a_{\rm eq} $, the mass reaches its constant value after equality. In this regime, the growth factor $D_{\rm iso}(k,a)$ is modified, and our analysis is not directly applicable. The region where the relativistic evolution persists even after equality is relevant for even lower masses, not shown in this figure.

More generally, the power spectrum is affected indirectly through the oscillation time $a_m$. For fixed $m$ and $f_a$, $a_m$ increases—i.e., the onset of oscillations is delayed. This shifts the characteristic momentum $k_\star$ to smaller values, as discussed in Appendix \ref{section:Cut-off}. Because the power spectrum scales as $P_\phi \propto k_\star^{-3}$, the amplitude increases, leading to mildly stronger constraints for a fixed relic density.

%% file: chapter3.tex
\chapter{Gravitational Waves from Resonant Transitions of Tidally Perturbed Gravitational Atoms} \label{sec:ga}

In this Chapter, we derive the form of the gravitational wave signal that is induced by transitions of the gravitational atom in a black hole binary. It is based on the paper \cite{Kyriazis:2025fis} that was published in the Journal of High Energy Physics in collaboration with Dr. Fengwei Yang. A work on superradiance and the effect of the axion-electron interaction on the thermal spectrum of the accretion disk in an X-ray binary system was also recently posted in \cite{Kyriazis:2026vkz}. 

The Chapter is structured as follows: in \Cref{section:overview}, we review in brief some basic properties of the GA, the Landau-Zener transition of a two-state system, and the effects of the decay rate of the second mode. In \Cref{section:gr waves}, we present the formalism for the computation of the GW strain waveform and frequency spectrum produced during a given transition and compare our signal to that of the inspiral and of the annihilations of the GA. In \Cref{section:application}, we discuss the detectability of the GW signal from a GA-binary system and show the SNR for various model parameters.

\section{Gravitational Atoms in Isolation and in Binaries} \label{section:overview}
\subsection{The Gravitational Atom }
In this section, we give a brief overview of the GA, establishing also the conventions we will be using throughout. An important quantity is the ``fine-structure'' constant of a GA, defined as the ratio of the gravitational radius $r_{g}$ of the black hole to the Compton wavelength $\lambda_{c}$ of the boson:
\begin{equation}
    \alpha \equiv \frac{r_{g}}{\lambda_{c}} = \mu M,
\end{equation}
where we use the Planck unit throughout the paper, $G=c=\hbar = 1 $, $\mu$ is the boson mass and $M$ is the host black hole mass.  
In the case where the particle is non-relativistic we can approximate $\omega \simeq \mu$. The angular velocity of the Kerr black hole is given by $\Omega_{H} = \frac{\tilde{a}}{2 r_{+}}$, where $\tilde{a} = \frac{a}{M} \leq 1$ is its dimensionless spin and $r_{+} = M + \sqrt{M^{2} - a^{2}}$ is its outer horizon \cite{Carroll:2004st}. The condition of superradiance \Eq{eqn:superradiance condition} can be re-expressed as an inequality for $\alpha$:
\begin{equation}
\label{eq:supercondition}
    \alpha<\frac{m}{2} \frac{\tilde{a}}{1 + \sqrt{1 - \tilde{a}^{2}}}.
\end{equation}
Thus, \Eq{eq:supercondition} determines the upper bound of the fine-structure constant $\alpha$ given the spin of the black hole $\tilde{a}$ and the azimuthal quantum number $m$ of the GA.
For maximally spinning black holes, $\tilde{a}=1$ and the $m=1$ states can only grow for $\alpha<0.5$. The saturated spin of the black hole can be obtained by rearranging \Eq{eq:supercondition},
\begin{equation}
    \tilde{a}_{\textrm{crit}} = \frac{4 m \alpha}{m^{2} + 4 \alpha^{2}}.
\label{eqn:saturated}\end{equation}

The equation of motion of the non-relativistic scalar field produced by superradiance is described by a Schr\"{o}dinger-like equation in the limit $r \gg M$ and $\alpha \ll 1$ \cite{Detweiler:1980uk}:
\begin{equation}
\label{eq:Schrodinger_eq}
    i \partial_{t} \psi(t,\vec{x}) =  \left( - \frac{\nabla^{2}}{2 \mu} - \frac{\alpha}{r} \right) \psi(t,\vec{x}).
\end{equation}
The size of the cloud is characterized by its Bohr radius $r_{c} = \frac{M}{\alpha^{2}}$. The bound states are given by: 
\begin{equation}
    \psi_{nlm} (t,\vec{x}) = e^{-i (\omega_{nlm} - \mu) t} R_{nl}(r) Y_{lm} (\theta,\varphi),
\label{eqn:ansatz}\end{equation}
where $n,l,m$ are the principal, angular, and azimuthal quantum numbers respectively, $\omega_{nlm}$ is the eigenfrequency of the eigenstate $|nlm\rangle$, which is generally complex due to the purely incoming boundary conditions at the black hole's outer horizon, $R_{nl}(r)$ is the hydrogenic radial wavefunction, and $Y_{lm}(\theta,\varphi)$ are the spherical harmonics.  
This non-relativistic approach holds very well if $\alpha \lesssim 0.1$, while it starts to break down for larger values of $\alpha$ and numerical approaches are required to solve the equation of motion \cite{spectra}.\footnote{By numerically solving the full relativistic equation of motion, one can find that the relativistic wavefunctions and eigenenergies (the real part of the eigenfrequencies) align with the non-relativistic ones if $\alpha\lesssim0.3$ while the decay rate (the imaginary part of the eigenfrequencies) starts to deviate from the non-relativistic one when $\alpha\gtrsim0.1$ (see also Appendix \ref{sec:relativistic}).}
\begin{figure}
    \centering
    \includegraphics[width=0.8\linewidth]{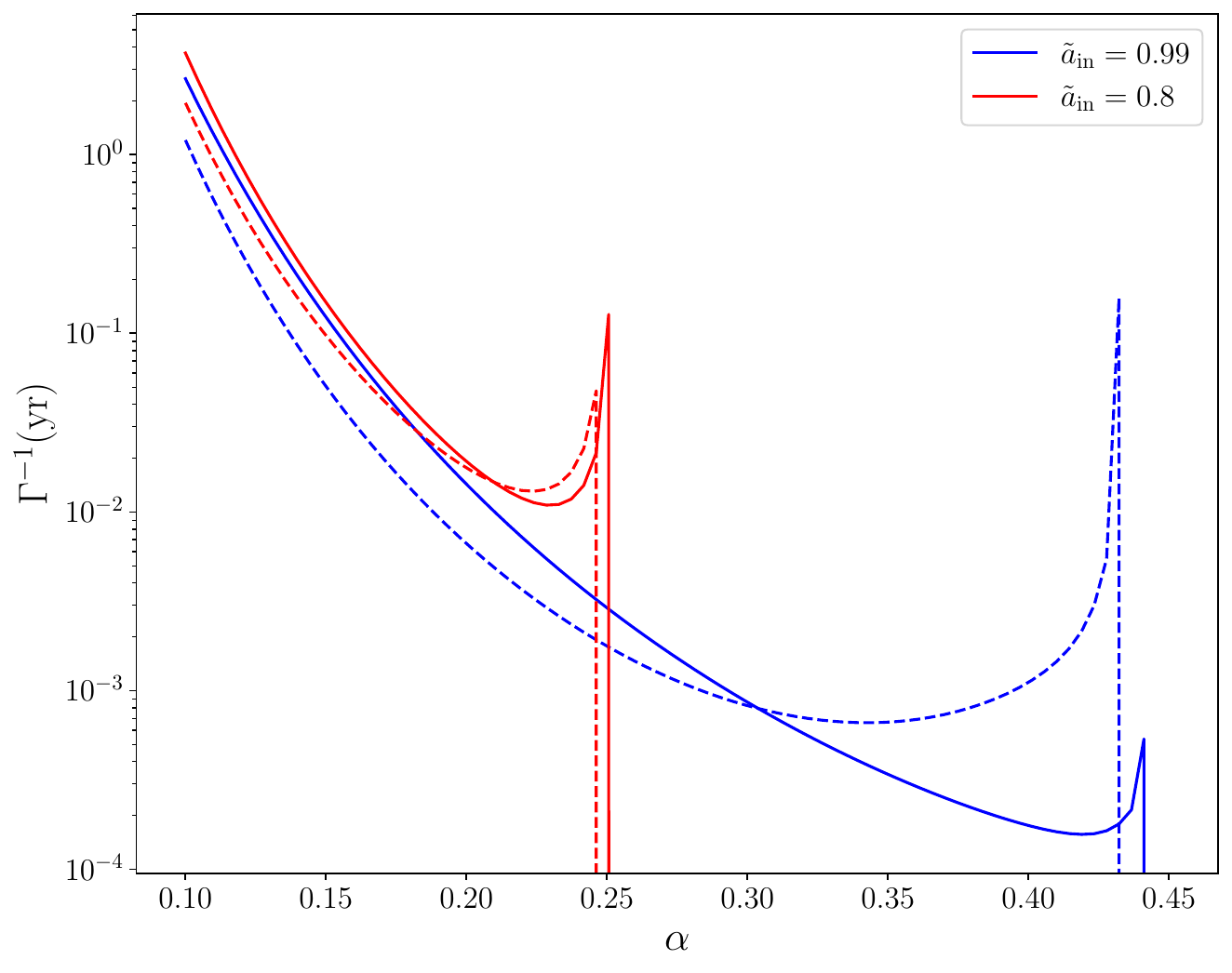}
    \caption[Superradiant time-scales for the $211$ state]{Superradiant timescale in years for the $|211\rangle$ state as a function of the $\alpha$ parameter.}
    \label{fig:Superrad rate}
\end{figure}
The real part of the eigenfrequencies gives the energy levels of the gravitational atom, while the imaginary part gives the instability rates \cite{Detweiler:1980uk,Arvanitaki_2011,spectra}:
\begin{align}
    \label{eqn:energy level}& \omega_{R,nlm} =  \mu \left( 1 - \frac{\alpha^{2}}{2n^{2}} - \frac{\alpha^{4}}{8n^{4}} + \left(\frac{2}{n} - \frac{6}{2l+1} \right) \frac{\alpha^{4}}{n^{3}} + \frac{16\tilde{a} m \alpha^{5}}{n^{3} 2l (2l+1)(2l+2)}  +\mathcal{O}(\alpha^6)\right), \\ 
    \label{eqn:sr rates}& 
    \Gamma_{nlm}\equiv\omega_{I,nlm} = \frac{2 r_{+}}{M} C_{nlm} (\alpha,\tilde{a}) (m \Omega_{H} - \mu) \alpha^{4l +5},
\end{align}
where $C_{nlm}$ are coefficients that can be readily found in \cite{spectra}.
The first two terms in \cref{eqn:energy level} are the same as those of the regular hydrogen atom. The fourth term is what will be relevant for the so-called fine transitions $(\Delta n =0, \Delta l \neq 0)$ while the fifth term, in which the black hole's spin breaks the degeneracy of the third quantum numbers $m$, is relevant for the hyperfine transitions $(\Delta n =0, \Delta l = 0, \Delta m \neq 0)$.
Since we are in the limit $\alpha \ll 1$ and $\Gamma$ strongly depends on $l$, \cref{eqn:sr rates} implies that the fastest growing superradiant levels are those with $l = m $, with the $l=m=1$ being the fastest. An estimate for the growth timescale of the fastest growing mode, around a maximally spinning black hole, is
\begin{equation}
    \Gamma^{-1}_{211} \simeq 0.3 \,\text{days} \left( \frac{M}{150 M_{\odot}} \right) \left( \frac{0.3}{\alpha}\right)^{9} \left( \frac{0.6}{1-2 \alpha}\right)^{3}.
\label{eqn:supperad growth timescale}\end{equation}

\Cref{fig:Superrad rate} shows the superradiant timescale for the $|211\rangle$ state for two different initial black hole spins, $\tilde{a}_{\rm in}=0.8,\,0.99$, comparing the relativistic result (solid lines) using the numerical code in \cite{hoof2024gettingblackholesuperradiance,hoof_git}, with the non-relativistic result (dashed lines) in \cref{eqn:sr rates}. The black hole mass is set to $150 M_{\odot}$. The results agree up to $\mathcal{O}(1)$ factors for $\alpha\lesssim 0.25$ with $\tilde{a}_{\rm in}=0.8$ and for $\alpha\lesssim 0.30$ with $\tilde{a}_{\rm in}=0.99$. In general, they diverge for large values of $\alpha$ and large spins. We also note that the smaller the black hole spin is, the smaller the range of $\alpha$ for which the superradiant condition is satisfied. Based on this result, we restrict the analysis to $\alpha<0.42$ in what follows.

\subsection{The perturbed Gravitational Atom}
In this section, we start to consider the GA in a binary system setup and summarize the results from the literature that are relevant to the later analysis. 
The companion in the binary system induces a tidal field $V_{\ast}(t,\vec{r})$ on the GA that will cause it to undergo transitions between states. We will assume that the companion is orbiting at a large distance from the host black hole so that we can treat the tidal field perturbatively. The exact form of the mixing between the states can be found in \cite{Baumann_2019,Baumann_2020,axion_cloud_backreaction} and is summarized in Appendix \ref{appendix: tidal} for reference. For our purposes, it is important to note that if the companion is far from the GA, the dominant multipole moment of the tidal field is the {\it quadrupole} moment, $l_*=2$. This leads to mixing between two states and the following selection rules are derived: 
\begin{align}
\label{eq:selection_rule}
    & m_{f} - m_{i} = m_{\ast}, \\
    & |l_{i} - l_{f}| \leq 2 \leq l_{i} + l_{f}, \\
    & l_{i} + l_{f} = 2p -2, p \in \mathbb{Z},
\end{align}
where $i,f$ represent the initial and final states respectively, and $m_*$ is the azimuthal number of the spherical harmonics of the tidal field quadrupole.
Throughout the paper, we make the following two assumptions to simplify the analysis and facilitate the analytical implementations: 
\begin{itemize}
    \item The orbit is quasi-circular. 
    \item The companion is on the same equatorial plane as the cloud.
\end{itemize}
With these assumptions, only states that satisfy $m_{\ast}= \pm 2 = m_{f} - m_{i}$ couple to each other as demanded by the selection rule, and the mixing between two different states takes a simple form\,\footnote{Relaxing the assumption regarding the orbit being on the GA's equatorial plane introduces higher harmonics in \Eq{eqn:matrix element}.}:
\begin{equation}
\langle \psi_{f} |V_{\ast} (t,\vec{r}) | \psi_{i} \rangle  = \eta e^{-i m_{\ast} \varphi_{\ast}},
\label{eqn:matrix element}\end{equation}
where $\eta$ denotes the amplitude of the mixing with its explicit expression given in \Eq{eqn:eta} and $\varphi_{\ast}$ is the phase of the binary, whose derivative is the frequency of the orbit, $\dot{\varphi_*}(t) = \pm \Omega(t)$
\cite{Baumann_2019,Baumann_2020}. The plus (minus) sign is for co(counter)-rotating orbits. The matrix element in \Eq{eqn:matrix element} is akin to a periodic, external driving force, and a resonance is expected to take place when the frequency of this force is equal to the energy splitting of the two states $|m_*|\Omega = \Delta \omega_{R}\equiv\omega_{R,i}-\omega_{R,f}$. \par

The orbit of the companion evolves very slowly during the transition\footnote{ Back-reaction effects affect the transition in two ways: 1. they shift the resonance frequency of the LZ transition \cite{axion_annihilation}; 2. they vary the change rate of the orbital frequency. For the $|211\rangle$ hyperfine transition we considered throughout the paper, the first effect renders the transition non-adiabatic for equal mass binaries, while we will show in \Cref{section:application} that our signal is mostly relevant for intermediate mass ratios. The second effect stalls the orbital frequency around the hyperfine frequency, increasing the duration of the resonance \cite{Baumann_2020}. When the decay rate is included, the resonance may break if the occupation number of the first state drops below a threshold \cite{resonant_history}. Even though back-reaction should be included systematically, we have checked that the resonant breaking effect is small for the $|211\rangle \rightarrow |21 \textrm{-}1\rangle$ transition, in the parameter space of interest.} due to the emission of GWs, so that the frequency can be linearized as a function of time $t$ \cite{Baumann_2020}:
\begin{equation}
    \Omega(t) =\Omega_{0} + \gamma t,
\label{eqn:linear freq}\end{equation}
where $\Omega_{0}$ is the reference frequency at $t=0$ and $\gamma$ is the change rate of the orbital frequency due to GW emission,
\begin{equation}
    \frac{\gamma}{\Omega^{2}_{0}} = \frac{96}{5} \frac{q}{(q+1)^{1/3}} (M \Omega_{0})^{5/3},
\label{eqn: gamma}\end{equation}
where $q$ is the mass ratio of the companion.
The linear approximation of the orbital frequency is valid for the time within $-\Omega_0/\gamma \lesssim t \lesssim \Omega_0/\gamma$.
The Hamiltonian that governs the evolution of the coefficients of the initial and final states of a two-state system, $c_{i}(t)$ and $c_{f} (t)$, that are mixed via the perturbing potential $V_*$, is given by
\begin{equation}
\mathcal{H} = 
    \begin{pmatrix}
 -\Delta \omega_{R}/2   &  \eta e^{i \Delta m \varphi_*(t)}  \\
    \eta e^{- i \Delta m \varphi_*(t)}  & \Delta \omega_{R}/2 - i |\Gamma|
\end{pmatrix},
\end{equation}
where $\Delta \omega_R = \omega_f-\omega_i$, 
$\Delta m = m_f-m_i$, and $\Gamma$ is the instability rate of the final state which characterizes the final-state decay due to black hole absorption (see \Eq{eqn:sr rates}).
By performing a unitary transformation of $\mathcal{H}$ to the co-rotating (or dressed) frame, one can eliminate the fast oscillations of the off-diagonal elements, and the coefficients of two states transform accordingly
\begin{equation}
\begin{pmatrix}
    c_{i}(t) \\
    c_{f}(t)
\end{pmatrix}
= 
    \begin{pmatrix}
          e^{i \Delta m \varphi_*(t)/2} &  0  \\
          0  & e^{- i \Delta m \varphi_*(t)/2} 
    \end{pmatrix}
    \begin{pmatrix}
    d_{i}(t) \\
    d_{f}(t)
\end{pmatrix}.
\label{eqn:cs}\end{equation}
Since it has been used widely in the literature, we provide the Hamiltonian in this dressed frame directly in Appendix \ref{appendix:Hamiltonian}. When setting the reference orbital frequency to match the energy splitting of the transition $\Omega_{0} = {\pm} \frac{\Delta \omega_{R}}{\Delta m}$ and applying the linear orbital frequency approximation, in \Eq{eqn:linear freq}, the Hamiltonian $\mathcal{H}$ is reduced to the well known Landau-Zener system \cite{Landau,Zener}. The main point is that a transition will take place at frequency $\Omega_{0}$. In this paper, we will primarily deal with hyperfine and fine transitions, whose orbital frequency can be found from \Eq{eqn:energy level},
\begin{align}  
    \label{eqn:Omega hyperfine}\Omega_{0,\textrm{hyp}}  
    &=\frac{64 m_i\alpha^{7}}{Mn^{3} 2l (2l+1)(2l+2)(m_i^2+4\alpha^2)}, 
    \\  
   \label{eqn:Omega fine} \Omega_{0,\textrm{fine}} &= \frac{\Delta l}{\Delta m}\frac{12 \alpha^{5}}{M n^{3} (2l_i+1)(2l_f+1)}.
\end{align}
where we assumed for hyperfine transitions that the black hole's spin is given by \Eq{eqn:saturated}, and the subscript of $l$ and $m$ represents the associated initial or final state.
An advantage of working with hyperfine and fine transitions is that they take place when the companion is far away from the GA, and so keeping only the quadrupole term in the tidal field is justified \cite{hyperfine_trans_are_favoured}. We will consider hyperfine transitions from the $|211\rangle$ state and fine transitions from the $|322\rangle$ state. The selection rules that we discussed above demand $|211\rangle\rightarrow |21\text{-}1\rangle$ and $|322\rangle \rightarrow |300\rangle$. Since $\Delta m = -2$ and $\Delta E <0$, only co-rotating orbits can induce these transitions, which we will consider throughout the paper. \par
In addition, the orbital frequency of these transitions happens to fall in the $\rm mHz - \rm Hz$ range, depending on the particular transition, black hole mass and $\alpha$. For example, the frequency for the $|211\rangle\rightarrow |21\text{-}1\rangle$ transition, $f_0\equiv\Omega_0/(2\pi)$:
\begin{equation}
   f_{0,211} \simeq 10 {\rm mHz}\,\left(\frac{\alpha}{0.3} \right)^7 \left(\frac{150 M_\odot}{M}\right) \left(\frac{1.36}{1+4 \alpha^{2}} \right).
\end{equation}
Similarly, for the fine transition $|322\rangle \rightarrow |300\rangle$, 
\begin{equation}
 f_{0,322} \simeq 46  \textrm{mHz} \left(  \frac{\alpha}{0.3}\right)^{5} \left( \frac{150 M_{\odot}}{M}\right).
\end{equation}

\par
An important parameter that characterizes the transition is the {\it adiabaticity} parameter:
\begin{equation}
    z \equiv \frac{\eta^{2}}{|\Delta m|\gamma}.
\label{eqn: adiabaticity}\end{equation}
For the $|211\rangle \rightarrow | 21\textrm{-}1 \rangle$ hyperfine transition, the typical value of $z$ is 
\begin{equation}
   z \simeq 0.7  \left( \frac{1.36}{1+4\alpha^{2}}\right)^{1/3} \left(\frac{q}{1/150}\right) \left( \frac{0.3}{\alpha}\right)^{11/3}.
\label{eqn:z hyperfine}
\end{equation}

\begin{figure*}[t!]
    \centering
    \begin{subfigure}[t]{0.8\textwidth}
        \centering
        \includegraphics[width=0.99\linewidth]{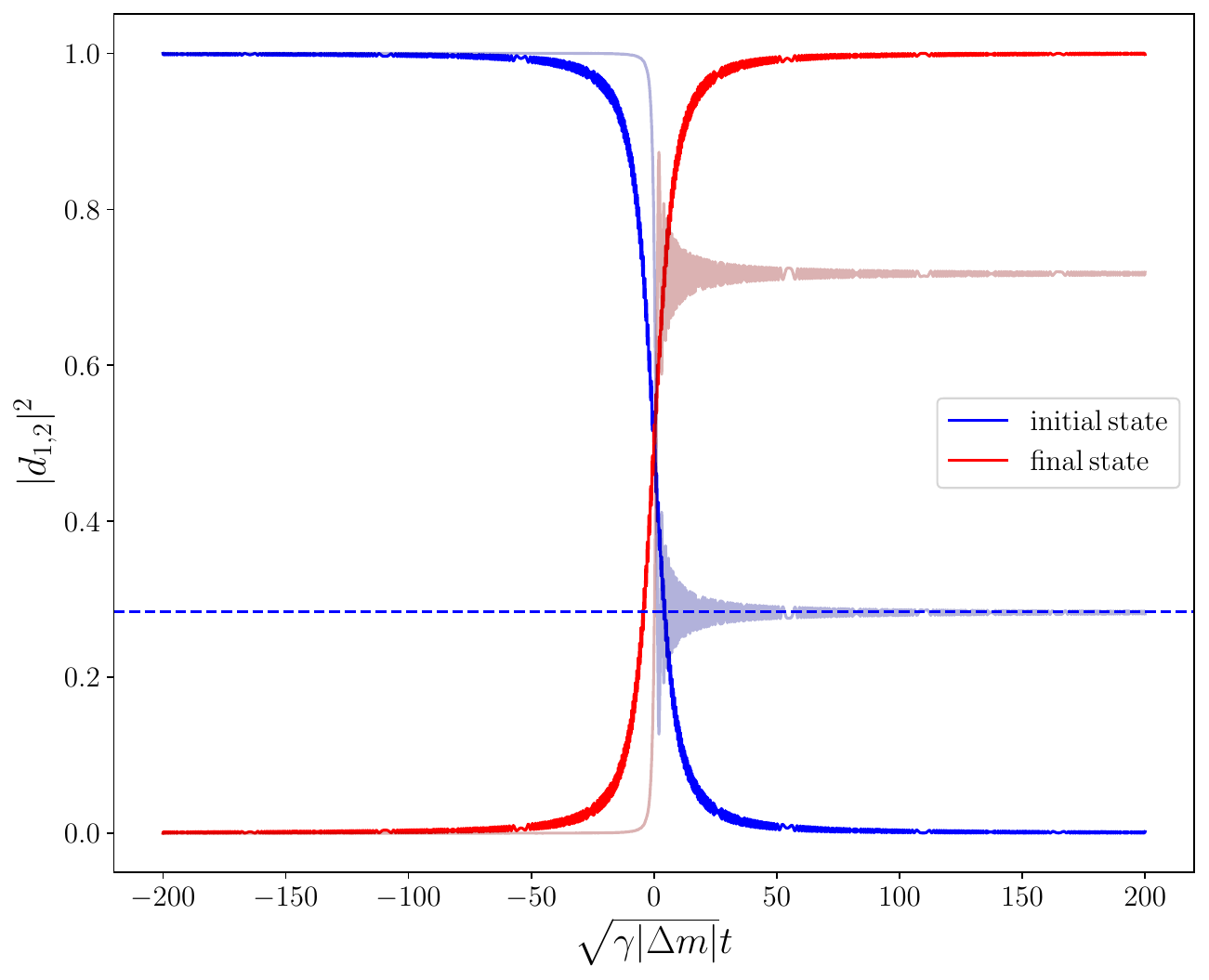}
    \end{subfigure}%
    ~
\caption[Evolution of the occupation numbers of the two states]{Evolution of the occupation numbers for adiabatic and non-adiabatic transitions for $z=20$ (bold) and $z=0.2$ (pale).}
   \label{fig:hyperfine}
\end{figure*}

The remaining population of the first state long after the transition is given by $|d_{1} (t \rightarrow \infty)|^{2} = e^{-2 \pi z}$. If the adiabaticity $z \gg 1$, the transition is adiabatic: the two states exchange populations after the transition. When $z \lesssim 1$, the transition is non-adiabatic and only a fraction of the bosons is transferred. The evolution of the states in these two cases is shown in \Fig{fig:hyperfine}. \par
\par
In the co-rotating frame, the Hamiltonian is transformed to a simpler form $\mathcal{\bar{H}}$ (see \Eq{eqn:Hamiltonian}) and the coefficients $d_1$ and $d_2$ evolve according to
\begin{equation}
\label{eq:eomOfd}
    i\frac{dd_a}{dt}=\sum_b \mathcal{\bar{H}}_{ab}(t)d_b,~~a,b=1,2.
\end{equation}
The solutions with the initial conditions $|d_{1} (t \rightarrow -\infty)|^{2} = 1 $ and $|d_{2} (t \rightarrow -\infty)|^{2} = 0$, ignoring the decay rate of the final state $\Gamma=0$, have an analytical form in terms of the parabolic cylinder function \cite{D},
\begin{align}
\label{eq:coefficient_d1}
    & d_{1}(t) = e^{- \frac{\pi z}{4}} D_{iz} (e^{\frac{3 i \pi}{4}} \sqrt{|\Delta m| \gamma} t), \\
    \label{eq:coefficient_d2}
    & d_{2}(t) = \sqrt{z} e^{-\frac{\pi z}{4}} D_{iz-1} (e^{\frac{3i\pi}{4}} \sqrt{|\Delta m| \gamma} t).
\end{align}

\subsection{The Decay Rate of the Final State}\label{sec: decay rate}
So far, we have ignored the decay of the final state into the black hole. However, if the mixing between the states is subdominant compared to the decay rate $\Gamma$, i.e. $\eta \ll |\Gamma|$, the evolution of the coefficients $d_{1}$ and $d_{2}$ will be significantly different from what we laid out above.  In \Eq{eqn:decay211} we evaluate the ratio of the mixing strength $\eta$ with the decay rate $|\Gamma|$ \cite{extreme_mass_ratio}. The latter is calculated assuming that the black hole is at the saturated spin of the corresponding initial state and we may approximate it as $|\Gamma_{21{\text-}1}| \simeq \alpha^{10}/(6 M)$, where we have used the analytical expressions that can be found in \cite{Detweiler:1980uk,Arvanitaki_2011,spectra} in the limit $\alpha \ll 1$. The relevant ratios are
\begin{equation}                         
 \label{eqn:decay211}   \frac{\eta_{211 \rightarrow 21{\text-}1}}{|\Gamma_{21{\text-}1}|} \simeq 6 \times 10^{-3} \left( \frac{q}{1/150} \right) \left( \frac{\alpha}{0.3}\right) \left( \frac{1.36}{1+4\alpha^{2}} \right)^{2} ,
\end{equation} 
and therefore, the decay of the second mode needs to be taken into account for a wide range of parameters and especially when small mass ratios are considered (this ratio becomes even smaller with the general relativistic computation of $|\Gamma|$ \cite{hoof2024gettingblackholesuperradiance,witte2025steppingsuperradianceconstraintsaxions,Siemonsen:2022yyf}). There are analytical results for the coefficients $d_{1}$ and $d_{2}$ in terms of the parabolic cylinder functions that can be readily used \cite{decayrate1,decayrate2}:
\begin{align}
  \label{eqn:d1 general eta Gamma} & d_{1}(t) = e^{-\frac{|\Gamma| t}{2} - \frac{\pi z}{4}} D_{iz} \left(e^{\frac{3i\pi}{4}}\left(\sqrt{\gamma |\Delta m|} t -i \frac{|\Gamma|}{\sqrt{\gamma |\Delta m|}} \right) \right), \\ & \label{eqn: d2 general eta Gamma}
   d_{2}(t) = e^{-\frac{|\Gamma| t}{2} - \frac{\pi z}{4}} \sqrt{z} D_{iz-1} \left(e^{\frac{3i\pi}{4}}\left(\sqrt{\gamma |\Delta m|} t -i \frac{|\Gamma|}{\sqrt{\gamma |\Delta m|}} \right) \right).
\end{align}
Specializing to the case where $\eta \ll \Gamma$, we can simplify these to \cite{extreme_mass_ratio}:
\begin{align}
    \label{eqn: d1^2 gamma>> 1/dt} &
    |d_{1} (t )|^{2} = \text{exp}\left[-z \left( \pi+2\text{tan}^{-1} \left( \frac{|\Delta m| \gamma t}{|\Gamma|}\right)\right) \right],   \\
    \label{eqn: d2 gamma>> 1/dt} & 
    d_{2} ( t) = - \frac{i\eta}{i|\Delta m| \gamma t + |\Gamma|} d_{1}( t). 
\end{align}
\begin{figure}[t!]
    \centering
        \includegraphics[width=0.74\linewidth]{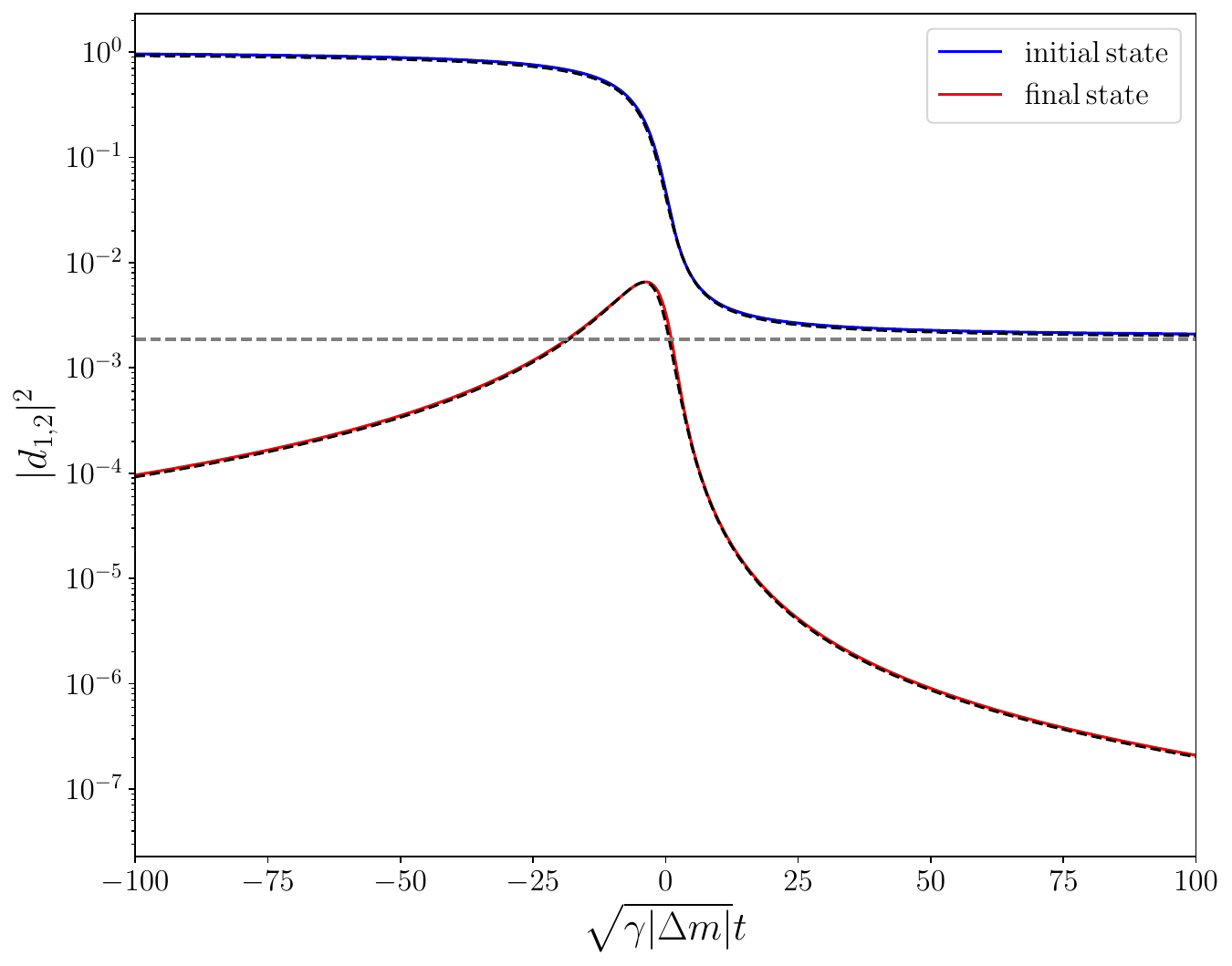}
    \caption[Comparison between numerical and analytical results.]{Comparison of the numerical solution (solid) and the analytical results (dashed) of the time evolution of the states for $z=1$ and $\frac{\eta}{|\Gamma|} = 0.25$.}
    \label{fig:d1 large Gamma evolution}
\end{figure}

In the limit $|d_{1} \left( t \rightarrow \infty \right)|^{2} \rightarrow e^{-2 \pi z}$, we recover the same result as in the Landau-Zener system for the initial state. In the $t \rightarrow \pm \infty$ limit, the occupation number of the second state is zero, while at $t=0$ it is suppressed by the factor $ \frac{\eta}{|\Gamma|}$, a reflection of the fast decay rate of the bosons. In \Fig{fig:d1 large Gamma evolution}, we plot the numerical and the analytical solutions \Eq{eqn: d1^2 gamma>> 1/dt} and \Eq{eqn: d2 gamma>> 1/dt} for a certain choice of the ratio $\frac{\eta}{|\Gamma|}$. The agreement is very good between the two. We notice that the oscillations in \Fig{fig:hyperfine} that are observed in the $z \lesssim 1$ case are now suppressed. \par

\section{Gravitational waves from transitions }\label{section:gr waves}
In this section, we present the key results of our work, which is the derivation of the GW strain from the tidal perturbation of the GA by the companion object, along with the frequency spectrum. The signal's peak frequency is set by the energy splitting of the level transition, the adiabaticity parameter $ z$ and the decay rate of the second state $|\Gamma|$; the frequency change rate is determined by the orbital frequency change rate of the binary; the amplitude modulation of the strain waveform $h(t)$ is determined by the LZ transition and the evolution of the coefficients $d_{1}$ and $d_{2}$. \par
To start, we employ the quadrupole formula of GW emission, which is derived in detail in Appendix \ref{sec:quadrupole}:
\begin{equation}
    h^{\rm TT}_{ij}({\bf x},t)\simeq\frac{2}{r}\Lambda_{ij,lm}(\hat{\bf n})\ddot{M}_{lm},
\label{eqn:quadrupole formula h}\end{equation}
where $M_{ij}$ are the quadrupole moments of the GW source, given by:
\begin{equation}
    M_{i j } = \int d^{3}x \rho(t,\vec{x}) x_{i} x_{j},
\label{eqn:quadrupole}\end{equation}
$\Lambda_{ij,lm}$ is a tensor that depends only on the GW propagation direction $\hat{\bf n}$ and projects the quadrupole moments of the source onto the (TT) gauge, and $r$ is the distance to the source. We work in the TT gauge and omit the superscript of the strain in the following. \par

The quadrupole formula is valid in the limit where the wavelength of the emitted radiation is much larger than the size of the source, which can be justified once we derive the frequency of the GW signal, and is valid throughout our parameter space. \par

For a real scalar field $\phi$ with mass $\mu$, the energy density is given by
\begin{equation}
    \rho(t,\vec{x}) = \frac{\dot{\phi}^{2}}{2} + \frac{(\nabla \phi)^{2}}{2} + \frac{\mu^{2} \phi^{2}}{2}.
\end{equation}
In the non-relativistic limit $\alpha \ll 1$, we express the field $\phi$ in terms of the wavefunction $\psi$ by separating the fast-oscillation mode $e^{-i\mu t}$ from it,  
\begin{equation}
    \phi(t,\vec{x}) = \frac{1}{\sqrt{2 \mu}} \left(  \psi( t,\vec{x})e^{-i \mu t} + c.c \right),
\end{equation}
for which it holds that $|\partial_{t} \psi| \ll \mu |\psi|$ and $|\nabla \psi| \ll \mu |\psi|$. From \Eq{eqn:ansatz}, one can find $|\partial_{t} \psi| \sim \mu \alpha^{2} $ and $|\nabla \psi| \sim \mu \alpha |\psi|$ and therefore the inequalities are satisfied when $\alpha \ll 1$. In these limits, the energy density of $\phi$ is dominated by 
\begin{equation}
    \rho(t,\vec{x}) \simeq \mu |\psi (t,\vec{x})|^{2} + \,...
\end{equation}
where the dots include terms of $\mathcal{O}(\alpha^{2})$ or higher.
\par

The state of the GA $\psi(t,\vec{x})$ is in general a superposition of the two states that participate in the transition
\begin{equation}
     \psi(t,\vec{x}) = \sqrt{N} \left( c_{1}(t) \psi_{1}(\vec{x}) + c_{2}(t) \psi_{2}(\vec{x}) \right),
\end{equation}
where $N$ is the number of axions. The energy density is given by:
\begin{equation}
    \rho(t,\vec{x})=  M_{c} \left( |c_{1}(t)|^{2} |\psi_{1}|^{2} + |c_{2}(t)|^{2}|\psi_{2}|^{2} + 2 \Re\left( c^{\ast}_{1}(t) c_{2}(t) \psi^{\ast}_{1} \psi_{2} \right) \right),
\label{eqn:density}\end{equation}
where $M_{c} = \mu N$ is the mass of the cloud. We stress that the usual normalization condition $|c_{1}(t)|^{2} + |c_{2}(t)|^{2} = 1$ does not apply here because of the decay rate of the second state, in which final-state bosons are dissipated by the black hole absorption.

\par
For the hyperfine $|211\rangle \rightarrow |21\text{-}1\rangle$ transition, 
the relevant quadrupole moments are
\begin{align}
   & M_{11}(t) = 12M_{c} r^{2}_{c} \left(1 - \Re \left[d^{\ast}_{1}(t_{\rm re}) d_{2}(t_{\rm re}) e^{-2i\Delta m \varphi(t_{\rm re})}  \right] \right), \\ &
   M_{22}(t) = 12M_{c} r^{2}_{c} \left(1 + \Re \left[d^{\ast}_{1}(t_{\rm re}) d_{2}(t_{\rm re}) e^{-2i\Delta m \varphi(t_{\rm re})}  \right] \right), \\ &
   M_{12}(t) = -12 M_{c} r^{2}_{c} \Im \left[d^{\ast}_{1}(t_{\rm re}) d_{2}(t_{\rm re}) e^{-2 i \Delta m \varphi(t_{\rm re})} \right],  \\ &
   M_{33}(t) = 6 M_{c} r^{2}_{c} \left( |d_{1}(t_{\rm re})|^{2} + |d_{2}(t_{\rm re})|^{2} \right) , 
\end{align}
where $\Re[\,]$ denotes the real part while $\Im[\,]$ denotes the imaginary part, we used the wavefunctions of the hydrogen atom \cite{Shankar:102017}, the unitary matrix from \Eq{eqn:cs} to express the coefficients in the co-rotating frame, and defined the retarded time $t_{\rm re} \equiv t - r$, with $r$ the distance between the GA and the observer. The $M_{13}$ and $M_{23}$ coefficients are zero due to the integral over the azimuthal angle.  \par
The second time derivatives of the above quadrupole moments will determine the GW strain from the transition, which eventually are dependent on the second time derivatives of $d_1$ and $d_2$. The equations of motion for $d_1$ and $d_2$ in \Eq{eq:eomOfd} can be used to re-express their second time derivatives in terms of the known solutions $d_1$ and $d_2$. The leading order term of $\ddot{M}_{ij},\,i,j=1,2,3$, can also be obtained in the following way: in the $\eta \ll |\Gamma|$ limit, when a time derivative acts on $d_{1,2}$ given by \Eq{eqn: d1^2 gamma>> 1/dt} and \Eq{eqn: d2 gamma>> 1/dt}, the result is of order $ \mathcal{O} \left( \frac{\gamma}{|\Gamma|} d_{1,2}  \right)$. When the time derivative acts on the exponential function $e^{-2i\Delta m\varphi(t_{\rm re})}$, it brings down a factor of $\Omega \sim \Omega_{0}$. By factoring out $\Omega^{2}_{0}$, the remaining terms are of order $\mathcal{O} \left( \frac{\gamma^{2}}{\Omega^{2}_{0} |\Gamma|^{2}}, \frac{\gamma}{\Omega_{0} |\Gamma|} \right)$, which scale like $\mathcal{O} \left( \alpha^{52/3}, \alpha^{26/3} \right)$ for the $|211\rangle \rightarrow |21\text{-}1\rangle$ transition, so they can be ignored. In the leading order, the relevant second-order derivatives are
\begin{align}
      & \ddot{ M}_{11}(t)   = - \ddot{ M}_{22}(t) = 12 M_{c} r^{2}_{c} \Omega_0^{2} \Re \left[ e^{-2 i \Delta m \varphi(t_{\rm re})} Q(t_{\rm re}) \right], \\ 
    & \ddot{ M}_{12} (t)  = 12 M_{c} r^{2}_{c} \Omega_0^{2} \Im \left[ e^{-2 i \Delta m \varphi(t_{\rm re})} Q(t_{\rm re}) \right], \\ & \label{eqn:Q}
    Q(t) = 4 |\Delta m|^{2} d^{\ast}_{1}(t_{\textrm{re}}) d_{2}(t_{\textrm{re}}) +\mathcal{O}(\frac{\gamma}{\Omega_{0} |\Gamma|}),
\end{align}
where we have ignored $\ddot{M}_{33}$ because it is suppressed by $\mathcal{O}\left( \frac{\gamma^{2}}{\Omega^{2}_{0}|\Gamma|^{2} }\right)$ relative to the other components. $Q(t)$ can be found analytically by \Eq{eqn: d1^2 gamma>> 1/dt} and \Eq{eqn: d2 gamma>> 1/dt} in the $\eta \ll |\Gamma|$ limit. In this case, it is straightforward to find that the maximum value of $|Q|$ occurs at $t_{\rm max} = - \frac{2 z |\Gamma|}{|\Delta m|\gamma} = - \frac{ z |\Gamma|}{\gamma} $. This result persists for the more general formula as long as $\eta \lesssim |\Gamma|$. 
\begin{figure}
\centering
\begin{subfigure}{.50\textwidth}
  \centering
  \includegraphics[width=0.98\linewidth]{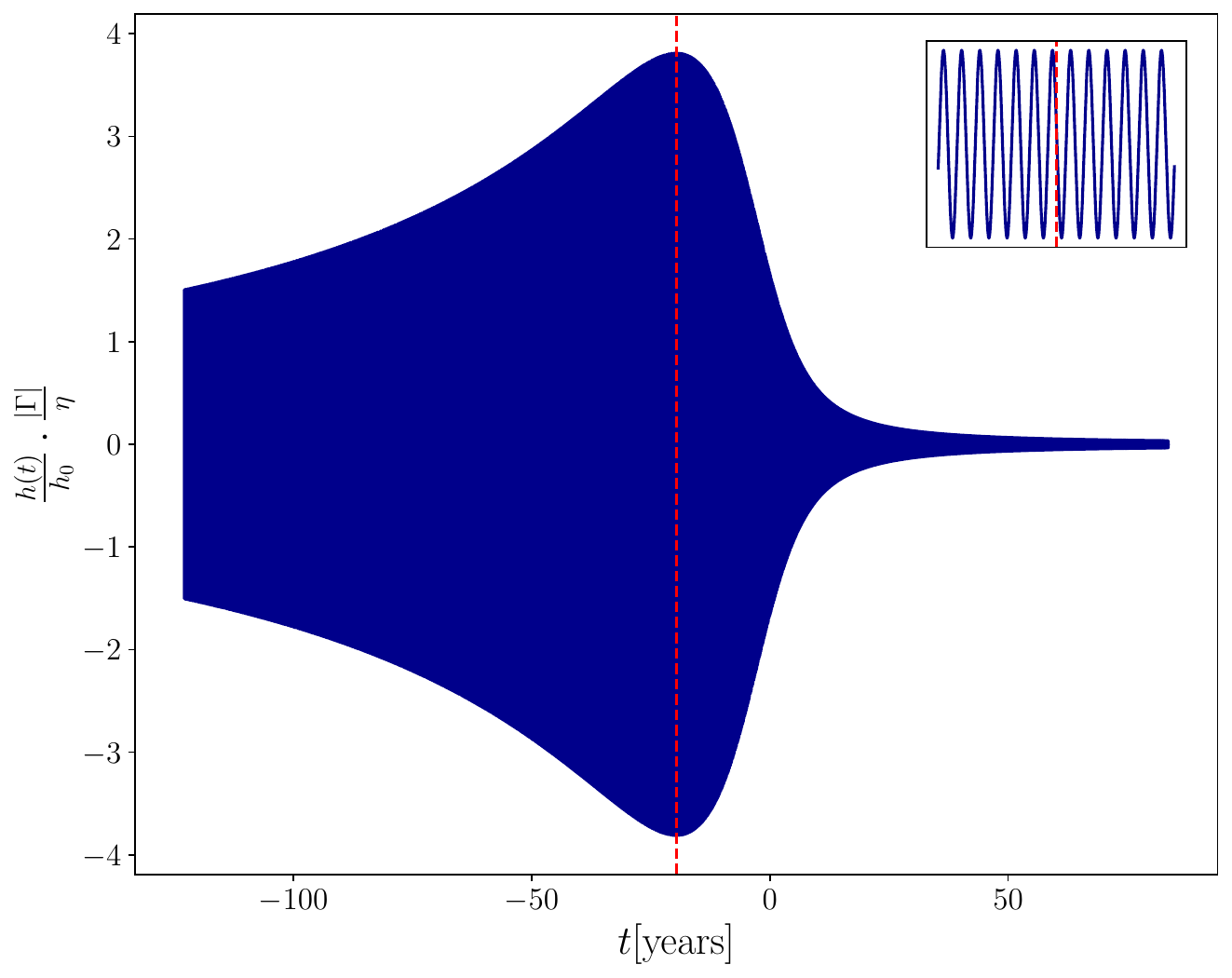}
  \end{subfigure}%
  \hfill
  \begin{subfigure}{.50\textwidth}
  \centering
  \includegraphics[width=1.1\linewidth]{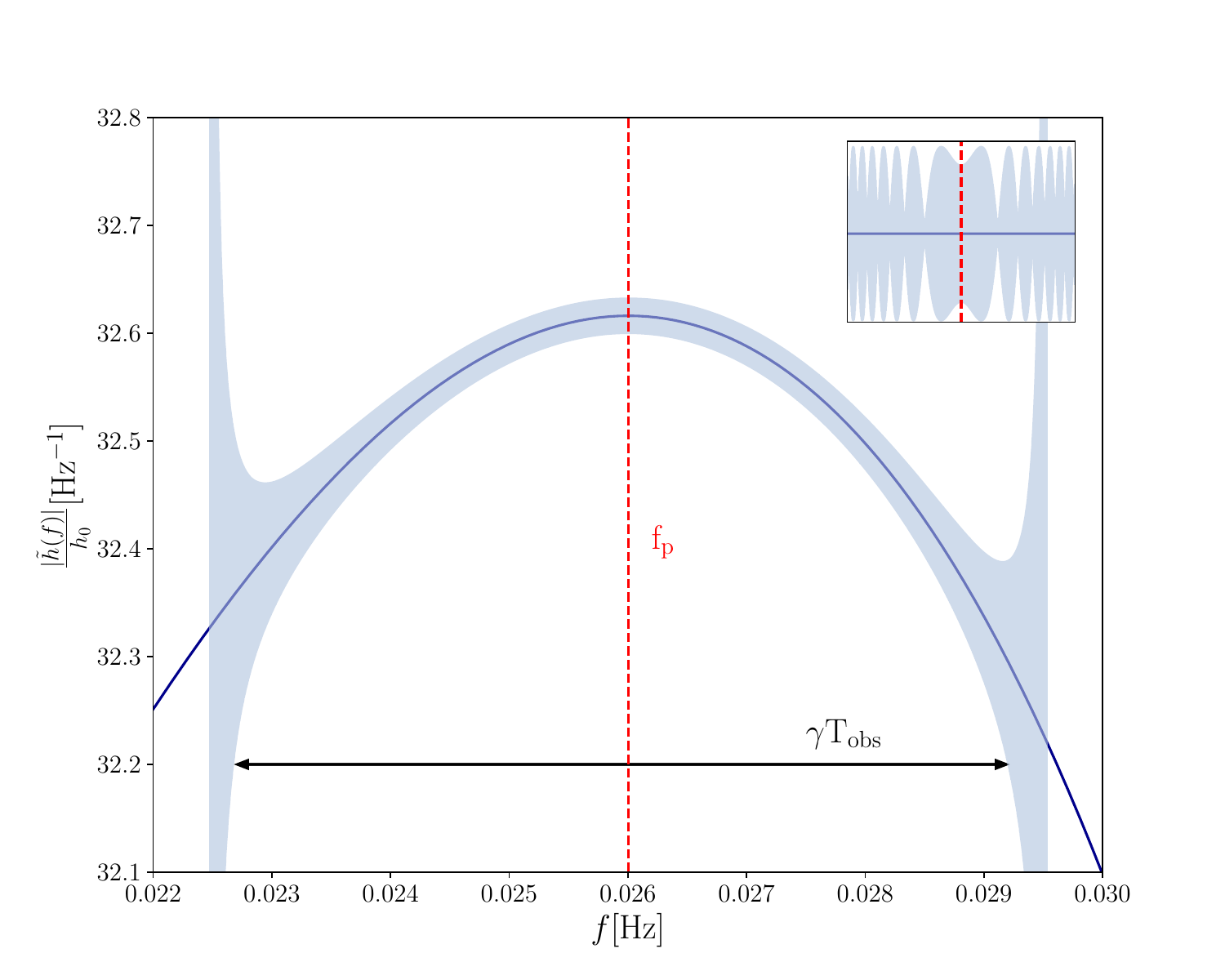}
\end{subfigure}%
 \caption[Time evolution and frequency spectrum of the gravitational wave signal.]{\label{fig: strain freq time} (left) The plus polarization of the GW signal \cref{eqn: plus strain} versus time for the $|211\rangle \rightarrow |21\textrm{-}1\rangle$ transition. We have chosen $\alpha = 0.3$, $q=1/150$, and $M = 150 M_{\odot}$. The red dashed line is set at $t_{\rm max} = - z |\Gamma|/\gamma$. The inset shows the oscillations around $t_{\rm max}$. (right) The frequency spectrum $|\tilde{h}(f)|/h_{0}$ versus frequency for the same parameters. The solid dark blue line is given by the stationary phase approximation, while the pale blue line is obtained by the short-time Fourier transform. The red dashed vertical line shows the peak frequency of the signal \cref{eqn:peak freq}, while the inset shows the fast oscillations around that frequency. The black arrow shows the frequency width, which is equal to $\gamma T_{\rm obs}$. } 
\end{figure} 

The strain of the plus- and cross-polarized GW is thus given by \cite{Maggiore:2007ulw},
\begin{align}
     \label{eqn: plus strain}h_{+,211}(t) &= h_{0} \frac{1+\cos^{2}(\iota)}{2} \Re \left[ e^{-2 i\Delta m \varphi(t_{\rm re})} Q(t_{\rm re}) \right], 
     \\    h_{\times,211} (t) & = h_{0} \cos(\iota) \Im \left[ e^{-2 i \Delta m \varphi(t_{\rm re})} Q(t_{\rm re}) \right],
\end{align}
where $\iota$ is the angle between the line-of-sight of the observer and normal to the orbital plane of the system. The signal is therefore composed of a fast oscillating exponential term that contains information about the phase of the binary and an amplitude modulation $Q$ that contains the details of the transition.  
The factor $h_{0}$ is given by
\begin{equation}
    h_{0} = \frac{24  M_{c} r^{2}_{c} \Omega^{2}_{0}}{r} = 24 \frac{q_{c} M}{r} \frac{1}{\alpha^{4}} (M \Omega_{0})^{2},
\label{eqn:amp}\end{equation}
where $q_{c}  \equiv \frac{M_{c}}{M}$ is defined as the GA mass ratio. $q_c$ can in general be computed numerically using the SupeRrad package \cite{Siemonsen:2022yyf,May:2024npn}. For example, $q_{c} \simeq 0.085$ for $\alpha=0.3$ and $\tilde{a}_{\rm in} = 0.99$ at saturation. The characteristic value of $h_{0}$ for the benchmark parameters is
\begin{equation}
    h_{0} = 5 \times 10^{-23} \left( \frac{q_{c}}{0.085} \right) \left( \frac{M}{150 M_{\odot}} \right) \left( \frac{100 \textrm{kpc}}{r} \right) \left( \frac{\alpha}{0.3}\right)^{10}.
    \label{eqn:amp-scaling}
\end{equation}

The GW strain waveform is shown in \Fig{fig: strain freq time} (left) for the benchmark parameters, $\alpha = 0.3$, $q=1/150$, and $M = 150 M_{\odot}$. The GW signal is quasimonochromatic with its oscillation frequency set by $2|\Delta m|f_0$, where $f_0$ is the orbital frequency of the binary, because of the exponential factor $e^{-2i \Delta m \varphi(t)}$, while the amplitude of the strain is slowly modulated by $Q(t)$. \par
When the decay rate is included, the perturbation starts to act roughly at time $ t_{I} \simeq -\frac{\Gamma}{2 \gamma} (1+2 z)$ \cite{extreme_mass_ratio}. If $t_{I}$ is earlier than the time when the binary enters the resonant band, the linear approximation that we employed in \Eq{eqn:linear freq} no longer holds and we would have to take into account the non-linear evolution of the binary. The condition for our analysis to be self-consistent is therefore: 
\begin{figure}[t!]
    \centering
    \includegraphics[width=0.74\linewidth]{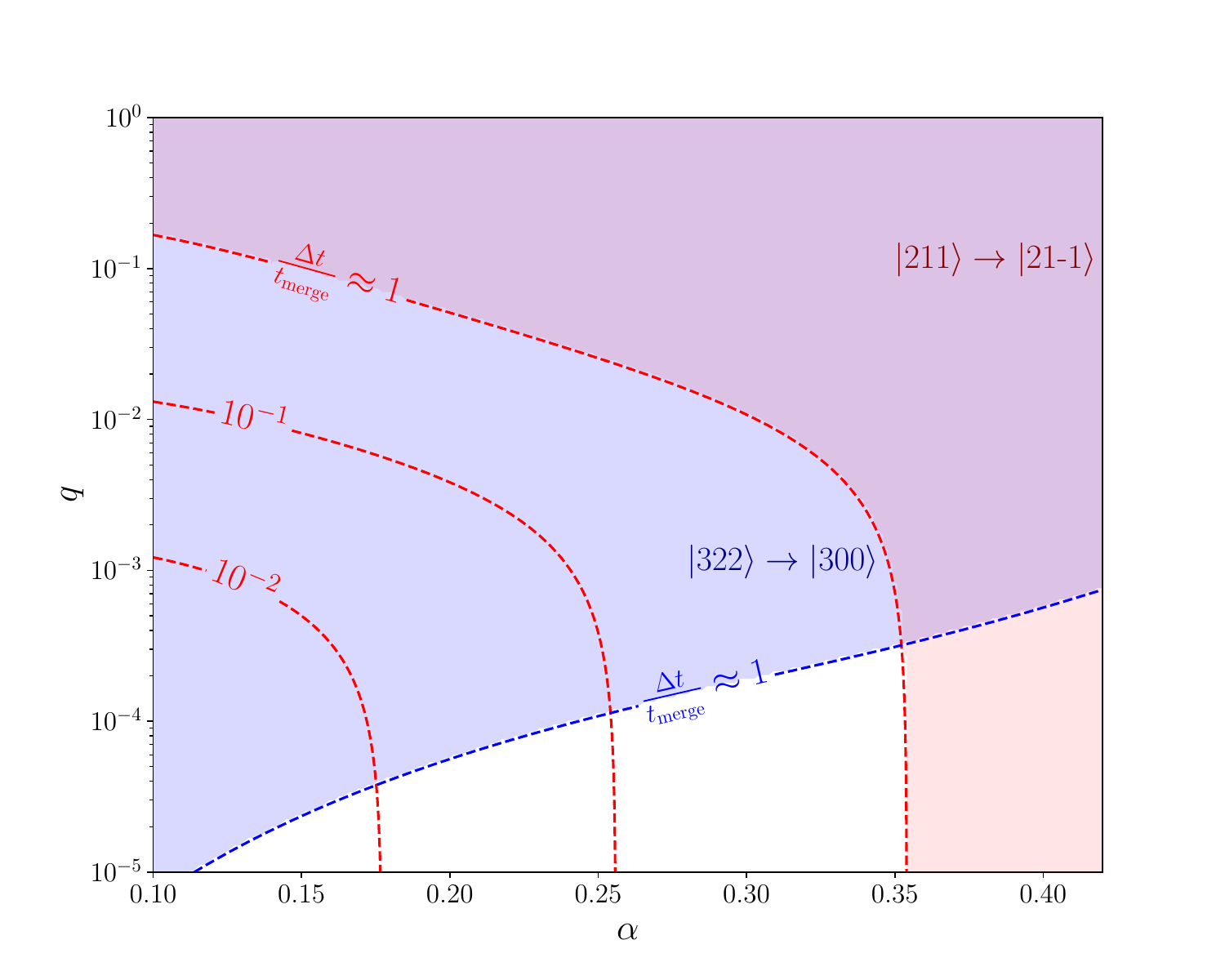}
    \caption{ Contours of $\Delta t/t_{\rm merger}$ in the $\alpha-q$ plane for two different transitions.
   }
    \label{fig:region with large timescale}
\end{figure}
\begin{equation}
    \frac{\gamma |t_{I}|}{\Omega_{0}} \simeq \frac{|\Gamma| (1+2z)}{2 \Omega_{0}} \lesssim \mathcal{O}(1).
\label{eqn:large timescale condition}\end{equation}
Equivalently, we require the duration of the transition, $\Delta t \simeq 2 |\Gamma| (1+2z)/\gamma$, to be smaller than the merger timescale when the binary is at the resonant frequency $\Omega_{0}$:
\begin{equation}
    t_{\rm merge} = \frac{5}{256} \frac{M}{(M \Omega_{0})^{8/3}} \frac{(1+q)^{1/3}}{q} \simeq 17\, \text{yrs} \left( \frac{M}{150 M_{\odot}}  \right) \left( \frac{1/150}{q} \right) \left( \frac{\alpha}{0.3}\right)^{-56/3},
\end{equation}
where $\Omega_0$ is set by $\Omega_{0,{\rm hyp}}$, in \Eq{eqn:Omega hyperfine}, and it gives the same order of magnitude as given by \Eq{eqn:large timescale condition}. In \Fig{fig:region with large timescale}, we plot the contours of $\Delta t/t_{\rm merger}$ in an $\alpha - q$ plane for the $|211\rangle \rightarrow |21\text{-}1\rangle$ and $|322\rangle \rightarrow |300\rangle$ transitions. We have used the relativistic computation of the decay rate for the final state \cite{hoof_git,hoof2024gettingblackholesuperradiance}. The shaded regions represent the case where $\Delta t > t_{\rm merger}$. For the $|211\rangle \rightarrow |21\text{-}1\rangle$ transition, the linear approximation of the orbital frequency is not valid for $\alpha>0.35$ and all values of $q$, while for smaller $\alpha$ values, the ratio is $q$ dependent. Based on this result, we limit our attention in what follows to $\alpha < 0.35$ and mass ratios that satisfy the condition $\Delta t< t_{\rm merger}$.
\par
The fine transition $|322\rangle \rightarrow |300\rangle$ covers a large parameter space because the $|300\rangle$ state has a large decay rate compared to the $|21{\text-}1\rangle$ state. The parameter space for this transition shrinks further when the effect of the termination of superradiance due to the off-resonant mixing of the two states is considered, which excludes mass ratios that satisfy $q \gtrsim 10^{-4}$ (this affects all fine transitions while the $|211\rangle \rightarrow |21{\text-}1\rangle$ is unaffected)\,\cite{Tong_2022}. For the rest of the paper, we will only consider the $|211\rangle \rightarrow |21{\text-}1\rangle$ transition in the parameter space that was mentioned and leave the treatment of the $|322\rangle$ state for future work.
\par
In \Fig{fig: strain freq time} (right), we show the frequency spectrum of the GW signal, derived using the stationary phase approximation (see details in Appendix \ref{sec:freq spectrum}), STFT of the strain. The frequency width accessible by the Fourier transform is equal to $\gamma T_{\rm obs}$ (we have chosen an unrealistically long observation time in the figure for display purposes). We note the good agreement between the two, except for the edges of STFT results due to the windowing effect \cite{Droz:1999qx}.
The peak frequency, that is shown as a vertical dashed line, is given by (derived in Appendix \ref{sec:freq spectrum})
\begin{equation} 
    f_{\textrm{p}} = f_{\textrm{c}} - \frac{2 z |\Gamma|}{\pi}.
\label{eqn:peak freq}\end{equation}
Either in the case $z\ll 1$ or when the final-state dacay timescale is much larger than the orbital period at resonance, the peak frequency reduces to $f_{\rm p} = f_{\rm c} = 4 f_{0}$. The conclusion is that the signal is monochromatic, something that is anticipated, given that we have linearized the orbital frequency that drives the resonance around the corresponding frequency, offset however by the decay rate. In the shaded regions of \Fig{fig:region with large timescale}, a signal that spans many frequency bins is expected, a possibility that we leave for future work.  \par
We may now assess the validity of the quadrupole approximation that we have used throughout. In the case where $z \ll 1$, the comparison of the wavelength of radiation to the size $r_{\rm c}$ of the system gives:

\begin{equation}
    \frac{\lambda}{r_{\rm c}}  = \frac{\alpha^{2}}{4 M f_{0}} \simeq 2 \times10^{3} \left( \frac{0.3}{\alpha}\right)^{5},
\end{equation}
and therefore the approximation is excellent. It becomes even better for larger values of $z$, where the peak frequency is smaller.
\subsection{Comparison to Inspiral and Annihilation GWs}
\begin{figure}
    \centering
    \includegraphics[width=0.8\linewidth]{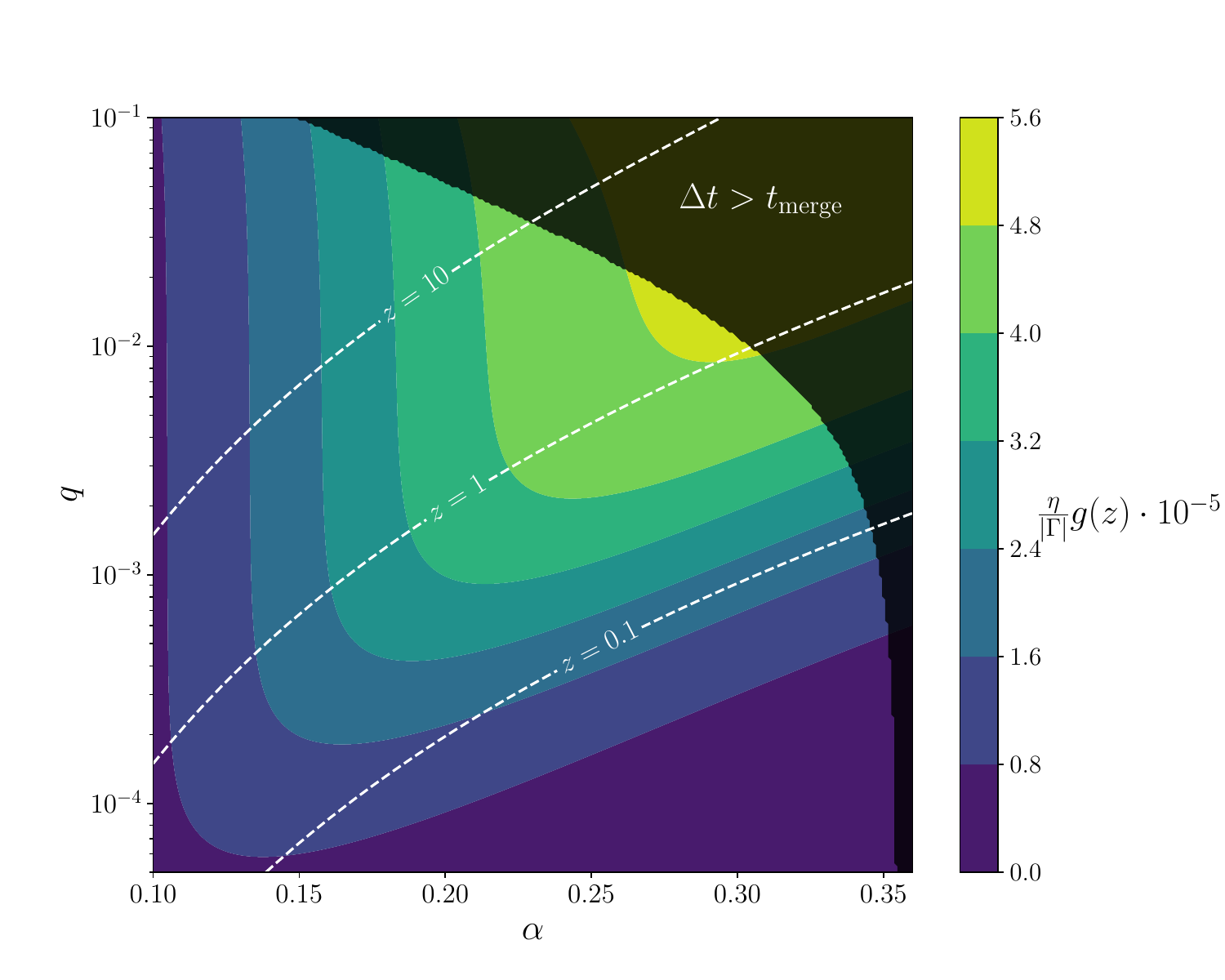}
    \caption[The factor $\frac{\eta}{|\Gamma|} g( z)$ in the $\alpha-q$ plane]{The factor $\frac{\eta}{|\Gamma|} g( z)$ in the $\alpha-q$ plane. Dashed lines show three values of the adiabaticity parameter $z$. The upper right corner is not considered because of the duration time being larger than the merger time.}
    \label{fig:g vs q}
\end{figure}

It is intriguing to compare the strain amplitude of the GW signal we have presented above to that of the other GW signals that originate from the inspiral itself and the annihilation of the GA. \par
Firstly, the order of magnitude of the maximum amplitude of the transition's GW signal is found by evaluating $| h_{\textrm{+}}|$ at $t_{\rm max}$ and averaging over the fast oscillations:
\begin{equation}
    h_{\textrm{max}}  = 4|\Delta m|^{2} h_{0} \frac{\eta}{|\Gamma|} g(z),
\end{equation}
where $ g( z)$ is a function of the adiabaticity parameter:
\begin{equation}
   g(z) = \frac{e^{-\pi z  + 2  z \tan^{-1}(2  z)}}{\sqrt{1+4z^{2}}},
\label{eqn: g(z)}\end{equation}.
The quantity $\eta g(z)/|\Gamma|$ is plotted in \Fig{fig:g vs q} in the $\alpha-q$ plane. We notice that it is maximized for $\alpha \sim 0.3$ and $q \sim 10^{-2}$. 

The inspiral GW strain amplitude is given by \cite{Maggiore:2007ulw}, 
\begin{equation}
    h_{0,\rm b} = \frac{4}{r} \frac{q}{(1+q)^{1/3}} M^{5/3} \Omega^{2/3}_{0}.
\label{eqn:binary gw}\end{equation}
For small $ z$, the transition GW's frequency $f_{\rm p}$ is expected at twice the inspiral GW's frequency $f_{\rm insp}$ from \Eq{eqn:peak freq}, i.e., $f_{\rm p}=2f_{\rm insp} = 4 f_{0}$. \par
\begin{figure}
    \centering
    \includegraphics[width=0.9\linewidth]{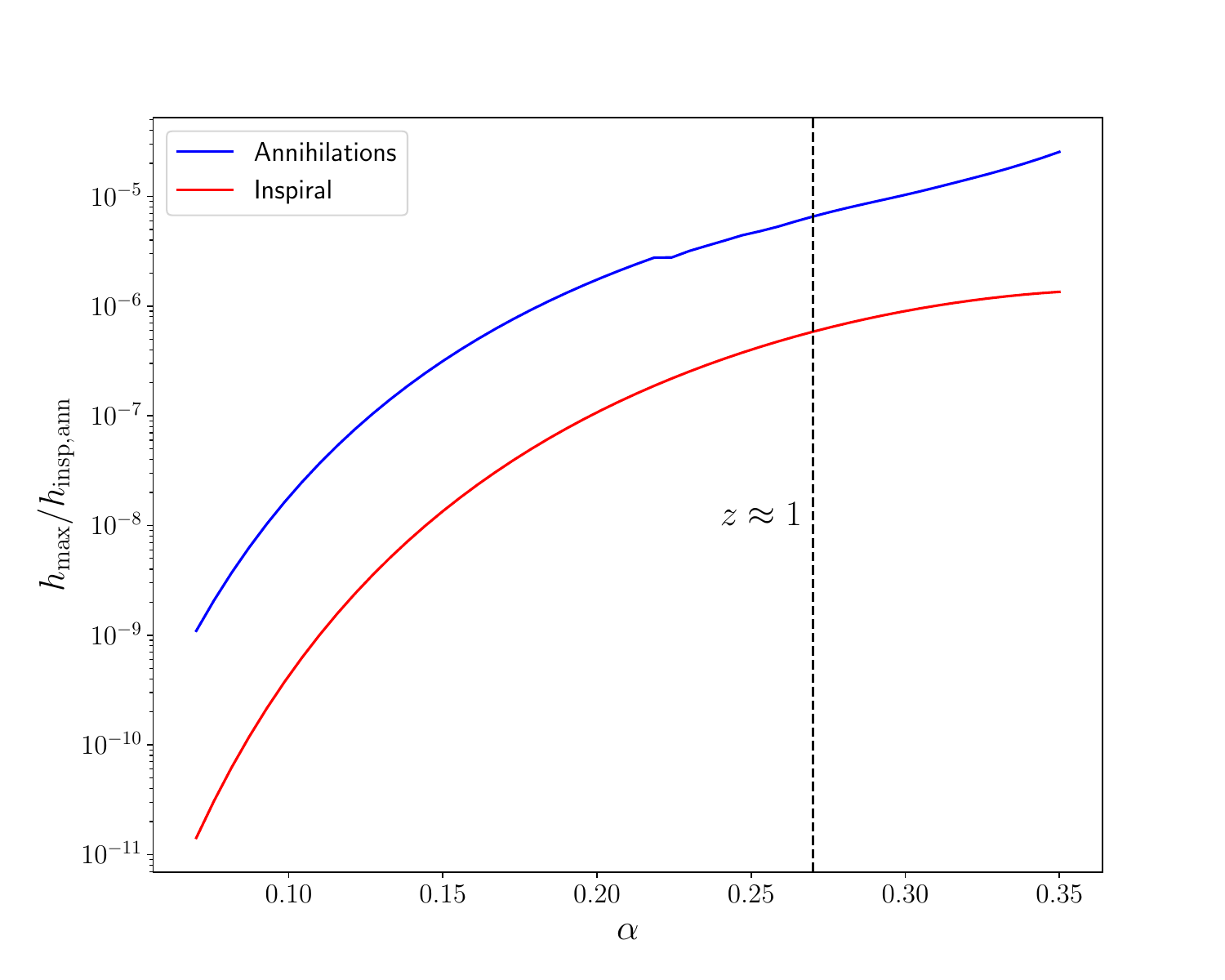}
    \caption[Comparison with binary and annihilation signal]{The ratio between $h_{\rm max}$  and the amplitudes of the GWs from the inspiral and the annihilations of the GA for $q = 1/150$ and $\tilde{a}_{\rm in}=0.99$. The vertical dashed line shows the $\alpha$ value where the adiabaticity $z \simeq 1$. For smaller $\alpha$, $z>1$ and the signal is suppressed by the exponential function in \Eq{eqn: g(z)}. }
    \label{fig:amp ratio}
\end{figure}

We also consider the GWs emitted by the annihilation of the bosons in GA, with frequency equal to
\begin{equation}
    f_{\rm ann} =  \frac{\mu}{\pi} \simeq 145 \,\textrm{Hz} \left( \frac{\alpha}{0.3} \right) \left( \frac{150 M_{\odot}}{M} \right).
\label{eqn:ann frequncy}\end{equation}
The SuperRad package allows us to calculate the GW strain amplitude for given $\alpha$, $M$ and $\tilde{a}_{\rm in}$ \cite{Siemonsen:2022yyf,May:2024npn}. We calculate the ratio of $h_{\rm max}$ with these two amplitudes in \Fig{fig:amp ratio} for fixed $q=1/150$ as a function of $\alpha\in[0.07,\,0.35]$. We observe that the amplitude of the transition GW signal is much smaller than the amplitudes of the other two signals. The ratio becomes very small in the small $\alpha$ limit, since $z \gg 1$ in this limit (see \Eq{eqn:z hyperfine}) and the $g(z)$ function is exponentially suppressed. Varying the mass ratio within the parameter space of \Fig{fig:region with large timescale} does not change this picture.    

\section{Detectability of a Transition GW Signal\label{section:application}}

\subsection{Relevant Timescales for a Transition GW Signal}
The GA of initial mass $M_{c,0}$ continuously depletes its mass due to the annihilation of the bosons into gravitons.\,\footnote{This is only true for a real scalar field. A complex scalar field is expected to be long-lived \cite{Baumann_2019}.} The evolution of its mass as a function of time is given by 
\begin{equation}
    M_{c}(t) = \frac{M_{c,0}}{1+ t/\tau_{\rm ann}},
\end{equation}
where an analytical estimate for the annihilation timescale $\tau_{\rm ann}$ is given by \cite{Yoshino:2013ofa,Gravitaitonal_wave_searches,Arvanitaki_2015,LIGOScientific:2021rnv},
\begin{equation}
    \tau_{\rm ann} \simeq 9.7 \times 10^{5} \,\text{yrs} \left( \frac{M}{150 M_{\odot}} \right) \left( \frac{0.1}{\alpha}\right)^{15} \left( \frac{1}{\tilde{a}_{\rm in}} \right).
\label{eqn:annihilation timescale}\end{equation}
This formula underestimates the timescale for larger values of $\alpha$ and numerical results are needed.

The relevant timescales in our case are the annihilation timescale $\tau_{\rm ann}$, the superradiance timescale $\tau_{\rm sr}\equiv\Gamma_{211}^{-1}$, and the orbital evolution timescale during the resonant transition $\tau_{\rm orb}\equiv \Omega_0/\gamma$, assuming the orbital frequency is changed only through GW emission. If $\tau_{\rm sr}\ll \tau_{\rm ann}$, the $|211\rangle$ state can form before it completely dissipates by annihilation; otherwise, the $|211\rangle$ state cannot form. We find that the condition for the formation of $|211\rangle$ state is always satisfied in the parameter space of interest. Similarly, we require $\tau_{\rm sr}\ll\tau_{\rm orb}$ in order to build up the $|211\rangle$ state before entering the resonance band, which is also always satisfied. Furthermore, to have a long-lasting transition GW signal, we require $\tau_{\rm orb}\ll\tau_{\rm ann}$; otherwise, the $|211\rangle$ state can decay completely before the resonant transition happens. However, we numerically find that $\tau_{\rm orb}\gg\tau_{\rm ann}$ for most of the parameter space of interest, which places a strong constraint on the formation time of the $|211\rangle$ state: it needs to be formed right before the resonant transition happens in order to have detectable transition GW signals. The probability of this occurring would require a detailed study of the evolution history of binary systems given a realistic astrophysical environment and their populations, which is beyond the scope of this work. In the following, we assume the above formation time condition of the $|211\rangle$ state is satisfied.

Even though the $|211\rangle$ state can decay sufficiently fast by annihilation before the resonance takes place, there is still a possibility that higher superradiant states can build up and survive until the resonances occur. In particular, the growth timescale of the next superradiant state $|322\rangle$ is comparable to \Eq{eqn:annihilation timescale} \cite{Tomaselli:2024faa}, which means that this state will have grown by the time the $|211\rangle$ state has decayed by the annihilations. We have left the treatment of transitions for higher energy superradiant states for future work, and the formalism we have developed can be straightforwardly applied to these cases as well. 
\par 
In many previous works, given these uncertainties about the annihilation of the GA and the mass of the cloud, its mass has been treated as a free parameter when performing calculations \cite{Tomaselli:2024faa,resonant_history,extreme_mass_ratio}. In what follows, we have chosen to calculate the mass of the cloud at saturation using the SuperRad package \cite{Siemonsen:2022yyf,May:2024npn} for a given boson mass, black hole mass and initial spin, with the understanding that this serves as an upper bound for the mass of the GA.   

\begin{figure}
\centering
\begin{subfigure}{.6\textwidth}
  \centering
  \includegraphics[width=0.99\linewidth]{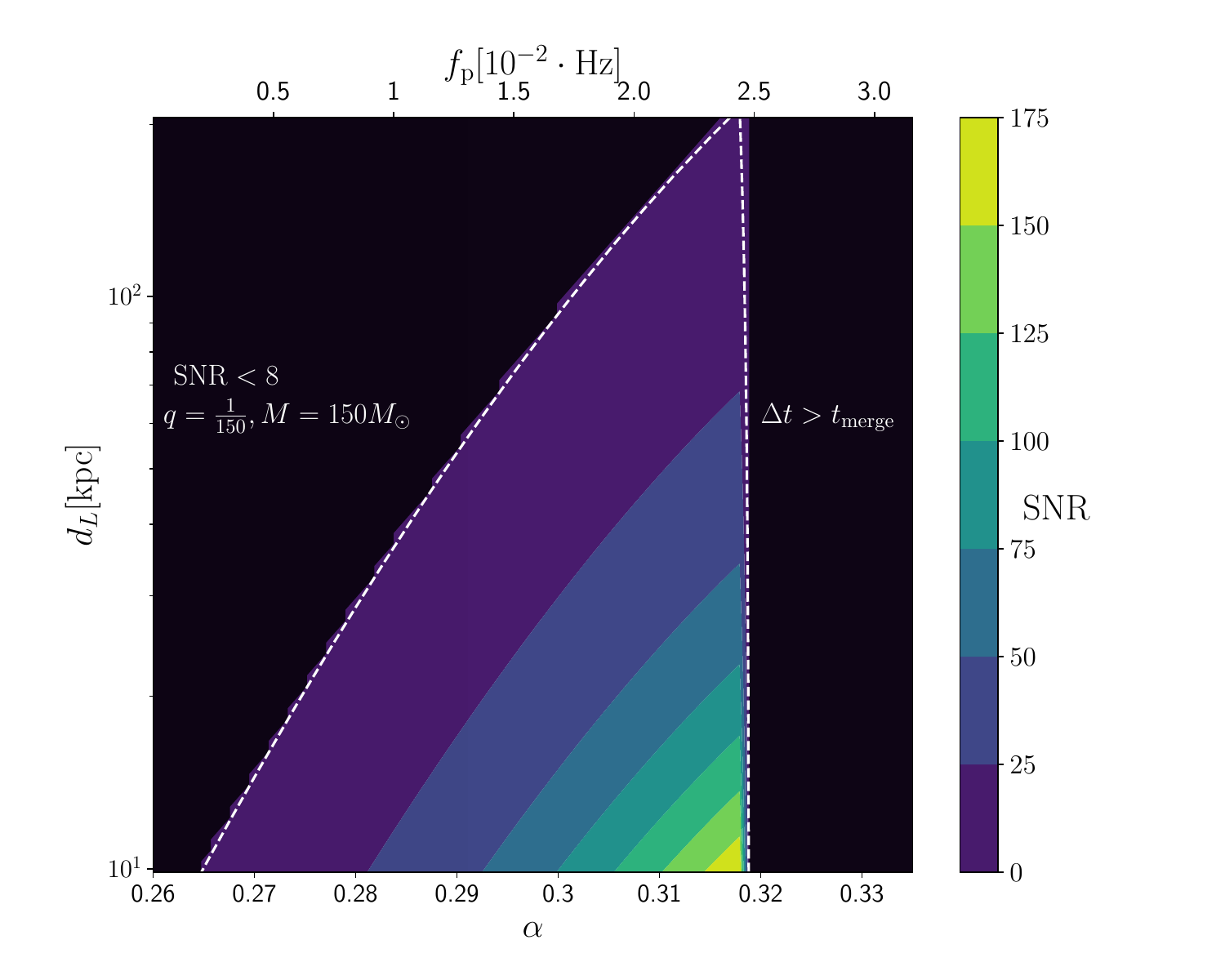}
  \end{subfigure}
  \hfill
  \begin{subfigure}{.49\textwidth}
  \centering
  \includegraphics[width=1.05\linewidth]{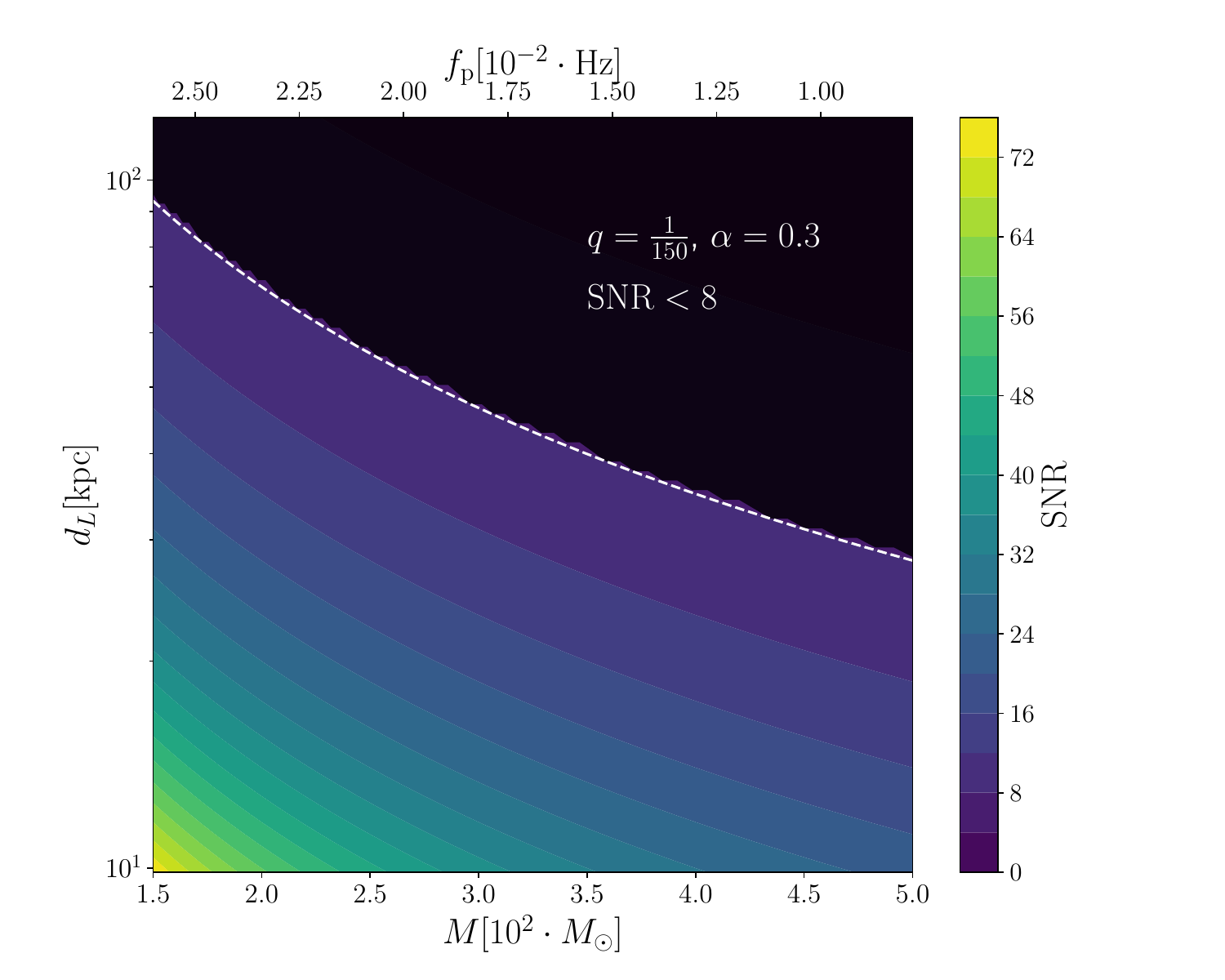}
\end{subfigure}%
 \begin{subfigure}{.49\textwidth}
  \centering
  \includegraphics[width=0.99\linewidth]{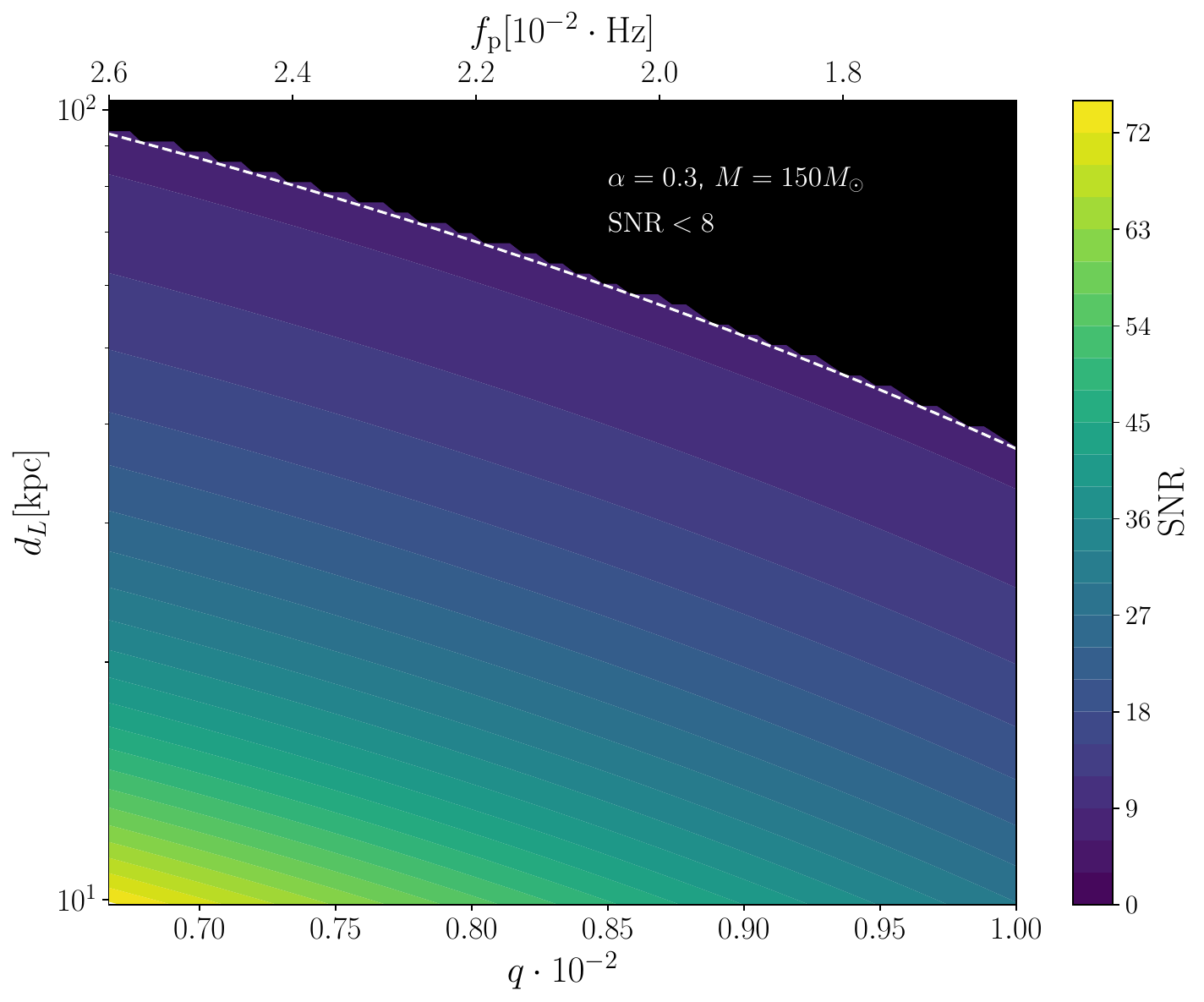}
\end{subfigure}
\caption[SNR contours for DECIGO and black hole spin $\tilde{a}_{\rm in} = 0.99$.]{SNR contours, using DECIGO's sensitivity curve, on the $\alpha - d_{L}$ (top), $M-d_{L}$ (bottom left) and $q-d_{L}$ (bottom right) planes for initial black hole spin $\tilde{a}_{\rm in} = 0.99$.  }
\label{fig:d_L-params decigo 099}
\end{figure}

\begin{figure}
\centering
\begin{subfigure}{.6\textwidth}
  \centering
  \includegraphics[width=0.99\linewidth]{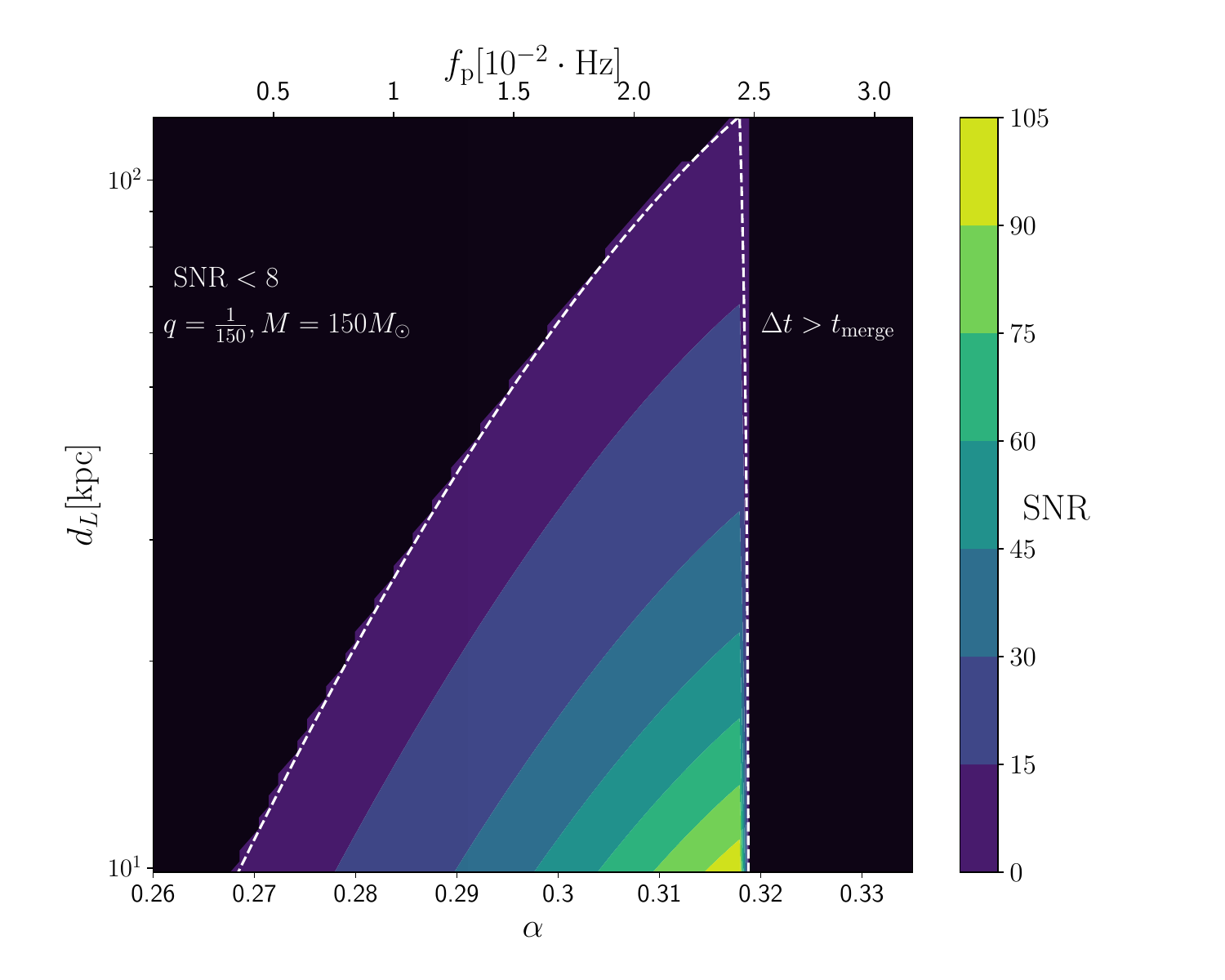}
  \end{subfigure}%
  \hfill
  \begin{subfigure}{.49\textwidth}
  \centering
  \includegraphics[width=1.05\linewidth]{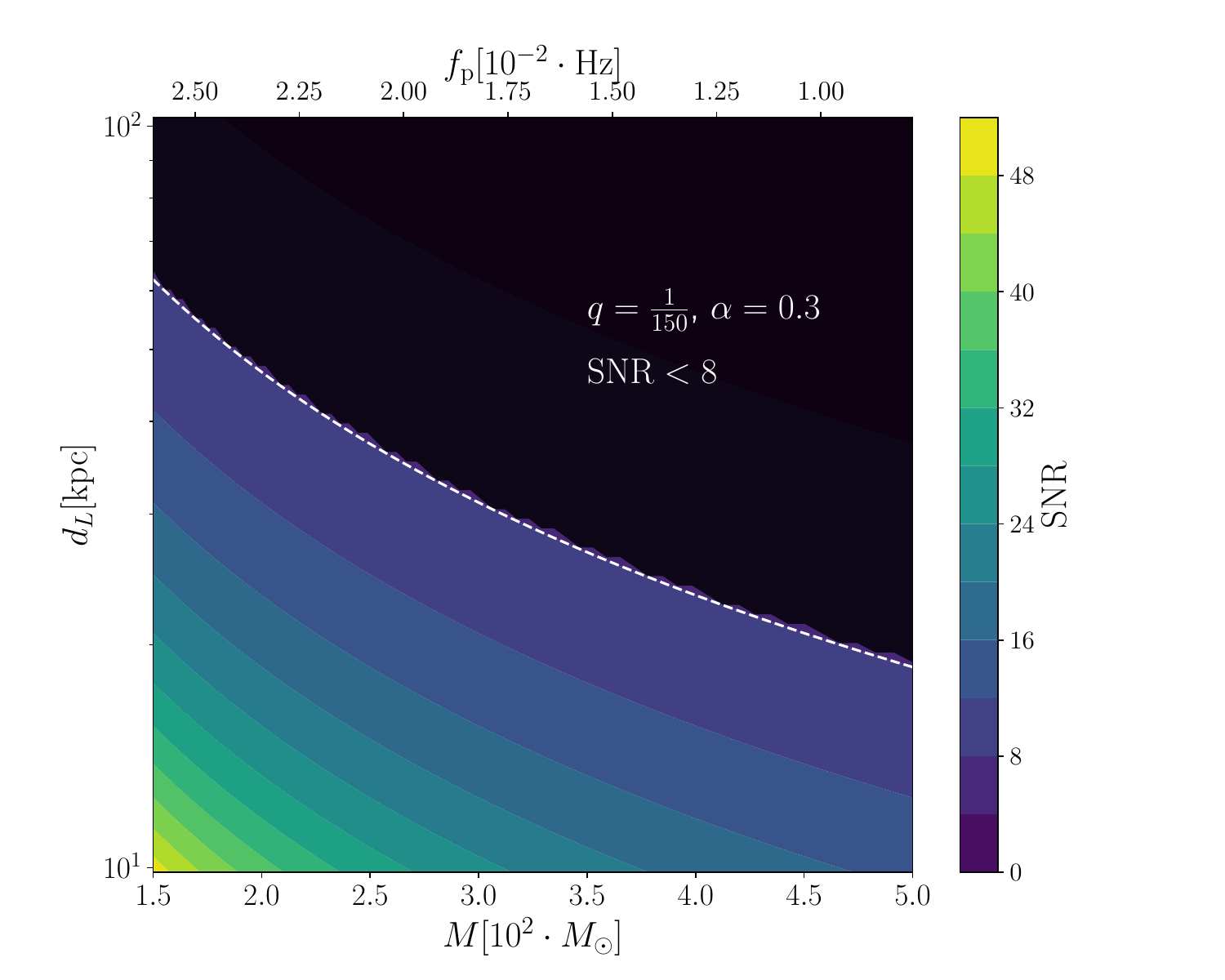}
\end{subfigure}%
 \begin{subfigure}{.49\textwidth}
  \centering
  \includegraphics[width=0.99\linewidth]{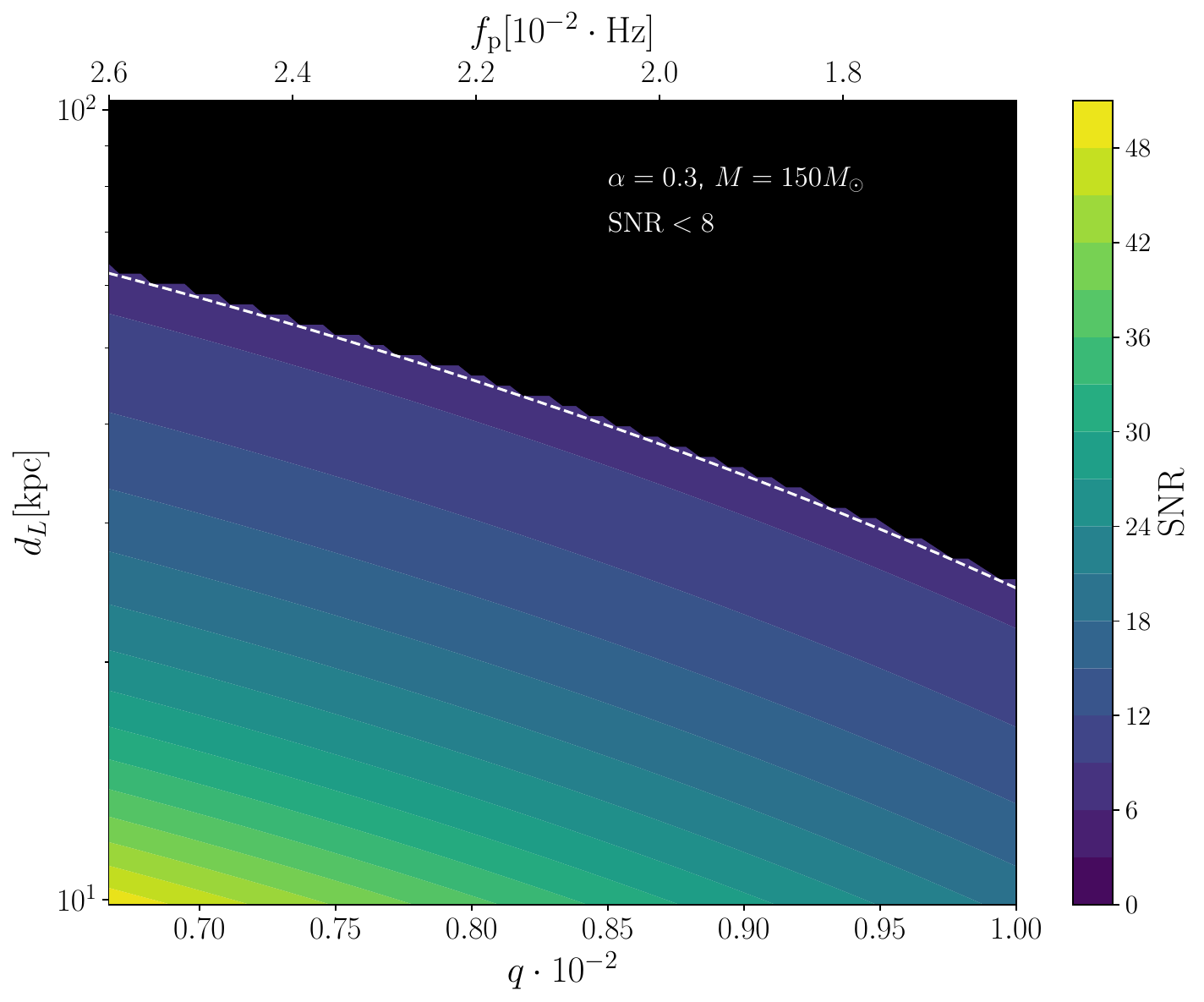}
\end{subfigure}
\caption[SNR contours for DECIGO and black hole spin $\tilde{a}_{\rm in} = 0.95$.]{SNR contours, using DECIGO's sensitivity curve, on the $\alpha - d_{L}$ (top), $M-d_{L}$ (bottom left) and $q-d_{L}$ (bottom right) planes for initial black hole spin $\tilde{a}_{\rm in} = 0.95$.}
\label{fig:d_L-params decigo 095}
\end{figure} 
\subsection{The Signal-To-Noise Ratio\label{sec:snr}}
\par 
A metric to estimate the detectability of our signal is the SNR, which for the monochromatic transition GW signal has the form \cite{LISA:2024hlh}: 
\begin{equation}
    {\rm SNR} = \mathcal{A} \frac{\eta}{|\Gamma|} h_{0} \sqrt{\frac{T_{\rm obs}}{ S_{n} ( f_{\rm p})}}  g( z) ,
\label{eqn:SNR mono}\end{equation}
where $S_{n}(f)$ is the noise power spectrum of a detector, $\mathcal{A}$ is a numerical prefactor that depends on the detector's geometry with $\mathcal{A} = \sqrt{512 \pi/5}$ for LISA and $\mathcal{A} = \sqrt{1024 \pi N_{\rm unit} / 15}$ for DECIGO, with $N_{\rm unit}$ the number of clusters in the detector \cite{DECIGO}, and $ g( z)$ is a function of the adiabaticity parameter that originates from the frequency spectrum in \Eq{eqn:Ip with expo} when it is evaluated at the peak frequency \Eq{eqn:peak freq} and it is given by \Eq{eqn: g(z)}.
 \par
We used \Eq{eqn:SNR mono} to compute the SNR for a variety of physical systems for given $\alpha$, black hole mass $M$, mass ratio $q$, initial black hole spin $\tilde{a}_{\rm in}$, and luminosity distance $ d_{ L}$\,\cite{Maggiore:2007ulw}.  We numerically determined the mass of the cloud at saturation using the SuperRad module\,\cite{May:2024npn,Siemonsen:2022yyf}, as discussed in the previous subsection. The decay rate of the second state was calculated numerically for all cases using the code in \cite{hoof2024gettingblackholesuperradiance,hoof_git} and assuming again that the black hole's spin has reached its saturated value. We used $ T_{\rm obs} = 4  \rm yrs$ for all calculations.
\par
Given the allowed parameter space for our analysis shown in \Fig{fig:region with large timescale}, we have focused on intermediate mass ratio binaries $q \sim 10^{-3} - 10^{-2}$ and IMBHs $M \geq 150 M_{\odot} $, which may be the massive host of such binary systems in dense nuclear and globular clusters \cite{MacLeod:2015bpa,askar2024intermediatemassblackholesstar}. In particular, if they form through successive mergers of lower mass black holes, they can also have large spins, making them ideal for the superradiance process \cite{Berti:2008af,Borchers:2025sid}. In fact, the LIGO event GW231123 has recently confirmed a discovery of a binary system with a black hole remnant mass of $190-265M_\odot$, resulting from a merger of two $\sim 100 M_\odot$ black holes with high spins $\sim 0.8-0.9$ \cite{LIGOScientific:2025rsn}. Regarding the detection prospects in future GW experiments, LISA's  is expected to detect $0.02-60$ such events per year, while for DECIGO that number is $6-3000$ per year, up to red-shift $z \approx 10$ (Figure 20 of \cite{Arca-Sedda:2020lso}). There are also hints that point to the existence of IMBHs in globular clusters in the Milky Way and in its satellite galaxies, though there is ongoing debate in the literature surrounding these findings \cite{10.1093/mnras/stw1779,Kiziltan:2017ijz,Abbate_2019, 2019MNRAS.487.2685E,2017ApJ...846...14B,2007MNRAS.376L..29G}. \par

In \Fig{fig:d_L-params decigo 099}, we plot the $\rm SNR$ contours for the DECIGO telescope in the $\alpha-d_{L}$, $M-d_{L}$ and $q - d_{L}$ planes for an initially maximally spinning black hole. In general, we were only able to find a detectable signal for systems within $100 \,\rm kpc$. In the $\alpha -  d_{ L}$ plot, we observe that the $\rm SNR$ increases as $\alpha$ increases, due to the strong dependence of GW strain amplitude on $\alpha$ (\Eq{eqn:amp-scaling}) and the fact that the peak frequency $f_{\rm p}=4f_0-2z|\Gamma|/\pi$ reaches the most sensitive frequency range of DECIGO as $\alpha $ increases. In our analysis, we cannot increase $\alpha$ past $\alpha \sim 0.32$ since in that case the duration of the signal would be comparable to the merger timescale for these parameters (see \Fig{fig:region with large timescale}).   
In the $ M- d_{ L}$ plot, we note that the strongest $\rm SNR$ is produced by the smallest black hole masses, since larger masses imply smaller peak frequency, outside of the experiment's reach. Finally, in the $ q -  d_{ L}$ plot, we observe that the low mass ratios are favored for the particular choice of $\alpha$ and black hole masses.
 \par
In \Fig{fig:d_L-params decigo 095}, we show the same plots for an initially smaller black hole spin, $\tilde{a}_{\rm in} = 0.95$, keeping all the other parameters fixed. This results in a smaller mass for the GA, hence reducing the strength of the signal. Even smaller spins would restrict us to $\alpha<0.3$ (see \Fig{fig:Superrad rate}), further reducing the $\rm SNR$. 
\par
Regarding the LISA experiment, we did not find detectable SNR values for the benchmark parameters we considered above and chose not to report those results.\par
Overall, our results are not as promising as those reported in \cite{Peng:2025zca}. The combined effects of the decay rate, which suppresses the $\rm SNR$ by the factor $\frac{\eta}{|\Gamma|}$, the initial black hole spin, which for smaller values produces a less massive GA, and the annihilation of the GA into gravitons would make this a challenging signal to observe, even for binary systems close to the Earth, for which the merger rate is expected to be very small \cite{Cheung:2025grp,Arca-Sedda:2020lso}.   A more complete study that accounts for all these effects simultaneously is left for the future. Finally, in Appendix \ref{sec:relativistic}, we establish where the non-relativistic approximation that we have employed throughout fails significantly in comparison to the full relativistic result. The biggest discrepancy is observed in the decay rate, which deviates by an order of magnitude and would lead to a larger $\rm SNR$ if the less accurate non-relativistic result is used. This establishes the importance of accounting for relativistic effects in assessing the detectability of this signal. \par
Given the effectively monochromatic nature of the GW signals from level transitions (within the parameter space considered), even if such a signal were detected, it could be degenerate with other quasi-monochromatic astrophysical sources. It is worth further studies in combining the inspiral GW signals and transition GW signals to break the degeneracy.

%% file: chapter4.tex
\chapter{Conclusions and future directions}

In this dissertation, we have examined the consequences of the existence of very light axion-like particles in two different scenarios: the first was in cosmology and the influence these particles would have on cosmological observables when they are emitted by cosmic strings and the second was in astrophysics, and in particular in the astrophysics of black holes, where a cloud of these light axions can form around a spinning black hole. In the next two subsections, we summarize the key findings and discuss future directions of research.  

\section{Cosmic Strings}

In \Cref{sec:strings}, we have developed numerical and analytical techniques to derive constraints on pseudo Nambu-Goldstone bosons that are emitted by cosmic strings. The formalism builds on existing work that treats the scalar field as a summation of plane waves and expands it to treat the case of an expanding universe. In this framework, we derived the power spectrum explicitly and identified the associated transfer function that connects the super-horizon and the sub-horizon modes when the string network collapses. The general form of the former was given in \Eq{eqn:general power spectrum}, while the latter is plotted in \Fig{fig:transfer functions}. We exhaustively investigated the properties of the transfer function and how that depends on the parameters of the systems. The form of the spectrum depends very sensitively on the characteristic momentum of the particles $k_\star$, and in this model, that is identified with infrared cutoff of the theory, the inverse size of the co-moving horizon at the time of the collapse of the string network.

In our analysis, we have derived the form of the transfer functions for all scales, while also recovering the white noise plateau for scales larger than $k_\star$. We have done this numerically using a time evolving spectral index $q$ \cite{Gorghetto_2018} and we have also done this analytically assuming that the energy spectrum $\Omega_{\phi}(k)$ is independent of $k$, with the result given in \Eq{eqn:transfer small k} and \Eq{eqn:transfer large k}. 

Using the likelihood method and power spectrum data from five different experiments, we placed constraints in the parameter space of $m_{a}-f_{a}$ (\Fig{fig:constraints cosmic strings} and \Fig{fig:constraints cosmic strings n=1}), where $f_{a}$ is the energy scale of the U(1) symmetry whose spontaneous breaking gave rise to the pseudo Nambu-Goldstone boson with mass $m_{a}$. Compared to previous works that imposed a cut-off of the spectrum at $k = k_\star$~\cite{Gorghetto:2021fsn,Gorghetto:2025uls}, the constrained parameter space is now significantly expanded for a given observable.

We have also investigated the effects of  introducing the temperature dependence of the boson's mass. The main consequence is that the characteristic momentum $k_\star$ is shifted to lower values for fixed mass. As a result, there is enhanced power for given relic density and we are able to probe smaller values of $f_{a}$ with a given cosmological dataset. As we have pointed out, our analysis does not apply in the parameter space where the mass has not reached its temperature independent value before radiation-matter equality. One interesting research question would be to investigate how the density perturbations of this non-relativistic component evolve when its mass is still temperature dependent and reliably extend the constraints of \Fig{fig:constraints cosmic strings n=1} to masses $10^{-25} \textrm{eV} \lesssim m_{a} \lesssim 10^{-23} \rm eV$.     

Also, as we argued, in the parameter region $10^{-28} \textrm{eV} < m_{a} \lesssim 10^{-26} \textrm{eV} $ in \Fig{fig:constraints cosmic strings}, even though the string network has collapsed by the time of radiation matter equality, the particles remain relativistic. This means that there would be a component of dark radiation present during matter domination. An interesting question here is what kind of power spectrum this dark radiation would produce and how it would influence the observables we have discussed in this work. 
  
Another interesting avenue to explore further would be to consider masses lower than $\rm 10^{-28} eV$. In this mass range, the string network would be present after equality and the CMB anisotropies would be affected, an observation that has led to the constraint of the string tension $G \mu \lesssim 10^{-7}$, where G is Newton's constant and $\mu$ is the effective tension of cosmic strings~\cite{Charnock:2016nzm,Lizarraga:2016onn,Lopez-Eiguren:2017dmc,Pen:1997ae,Pogosian:1999np}. In particular, the fluctuations induced by cosmic strings are starkly different from the adiabatic fluctuations produced during inflation. While the latter exhibit the well-known acoustic peaks, the former are incoherent and the resulting spectrum features a single bump. It would be interesting to investigate if the constraints we have derived here are competitive to those derived from the anisotropies of the $\rm CMB$ due to the presence of the strings during recombination. 

\section{Superradiance}

In \Cref{sec:ga}, we developed a formalism to calculate the GW signal produced by the resonant transition of a GA in a binary system. In the perturbative regime where the companion's orbital distance is much larger than the size of the GA, the dominant multipole moment of the tidal field from the companion is the quadrupole moment $l_*=2$, which gravitationally mixes two GA states and thus triggers the transition.
\par
The two-state transition is modeled by the Landau-Zener system, which is characterized by the adiabaticity parameter that depends on the fine-structure constant $\alpha$, the mass ratio of the companion $q$, and the decay rate of the bosons in the final state back to the black hole. Under the assumption of a quasi-circular and equatorial orbit, the analytical formulae of the GW strain waveform and frequency spectrum are derived in the linearized orbital frequency evolution region. We validate the parameter region of $(\alpha,\,q)$ where the timescale of the transition is less than the merger time of the binary for the $|211\rangle \rightarrow |21 \text{-}1\rangle$ hyperfine and $|322\rangle \rightarrow |300\rangle$ fine transitions. In the present work, we restrict our attention to the $|211\rangle \rightarrow |21 \text{-}1\rangle$ hyperfine transition for mass ratios $q \lesssim 10^{-2} $ and $\alpha < 0.35$.    \par
The GW strain waveform can be considered as a continuous signal with amplitude modulated by the LZ transition. Its analytical formula is given in \Eq{eqn: plus strain} and it is plotted in \Fig{fig: strain freq time} for the benchmark parameters. We stress that the analytical waveform of the signal can serve as the template used by the match-filtering technique for future GW searches. The frequency spectrum of the strain is also derived analytically using the stationary phase approximation and its peak frequency is derived in terms of the fundamental parameters in \Eq{eqn:peak freq} and it is plotted in \Fig{fig: strain freq time}. In the parameter space of interest, the transition GW signal is always monochromatic.
\par
In terms of the detectability of the signal, we calculated the SNR for a variety of systems and determined that the DECIGO experiment would be suited for detecting it from systems that host IMBHs with $M \sim 150 M_{\odot}$, and $q \sim  1/150$, with $\alpha \sim 0.3$, which corresponds to boson masses $10^{-13} \rm eV$. The maximum distance that the signal can be detected at is $100 \,\rm kpc$. We also explored the possibility of a smaller initial black hole spin and found that it reduces the strength of the signal, as it implies a smaller mass for the GA.
\par
Finally, we commented on the validity of the non-relativistic approximation. We determined that the relativistic radial wavefunction is well approximated by its non-relativistic counterpart, even for moderately large values of $\alpha$, and the same holds true for the energy levels of the GA. We spotted the biggest discrepancy in the decay rate of the second state, where the relativistic result deviates up to an order of magnitude from the non-relativistic approximation for large $\alpha$. Given how sensitive the signal is to this decay rate, we high-lighted the importance of including the accurate result in the calculations.
\par
To streamline the analysis, several assumptions were made. We now discuss potential follow-up studies in the future to improve it:
\begin{itemize}
    \item An important continuation of the present study would be to extend its regime of validity to the entirety of the parameter space of \cref{fig:region with large timescale}, by including the non-linear evolution of the orbit throughout the transition.
    \item The simplifying assumption of 
quasicircular and equatorial orbits could be refined, especially for extreme mass ratio inspirals, which are expected to be highly eccentric. It has been found that if we relax these assumptions, transitions can occur at a variety of frequencies \cite{resonant_history,Berti:2019wnn}. The detectability of the GA transition signal in this case is something worth examining.
\item Taking into account the back-reaction of the transition to the orbit is another important topic. This is manifested through the conservation of angular momentum for the entire system.
The GWs that we have found will contribute to the total torque of the system, possibly leading to non-trivial dynamics of the orbit. This interplay between the back-reaction of the transition and the back-reaction of the emitted GWs might lead to unexpected and interesting behavior for the binary's orbit.
\item In terms of environmental effects, since the black holes that would produce the largest $\rm SNR$ are of intermediate mass that are typically found in dense globular clusters, incorporating the perturbations to the GA from nearby stars and quantifying their influence on the GA's evolution, perhaps in the spirit of \cite{Dandoy:2022prp}, would be an interesting direction.
\item Finally, a more complete treatment of the relativistic GA \cite{witte2025steppingsuperradianceconstraintsaxions}, the perturbations from the companion \cite{Li:2025ffh} and the effects those would have on the signal, would be an important addition to the present study.
\end{itemize}

%% file: referenceFile.bib
@article{2c5c-vz3h,
  title = {Echoes of global cosmic strings},
  author = {Dror, Jeff and Kyriazis, Antonios},
  journal = {Phys. Rev. D},
  volume = {114},
  issue = {3},
  pages = {035035},
  numpages = {14},
  year = {2026},
  month = {Aug},
  publisher = {American Physical Society},
  doi = {10.1103/2c5c-vz3h},
  url = {https://link.aps.org/doi/10.1103/2c5c-vz3h}
}

@article{LIGOScientific:2025rsn,
    author = "Abac, A. G. and others",
    collaboration = "LIGO Scientific, VIRGO, KAGRA",
    title = "{GW231123: a Binary Black Hole Merger with Total Mass 190-265 $M_{\odot}$}",
    eprint = "2507.08219",
    archivePrefix = "arXiv",
    primaryClass = "astro-ph.HE",
    reportNumber = "DCC: P2500026-v6",
    month = "7",
    year = "2025"
}

@article{Viel:2013fqw,
    author = "Viel, Matteo and Becker, George D. and Bolton, James S. and Haehnelt, Martin G.",
    title = "{Warm dark matter as a solution to the small scale crisis: New constraints from high redshift Lyman-{\ensuremath{\alpha}} forest data}",
    eprint = "1306.2314",
    archivePrefix = "arXiv",
    primaryClass = "astro-ph.CO",
    doi = "10.1103/PhysRevD.88.043502",
    journal = "Phys. Rev. D",
    volume = "88",
    pages = "043502",
    year = "2013"
}

@article{Garcia-Gallego:2026phh,
    author = "Garcia-Gallego, Olga and Ir{\v{s}}i{\v{c}}, Vid and Viel, Matteo and Haehnelt, Martin G. and Bolton, James S.",
   title = "{Post-inflationary axion constraints from the Lyman-$\alpha$ forest}",
    eprint = "2603.04401",
    archivePrefix = "arXiv",
    primaryClass = "astro-ph.CO",
    month = "3",
    journal="To appear",
    year = "2026"
}

@article{Buckley:2025zgh,
    author = "Buckley, Matthew R. and Du, Peizhi and Fernandez, Nicolas and Weikert, Mitchell J.",
    title = "{General constraints on isocurvature from the CMB and Ly-{\ensuremath{\alpha}} forest}",
    eprint = "2502.20434",
    archivePrefix = "arXiv",
    primaryClass = "astro-ph.CO",
    doi = "10.1088/1475-7516/2025/12/006",
    journal = "JCAP",
    volume = "12",
    pages = "006",
    year = "2025"
}

@article{LISA:2024hlh,
    author = "Colpi, Monica and others",
    collaboration = "LISA",
    title = "{LISA Definition Study Report}",
    eprint = "2402.07571",
    archivePrefix = "arXiv",
    primaryClass = "astro-ph.CO",
    month = "2",
    year = "2024"
}

@article{Arvanitaki_2011,
    author = "Arvanitaki, Asimina and Dubovsky, Sergei",
    title = "{Exploring the String Axiverse with Precision Black Hole Physics}",
    eprint = "1004.3558",
    archivePrefix = "arXiv",
    primaryClass = "hep-th",
    doi = "10.1103/PhysRevD.83.044026",
    journal = "Phys. Rev. D",
    volume = "83",
    pages = "044026",
    year = "2011"
}

@article{Baumann_2020,
    author = "Baumann, Daniel and Chia, Horng Sheng and Porto, Rafael A. and Stout, John",
    title = "{Gravitational Collider Physics}",
    eprint = "1912.04932",
    archivePrefix = "arXiv",
    primaryClass = "gr-qc",
    reportNumber = "DESY-19-221, DESY 19-221",
    doi = "10.1103/PhysRevD.101.083019",
    journal = "Phys. Rev. D",
    volume = "101",
    number = "8",
    pages = "083019",
    year = "2020"
}

@misc{AxionLimits,
  author       = {Ciaran O'Hare},
  title        = {cajohare/AxionLimits: AxionLimits},
  month        = jul,
  year         = 2020,
  publisher    = {Zenodo},
  version      = {v1.0},
  doi          = {10.5281/zenodo.3932430},
  howpublished = {\url{https://cajohare.github.io/AxionLimits/}}
}

@article{Clowe:2006eq,
    author = "Clowe, Douglas and Bradac, Marusa and Gonzalez, Anthony H. and Markevitch, Maxim and Randall, Scott W. and Jones, Christine and Zaritsky, Dennis",
    title = "{A direct empirical proof of the existence of dark matter}",
    eprint = "astro-ph/0608407",
    archivePrefix = "arXiv",
    reportNumber = "SLAC-PUB-12078",
    doi = "10.1086/508162",
    journal = "Astrophys. J. Lett.",
    volume = "648",
    pages = "L109--L113",
    year = "2006"
}

@book{Maggiore:2007ulw,
    author = "Maggiore, Michele",
    title = "{Gravitational Waves. Vol. 1: Theory and Experiments}",
    doi = "10.1093/acprof:oso/9780198570745.001.0001",
    isbn = "978-0-19-171766-6, 978-0-19-852074-0",
    publisher = "Oxford University Press",
    year = "2007"
}

@article{Arvanitaki_2015,
    author = "Arvanitaki, Asimina and Baryakhtar, Masha and Huang, Xinlu",
    title = "{Discovering the QCD Axion with Black Holes and Gravitational Waves}",
    eprint = "1411.2263",
    archivePrefix = "arXiv",
    primaryClass = "hep-ph",
    doi = "10.1103/PhysRevD.91.084011",
    journal = "Phys. Rev. D",
    volume = "91",
    number = "8",
    pages = "084011",
    year = "2015"
}

@article{Tong_2022,
    author = "Tong, Xi and Wang, Yi and Zhu, Hui-Yu",
    title = "{Termination of superradiance from a binary companion}",
    eprint = "2205.10527",
    archivePrefix = "arXiv",
    primaryClass = "gr-qc",
    doi = "10.1103/PhysRevD.106.043002",
    journal = "Phys. Rev. D",
    volume = "106",
    number = "4",
    pages = "043002",
    year = "2022"
}

@article{Siemonsen:2022yyf,
    author = "Siemonsen, Nils and May, Taillte and East, William E.",
    title = "{Modeling the black hole superradiance gravitational waveform}",
    eprint = "2211.03845",
    archivePrefix = "arXiv",
    primaryClass = "gr-qc",
    doi = "10.1103/PhysRevD.107.104003",
    journal = "Phys. Rev. D",
    volume = "107",
    number = "10",
    pages = "104003",
    year = "2023"
}

@article{Yoshino:2013ofa,
    author = "Yoshino, Hirotaka and Kodama, Hideo",
    title = "{Gravitational radiation from an axion cloud around a black hole: Superradiant phase}",
    eprint = "1312.2326",
    archivePrefix = "arXiv",
    primaryClass = "gr-qc",
    reportNumber = "KEK-TH-1694",
    doi = "10.1093/ptep/ptu029",
    journal = "PTEP",
    volume = "2014",
    pages = "043E02",
    year = "2014"
}

@article{Baumann_2019,
    author = "Baumann, Daniel and Chia, Horng Sheng and Porto, Rafael A.",
    title = "{Probing Ultralight Bosons with Binary Black Holes}",
    eprint = "1804.03208",
    archivePrefix = "arXiv",
    primaryClass = "gr-qc",
    reportNumber = "DESY-18-060, DESY 18-060",
    doi = "10.1103/PhysRevD.99.044001",
    journal = "Phys. Rev. D",
    volume = "99",
    number = "4",
    pages = "044001",
    year = "2019"
}

@article{spectra,
    author = "Baumann, Daniel and Chia, Horng Sheng and Stout, John and ter Haar, Lotte",
    title = "{The Spectra of Gravitational Atoms}",
    eprint = "1908.10370",
    archivePrefix = "arXiv",
    primaryClass = "gr-qc",
    doi = "10.1088/1475-7516/2019/12/006",
    journal = "JCAP",
    volume = "12",
    pages = "006",
    year = "2019"
}

@book{Carroll:2004st,
    author = "Carroll, Sean M.",
    title = "{Spacetime and Geometry}: {An Introduction to General Relativity}",
    doi = "10.1017/9781108770385",
    isbn = "978-0-8053-8732-2, 978-1-108-48839-6, 978-1-108-77555-7",
    publisher = "Cambridge University Press",
    month = "7",
    year = "2019"
}

@article{Detweiler:1980uk,
    author = "Detweiler, Steven L.",
    title = "{KLEIN-GORDON EQUATION AND ROTATING BLACK HOLES}",
    doi = "10.1103/PhysRevD.22.2323",
    journal = "Phys. Rev. D",
    volume = "22",
    pages = "2323--2326",
    year = "1980"
}

@article{self-interactions,
    author = "Baryakhtar, Masha and Galanis, Marios and Lasenby, Robert and Simon, Olivier",
    title = "{Black hole superradiance of self-interacting scalar fields}",
    eprint = "2011.11646",
    archivePrefix = "arXiv",
    primaryClass = "hep-ph",
    doi = "10.1103/PhysRevD.103.095019",
    journal = "Phys. Rev. D",
    volume = "103",
    number = "9",
    pages = "095019",
    year = "2021"
}

@article{Gravitaitonal_wave_searches,
    author = "Brito, Richard and Ghosh, Shrobana and Barausse, Enrico and Berti, Emanuele and Cardoso, Vitor and Dvorkin, Irina and Klein, Antoine and Pani, Paolo",
    title = "{Gravitational wave searches for ultralight bosons with LIGO and LISA}",
    eprint = "1706.06311",
    archivePrefix = "arXiv",
    primaryClass = "gr-qc",
    doi = "10.1103/PhysRevD.96.064050",
    journal = "Phys. Rev. D",
    volume = "96",
    number = "6",
    pages = "064050",
    year = "2017"
}

@article{axion_cloud_backreaction,
    author = "Takahashi, Takuya and Omiya, Hidetoshi and Tanaka, Takahiro",
    title = "{Axion cloud evaporation during inspiral of black hole binaries: The effects of backreaction and radiation}",
    eprint = "2112.05774",
    archivePrefix = "arXiv",
    primaryClass = "gr-qc",
    doi = "10.1093/ptep/ptac044",
    journal = "PTEP",
    volume = "2022",
    number = "4",
    pages = "043E01",
    year = "2022"
}

@article{Landau,
    author = "Landau, Lev Davidovich",
    editor = "ter Haar, D.",
    title = "{A theory of energy transfer. 2.}",
    doi = "10.1016/B978-0-08-010586-4.50014-6",
    journal = "Phys. Z. Sowjetunion",
    volume = "2",
    year = "1932"
}

@article{zener,
    author = "Zener, Clarence",
    title = "{Nonadiabatic crossing of energy levels}",
    doi = "10.1098/rspa.1932.0165",
    journal = "Proc. Roy. Soc. Lond. A",
    volume = "137",
    pages = "696--702",
    year = "1932"
}

@article{resonant_history,
    author = "Tomaselli, Giovanni Maria and Spieksma, Thomas F. M. and Bertone, Gianfranco",
    title = "{Resonant history of gravitational atoms in black hole binaries}",
    eprint = "2403.03147",
    archivePrefix = "arXiv",
    primaryClass = "gr-qc",
    doi = "10.1103/PhysRevD.110.064048",
    journal = "Phys. Rev. D",
    volume = "110",
    number = "6",
    pages = "064048",
    year = "2024"
}

@article{legacy,
    author = "Tomaselli, Giovanni Maria and Spieksma, Thomas F. M. and Bertone, Gianfranco",
    title = "{Legacy of Boson Clouds on Black Hole Binaries}",
    eprint = "2407.12908",
    archivePrefix = "arXiv",
    primaryClass = "gr-qc",
    doi = "10.1103/PhysRevLett.133.121402",
    journal = "Phys. Rev. Lett.",
    volume = "133",
    number = "12",
    pages = "121402",
    year = "2024"
}

@article{extreme_mass_ratio,
    author = "Takahashi, Takuya and Omiya, Hidetoshi and Tanaka, Takahiro",
    title = "{Evolution of binary systems accompanying axion clouds in extreme mass ratio inspirals}",
    eprint = "2301.13213",
    archivePrefix = "arXiv",
    primaryClass = "gr-qc",
    doi = "10.1103/PhysRevD.107.103020",
    journal = "Phys. Rev. D",
    volume = "107",
    number = "10",
    pages = "103020",
    year = "2023"
}

@article{Brito_2020,
    author = "Brito, Richard and Cardoso, Vitor and Pani, Paolo",
    title = "{Superradiance}: {New Frontiers in Black Hole
Physics}",
    eprint = "1501.06570",
    archivePrefix = "arXiv",
    primaryClass = "gr-qc",
    doi = "10.1007/978-3-319-19000-6",
    journal = "Lect. Notes Phys.",
    volume = "906",
    pages = "pp.1--237",
    year = "2015"
}

@article{_nal_2021,
    author = {\"Unal, Caner and Pacucci, Fabio and Loeb, Abraham},
    title = "{Properties of ultralight bosons from heavy quasar spins via superradiance}",
    eprint = "2012.12790",
    archivePrefix = "arXiv",
    primaryClass = "hep-ph",
    doi = "10.1088/1475-7516/2021/05/007",
    journal = "JCAP",
    volume = "05",
    pages = "007",
    year = "2021"
}

@article{hoof2024gettingblackholesuperradiance,
    author = "Hoof, Sebastian and Marsh, David J. E. and Sisk-Reyn\'es, J\'ulia and Matthews, James H. and Reynolds, Christopher",
    title = "{Getting More Out of Black Hole Superradiance: a Statistically Rigorous Approach to Ultralight Boson Constraints}",
    eprint = "2406.10337",
    archivePrefix = "arXiv",
    primaryClass = "hep-ph",
    month = "6",
    year = "2024"
}

@article{witte2025steppingsuperradianceconstraintsaxions,
    author = "Witte, Samuel J. and Mummery, Andrew",
    title = "{Stepping Up Superradiance Constraints on Axions}",
    eprint = "2412.03655",
    archivePrefix = "arXiv",
    primaryClass = "hep-ph",
    month = "12",
    year = "2024"
}

@article{Ionization,
    author = "Baumann, Daniel and Bertone, Gianfranco and Stout, John and Tomaselli, Giovanni Maria",
    title = "{Ionization of gravitational atoms}",
    eprint = "2112.14777",
    archivePrefix = "arXiv",
    primaryClass = "gr-qc",
    doi = "10.1103/PhysRevD.105.115036",
    journal = "Phys. Rev. D",
    volume = "105",
    number = "11",
    pages = "115036",
    year = "2022"
}

@article{sharp_signals,
    author = "Baumann, Daniel and Bertone, Gianfranco and Stout, John and Tomaselli, Giovanni Maria",
    title = "{Sharp Signals of Boson Clouds in Black Hole Binary Inspirals}",
    eprint = "2206.01212",
    archivePrefix = "arXiv",
    primaryClass = "gr-qc",
    doi = "10.1103/PhysRevLett.128.221102",
    journal = "Phys. Rev. Lett.",
    volume = "128",
    number = "22",
    pages = "221102",
    year = "2022"
}

@article{Kyriazis:2026vkz,
    author = "Kyriazis, Antonios and Yang, Fengwei and Zhou, Siyu",
    title = "{Heating Up the Black Hole X-ray Binary Accretion Disk by Superradiance}",
    eprint = "2608.18200",
    archivePrefix = "arXiv",
    primaryClass = "astro-ph.HE",
    month = "8",
    year = "2026"
}

@article{self_interaction_binary,
    author = "Takahashi, Takuya and Omiya, Hidetoshi and Tanaka, Takahiro",
    title = "{Self-interacting axion clouds around rotating black holes in binary systems}",
    eprint = "2408.08349",
    archivePrefix = "arXiv",
    primaryClass = "gr-qc",
    reportNumber = "RUP-24-12",
    doi = "10.1103/PhysRevD.110.104038",
    journal = "Phys. Rev. D",
    volume = "110",
    number = "10",
    pages = "104038",
    year = "2024"
}

@article{Guo:2024iye,
    author = "Guo, Ao and Zhang, Jun and Yang, Huan",
    title = "{Superradiant clouds may be relevant for close compact object binaries}",
    eprint = "2401.15003",
    archivePrefix = "arXiv",
    primaryClass = "gr-qc",
    doi = "10.1103/PhysRevD.110.023022",
    journal = "Phys. Rev. D",
    volume = "110",
    number = "2",
    pages = "023022",
    year = "2024"
}

@article{axion_annihilation,
    author = "Yang, Jing and Huang, Fa Peng",
    title = "{Gravitational waves from axions annihilation through quantum field theory}",
    eprint = "2306.12375",
    archivePrefix = "arXiv",
    primaryClass = "hep-ph",
    doi = "10.1103/PhysRevD.108.103002",
    journal = "Phys. Rev. D",
    volume = "108",
    number = "10",
    pages = "103002",
    year = "2023"
}

@article{superradiance_string_theory,
    author = "Mehta, Viraf M. and Demirtas, Mehmet and Long, Cody and Marsh, David J. E. and McAllister, Liam and Stott, Matthew J.",
    title = "{Superradiance in string theory}",
    eprint = "2103.06812",
    archivePrefix = "arXiv",
    primaryClass = "hep-th",
    doi = "10.1088/1475-7516/2021/07/033",
    journal = "JCAP",
    volume = "07",
    pages = "033",
    year = "2021"
}

@article{Yang_2023,
   title={Gravitational waves from axions annihilation through quantum field theory},
   volume={108},
   ISSN={2470-0029},
   url={http://dx.doi.org/10.1103/PhysRevD.108.103002},
   DOI={10.1103/physrevd.108.103002},
   number={10},
   journal={Physical Review D},
   publisher={American Physical Society (APS)},
   author={Yang, Jing and Huang, Fa Peng},
   year={2023},
   month=nov }

@article{Berti:2019wnn,
    author = "Berti, Emanuele and Brito, Richard and Macedo, Caio F. B. and Raposo, Guilherme and Rosa, Joao Luis",
    title = "{Ultralight boson cloud depletion in binary systems}",
    eprint = "1904.03131",
    archivePrefix = "arXiv",
    primaryClass = "gr-qc",
    doi = "10.1103/PhysRevD.99.104039",
    journal = "Phys. Rev. D",
    volume = "99",
    number = "10",
    pages = "104039",
    year = "2019"
}

@article{decayrate1,
  title = {Landau-Zener transition to a decaying level},
  author = {Akulin, V. M. and Schleich, W. P.},
  journal = {Phys. Rev. A},
  volume = {46},
  issue = {7},
  pages = {4110--4113},
  numpages = {0},
  year = {1992},
  month = {Oct},
  publisher = {American Physical Society},
  doi = {10.1103/PhysRevA.46.4110},
  url = {https://link.aps.org/doi/10.1103/PhysRevA.46.4110}
}

@article{decayrate2,
  title = {Pulsed excitation of a transition to a decaying level},
  author = {Vitanov, N. V. and Stenholm, S.},
  journal = {Phys. Rev. A},
  volume = {55},
  issue = {4},
  pages = {2982--2988},
  numpages = {0},
  year = {1997},
  month = {Apr},
  publisher = {American Physical Society},
  doi = {10.1103/PhysRevA.55.2982},
  url = {https://link.aps.org/doi/10.1103/PhysRevA.55.2982}
}

@article{Bo_kovi__2024,
    author = "Bo\v{s}kovi\'c, Mateja and Koschnitzke, Matthias and Porto, Rafael A.",
    title = "{Signatures of Ultralight Bosons in the Orbital Eccentricity of Binary Black Holes}",
    eprint = "2403.02415",
    archivePrefix = "arXiv",
    primaryClass = "gr-qc",
    reportNumber = "DESY-24-030",
    doi = "10.1103/PhysRevLett.133.121401",
    journal = "Phys. Rev. Lett.",
    volume = "133",
    number = "12",
    pages = "121401",
    year = "2024"
}

@article{hyperfine_trans_are_favoured,
    author = "Tong, Xi and Wang, Yi and Zhu, Hui-Yu",
    title = "{Gravitational Collider Physics via Pulsar\textendash{}Black Hole Binaries II: Fine and Hyperfine Structures Are Favored}",
    eprint = "2106.13484",
    archivePrefix = "arXiv",
    primaryClass = "astro-ph.HE",
    doi = "10.3847/1538-4357/ac36db",
    journal = "Astrophys. J.",
    volume = "924",
    number = "2",
    pages = "99",
    year = "2022"
}

@inproceedings{askar2024intermediatemassblackholesstar,
    author = "Askar, Abbas and Baldassare, Vivienne F. and Mezcua, Mar",
    title = "{Intermediate-Mass Black Holes in Star Clusters and Dwarf Galaxies}",
    eprint = "2311.12118",
    archivePrefix = "arXiv",
    primaryClass = "astro-ph.GA",
    month = "11",
    year = "2023"
}

@misc{D,
author={N. M. Temme },
title="{NIST Digital Library of Mathematical Functions}",
url={https://dlmf.nist.gov/12}}

@book{Shankar:102017,
      author        = "Shankar, Ramamurti",
      title         = "{Principles of quantum mechanics}",
      publisher     = "Plenum",
      address       = "New York, NY",
      year          = "1980",
      url           = "https://cds.cern.ch/record/102017",
}

@article{Dandoy:2022prp,
    author = "Dandoy, Virgile and Schwetz, Thomas and Todarello, Elisa",
    title = "{A self-consistent wave description of axion miniclusters and their survival in the galaxy}",
    eprint = "2206.04619",
    archivePrefix = "arXiv",
    primaryClass = "astro-ph.CO",
    doi = "10.1088/1475-7516/2022/09/081",
    journal = "JCAP",
    volume = "09",
    pages = "081",
    year = "2022"
}

@article{mehta2021superradianceexclusionslandscapetype,
    author = "Mehta, Viraf M. and Demirtas, Mehmet and Long, Cody and Marsh, David J. E. and Mcallister, Liam and Stott, Matthew J.",
    title = "{Superradiance Exclusions in the Landscape of Type IIB String Theory}",
    eprint = "2011.08693",
    archivePrefix = "arXiv",
    primaryClass = "hep-th",
    reportNumber = "KCL-PH-TH/2020-77",
    month = "11",
    year = "2020"
}

@article{deci_Hz,
    author = "Omiya, Hidetoshi and Takahashi, Takuya and Tanaka, Takahiro and Yoshino, Hirotaka",
    title = "{Deci-Hz gravitational waves from the self-interacting axion cloud around a rotating stellar mass black hole}",
    eprint = "2404.16265",
    archivePrefix = "arXiv",
    primaryClass = "gr-qc",
    doi = "10.1103/PhysRevD.110.044002",
    journal = "Phys. Rev. D",
    volume = "110",
    number = "4",
    pages = "044002",
    year = "2024"
}

@article{Collaviti_2024,
    author = "Collaviti, Spencer and Sun, Ling and Galanis, Marios and Baryakhtar, Masha",
    title = "{Observational prospects of self-interacting scalar superradiance with next-generation gravitational-wave detectors}",
    eprint = "2407.04304",
    archivePrefix = "arXiv",
    primaryClass = "gr-qc",
    doi = "10.1088/1361-6382/ad96ff",
    journal = "Class. Quant. Grav.",
    volume = "42",
    number = "2",
    pages = "025006",
    year = "2025"
}

@article{DellaMonica:2025zby,
    author = "Della Monica, Riccardo and Brito, Richard",
    title = "{Detectability of gravitational atoms in black hole binaries with the Einstein Telescope}",
    eprint = "2503.23419",
    archivePrefix = "arXiv",
    primaryClass = "gr-qc",
    reportNumber = "ET-0056A-25",
    month = "3",
    year = "2025"
}

@article{Zhang:2018kib,
    author = "Zhang, Jun and Yang, Huan",
    title = "{Gravitational floating orbits around hairy black holes}",
    eprint = "1808.02905",
    archivePrefix = "arXiv",
    primaryClass = "gr-qc",
    doi = "10.1103/PhysRevD.99.064018",
    journal = "Phys. Rev. D",
    volume = "99",
    number = "6",
    pages = "064018",
    year = "2019"
}

@article{May:2024npn,
    author = "May, Taillte and East, William E. and Siemonsen, Nils",
    title = "{Self-Gravity Effects of Ultralight Boson Clouds Formed by Black Hole Superradiance}",
    eprint = "2410.21442",
    archivePrefix = "arXiv",
    primaryClass = "gr-qc",
    month = "10",
    year = "2024"
}

@article{DECIGO,
    author = "Yagi, Kent and Seto, Naoki",
    title = "{Detector configuration of DECIGO/BBO and identification of cosmological neutron-star binaries}",
    eprint = "1101.3940",
    archivePrefix = "arXiv",
    primaryClass = "astro-ph.CO",
    doi = "10.1103/PhysRevD.83.044011",
    journal = "Phys. Rev. D",
    volume = "83",
    pages = "044011",
    year = "2011",
    note = "[Erratum: Phys.Rev.D 95, 109901 (2017)]"
}

@article{Peng:2025zca,
    author = "Peng, Si-Tong and Zhang, Jun",
    title = "{Gravitational Waves from Superradiant Cloud Level Transition}",
    eprint = "2504.00728",
    archivePrefix = "arXiv",
    primaryClass = "gr-qc",
    month = "4",
    year = "2025"
}

@phdthesis{Tomaselli:2024faa,
    author = "Tomaselli, Giovanni Maria",
    title = "{Gravitational atoms and black hole binaries}",
    eprint = "2412.12526",
    archivePrefix = "arXiv",
    primaryClass = "gr-qc",
    school = "Amsterdam U.",
    year = "2024"
}

@misc{hoof_git,
  author       = {Hoof,S.},
  howpublished         ={\url{https://github.com/sebhoof/bhsr}}
}

@article{MacLeod:2015bpa,
    author = "MacLeod, Morgan and Trenti, Michele and Ramirez-Ruiz, Enrico",
    title = "{The Close Stellar Companions to Intermediate Mass Black Holes}",
    eprint = "1508.07000",
    archivePrefix = "arXiv",
    primaryClass = "astro-ph.HE",
    doi = "10.3847/0004-637X/819/1/70",
    journal = "Astrophys. J.",
    volume = "819",
    number = "1",
    pages = "70",
    year = "2016"
}

@article{Berti:2008af,
    author = "Berti, Emanuele and Volonteri, Marta",
    title = "{Cosmological black hole spin evolution by mergers and accretion}",
    eprint = "0802.0025",
    archivePrefix = "arXiv",
    primaryClass = "astro-ph",
    doi = "10.1086/590379",
    journal = "Astrophys. J.",
    volume = "684",
    pages = "822--828",
    year = "2008"
}

@article{Borchers:2025sid,
    author = "Borchers, Angela and Ye, Claire S. and Fishbach, Maya",
    title = "{Gravitational-wave kicks impact spins of black holes from hierarchical mergers}",
    eprint = "2503.21278",
    archivePrefix = "arXiv",
    primaryClass = "astro-ph.HE",
    month = "3",
    year = "2025"
}

@article{Li:2025ffh,
    author = "Li, Dongjun and Weller, Colin and Bourg, Patrick and LaHaye, Michael and Yunes, Nicol{\'a}s and Yang, Huan",
    title = "{Extreme mass-ratio inspiral within an ultralight scalar cloud I. Scalar radiation}",
    eprint = "2507.02045",
    archivePrefix = "arXiv",
    primaryClass = "gr-qc",
    month = "7",
    year = "2025"
}

@article{Droz:1999qx,
    author = "Droz, Serge and Knapp, Daniel J. and Poisson, Eric and Owen, Benjamin J.",
    title = "{Gravitational waves from inspiraling compact binaries: Validity of the stationary phase approximation to the Fourier transform}",
    eprint = "gr-qc/9901076",
    archivePrefix = "arXiv",
    doi = "10.1103/PhysRevD.59.124016",
    journal = "Phys. Rev. D",
    volume = "59",
    pages = "124016",
    year = "1999"
}

@article{LIGOScientific:2021rnv,
    author = "Abbott, R. and others",
    collaboration = "LIGO Scientific, Virgo, KAGRA",
    title = "{All-sky search for gravitational wave emission from scalar boson clouds around spinning black holes in LIGO O3 data}",
    eprint = "2111.15507",
    archivePrefix = "arXiv",
    primaryClass = "astro-ph.HE",
    reportNumber = "P2100343",
    doi = "10.1103/PhysRevD.105.102001",
    journal = "Phys. Rev. D",
    volume = "105",
    number = "10",
    pages = "102001",
    year = "2022"
}

@article{Arca-Sedda:2020lso,
    author = "Arca-Sedda, Manuel and Amaro-Seoane, Pau and Chen, Xian",
    title = "{Merging stellar and intermediate-mass black holes in dense clusters: implications for LIGO, LISA, and the next generation of gravitational wave detectors}",
    eprint = "2007.13746",
    archivePrefix = "arXiv",
    primaryClass = "astro-ph.GA",
    doi = "10.1051/0004-6361/202037785",
    journal = "Astron. Astrophys.",
    volume = "652",
    pages = "A54",
    year = "2021"
}

@article{Cheung:2025grp,
    author = "Cheung, Mark Ho-Yeuk and Wadekar, Digvijay and Mehta, Ajit Kumar and Islam, Tousif and Roulet, Javier and Berti, Emanuele and Venumadhav, Tejaswi and Zackay, Barak and Zaldarriaga, Matias",
    title = "{Searching for intermediate mass ratio binary black hole mergers in the third observing run of LIGO-Virgo-KAGRA}",
    eprint = "2507.01083",
    archivePrefix = "arXiv",
    primaryClass = "gr-qc",
    month = "7",
    year = "2025"
}

@article{Kiziltan:2017ijz,
    author = {K{\i}z{\i}ltan, B{\"u}lent and Baumgardt, Holger and Loeb, Abraham},
    title = "{An intermediate-mass black hole in the centre of the globular cluster 47 Tucanae}",
    eprint = "1702.02149",
    archivePrefix = "arXiv",
    primaryClass = "astro-ph.GA",
    doi = "10.1038/nature21361",
    journal = "Nature",
    volume = "542",
    number = "7640",
    pages = "203--205",
    year = "2017"
}

@article{Abbate_2019,
doi = {10.3847/2041-8213/ab46c3},
url = {https://doi.org/10.3847/2041-8213/ab46c3},
year = {2019},
month = {oct},
publisher = {The American Astronomical Society},
volume = {884},
number = {1},
pages = {L9},
author = {Abbate, Federico and Possenti, Andrea and Colpi, Monica and Spera, Mario},
title = {Evidence of Nonluminous Matter in the Center of M62},
journal = {The Astrophysical Journal Letters}
}

@article{10.1093/mnras/stw1779,
    author = {Sollima, A. and Ferraro, F. R. and Lovisi, L. and Contenta, F. and Vesperini, E. and Origlia, L. and Lapenna, E. and Lanzoni, B. and Mucciarelli, A. and Dalessandro, E. and Pallanca, C.},
    title = {Searching in the dark: the dark mass content of the Milky Way globular clusters NGC288 and NGC6218},
    journal = {Monthly Notices of the Royal Astronomical Society},
    volume = {462},
    number = {2},
    pages = {1937-1951},
    year = {2016},
    month = {07},
    issn = {0035-8711},
    doi = {10.1093/mnras/stw1779},
    url = {https://doi.org/10.1093/mnras/stw1779}
}

@ARTICLE{2019MNRAS.487.2685E,
       author = {{Erkal}, D. and {Belokurov}, V. and {Laporte}, C.~F.~P. and {Koposov}, S.~E. and {Li}, T.~S. and {Grillmair}, C.~J. and {Kallivayalil}, N. and {Price-Whelan}, A.~M. and {Evans}, N.~W. and {Hawkins}, K. and {Hendel}, D. and {Mateu}, C. and {Navarro}, J.~F. and {del Pino}, A. and {Slater}, C.~T. and {Sohn}, S.~T. and {Orphan Aspen Treasury Collaboration}},
        title = "{The total mass of the Large Magellanic Cloud from its perturbation on the Orphan stream}",
      journal = {\mnras},
         year = 2019,
        month = aug,
       volume = {487},
       number = {2},
        pages = {2685-2700},
          doi = {10.1093/mnras/stz1371},
archivePrefix = {arXiv},
       eprint = {1812.08192},
 primaryClass = {astro-ph.GA},
       adsurl = {https://ui.adsabs.harvard.edu/abs/2019MNRAS.487.2685E}
}

@ARTICLE{2017ApJ...846...14B,
       author = {{Boyce}, H. and {L{\"u}tzgendorf}, N. and {van der Marel}, R.~P. and {Baumgardt}, H. and {Kissler-Patig}, M. and {Neumayer}, N. and {de Zeeuw}, P.~T.},
        title = "{An Upper Limit on the Mass of a Central Black Hole in the Large Magellanic Cloud from the Stellar Rotation Field}",
      journal = {\apj},
         year = 2017,
        month = sep,
       volume = {846},
       number = {1},
          eid = {14},
        pages = {14},
          doi = {10.3847/1538-4357/aa830c},
archivePrefix = {arXiv},
       eprint = {1612.00045},
 primaryClass = {astro-ph.GA},
       adsurl = {https://ui.adsabs.harvard.edu/abs/2017ApJ...846...14B}
}

@ARTICLE{2007MNRAS.376L..29G,
       author = {{Gualandris}, Alessia and {Portegies Zwart}, Simon},
        title = "{A hypervelocity star from the Large Magellanic Cloud}",
      journal = {\mnras},
         year = 2007,
        month = mar,
       volume = {376},
       number = {1},
        pages = {L29-L33},
          doi = {10.1111/j.1745-3933.2007.00280.x},
archivePrefix = {arXiv},
       eprint = {astro-ph/0612673},
 primaryClass = {astro-ph},
       adsurl = {https://ui.adsabs.harvard.edu/abs/2007MNRAS.376L..29G}
}

@article{Boddy:2025oxn,
    author = "Boddy, Kimberly K. and Dror, Jeff A. and Lam, Austin",
    title = "{Ultralight Dark Matter Statistics for Pulsar Timing Detection}",
    eprint = "2502.15874",
    archivePrefix = "arXiv",
    primaryClass = "hep-ph",
    doi = "10.1103/hgnx-w1dn",
    journal = "Phys. Rev. Lett.",
    volume = "135",
    number = "10",
    pages = "101001",
    year = "2025"
}

@misc{macinnis2025cmbhdprobedarkmatter,
      title={CMB-HD as a Probe of Dark Matter on Sub-Galactic Scales}, 
      author={Amanda MacInnis and Neelima Sehgal},
      year={2025},
      eprint={2405.12220},
      archivePrefix={arXiv},
      primaryClass={astro-ph.CO},
      url={https://arxiv.org/abs/2405.12220}, 
}

@article{Liu:2024pjg,
    author = "Liu, Rayne and Hu, Wayne and Xiao, Huangyu",
    title = "{Warm and fuzzy dark matter: Free streaming of wave dark matter}",
    eprint = "2406.12970",
    archivePrefix = "arXiv",
    primaryClass = "hep-ph",
    reportNumber = "FERMILAB-PUB-24-0296-T",
    doi = "10.1103/PhysRevD.111.023535",
    journal = "Phys. Rev. D",
    volume = "111",
    number = "2",
    pages = "023535",
    year = "2025"
}

@article{Harigaya:2025pox,
    author = "Harigaya, Keisuke and Hu, Wayne and Liu, Rayne and Xiao, Huangyu",
    title = "{Universal lower bound on the axion decay constant from free streaming effects}",
    eprint = "2507.01956",
    archivePrefix = "arXiv",
    primaryClass = "astro-ph.CO",
    reportNumber = "FERMILAB-PUB-25-0430-T",
    month = "7",
    year = "2025",
journal="",
}

@misc{amin2024lowerbounddarkmatter,
      title={A lower bound on dark matter mass}, 
      author={Mustafa A. Amin and Mehrdad Mirbabayi},
      year={2024},
      eprint={2211.09775},
      archivePrefix={arXiv},
      primaryClass={hep-ph},
      url={https://arxiv.org/abs/2211.09775}, 
}

@article{Pogosian:1999np,
    author = "Pogosian, Levon and Vachaspati, Tanmay",
    title = "{Cosmic microwave background anisotropy from wiggly strings}",
    eprint = "astro-ph/9903361",
    archivePrefix = "arXiv",
    reportNumber = "CWRU-P15-99",
    doi = "10.1103/PhysRevD.60.083504",
    journal = "Phys. Rev. D",
    volume = "60",
    pages = "083504",
    year = "1999"
}

@article{Pen:1997ae,
    author = "Pen, Ue-Li and Seljak, Uros and Turok, Neil",
    title = "{Power spectra in global defect theories of cosmic structure formation}",
    eprint = "astro-ph/9704165",
    archivePrefix = "arXiv",
    doi = "10.1103/PhysRevLett.79.1611",
    journal = "Phys. Rev. Lett.",
    volume = "79",
    pages = "1611--1614",
    year = "1997"
}

@inbook{Sikivie_2008,
   title={Axion Cosmology},
   ISBN={9783540735182},
   ISSN={0075-8450},
   url={http://dx.doi.org/10.1007/978-3-540-73518-2_2},
   DOI={10.1007/978-3-540-73518-2_2},
   booktitle={Axions},
   publisher={Springer Berlin Heidelberg},
   author={Sikivie, Pierre},
   year={2008},
   pages={19–50} }

@article{Gorghetto_2018,
   title={Axions from strings: the attractive solution},
   volume={2018},
   ISSN={1029-8479},
   url={http://dx.doi.org/10.1007/JHEP07(2018)151},
   DOI={10.1007/jhep07(2018)151},
   number={7},
   journal={Journal of High Energy Physics},
   publisher={Springer Science and Business Media LLC},
   author={Gorghetto, Marco and Hardy, Edward and Villadoro, Giovanni},
   year={2018},
   month=jul }

@misc{Gorghetto:2025uls,
    author = "Gorghetto, Marco and Trifinopoulos, Sokratis and Valogiannis, Georgios",
    title = "{Large-Scale Structure Probes of the Post-Inflationary Axiverse}",
    eprint = "2511.04734",
    archivePrefix = "arXiv",
    primaryClass = "astro-ph.CO",
    reportNumber = "DESY-25-147, CERN-TH-2025-193",
    month = "11",
    year = "2025"
}

@misc{Amin:2025ayf,
    author = "Amin, Mustafa A. and Delos, M. Sten and Yang, Kiaxin",
    title = "{Multi-species Dark Matter with Warmth and Randomness}",
    eprint = "2510.15046",
    archivePrefix = "arXiv",
    primaryClass = "astro-ph.CO",
    month = "10",
    year = "2025"
}

@article{Gorghetto:2022ikz,
    author = "Gorghetto, Marco and Hardy, Edward",
    title = "{Post-inflationary axions: a minimal target for axion haloscopes}",
    eprint = "2212.13263",
    archivePrefix = "arXiv",
    primaryClass = "hep-ph",
    doi = "10.1007/JHEP05(2023)030",
    journal = "JHEP",
    volume = "05",
    pages = "030",
    year = "2023"
}

@article{Benabou:2024msj,
    author = "Benabou, Joshua N. and Buschmann, Malte and Foster, Joshua W. and Safdi, Benjamin R.",
    title = "{Axion Mass Prediction from Adaptive Mesh Refinement Cosmological Lattice Simulations}",
    eprint = "2412.08699",
    archivePrefix = "arXiv",
    primaryClass = "hep-ph",
    reportNumber = "FERMILAB-PUB-24-0912-T",
    doi = "10.1103/6v21-d6sj",
    journal = "Phys. Rev. Lett.",
    volume = "134",
    number = "24",
    pages = "241003",
    year = "2025"
}

@misc{Chathirathas:2025aan,
    author = "Chathirathas, Kierthika and Schwetz, Thomas",
    title = "{How light can ALP dark matter be?}",
    eprint = "2511.15790",
    archivePrefix = "arXiv",
    primaryClass = "hep-ph",
    month = "11",
    year = "2025"
}

@article{Dror_2021,
   title={Cosmic axion background},
   volume={103},
   ISSN={2470-0029},
   url={http://dx.doi.org/10.1103/PhysRevD.103.115004},
   DOI={10.1103/physrevd.103.115004},
   number={11},
   journal={Physical Review D},
   publisher={American Physical Society (APS)},
   author={Dror, Jeff A. and Murayama, Hitoshi and Rodd, Nicholas L.},
   year={2021},
   month=jun }

@book{Maggiore:2018sht,
    author = "Maggiore, Michele",
    title = "{Gravitational Waves. Vol. 2: Astrophysics and Cosmology}",
    isbn = "978-0-19-857089-9",
    publisher = "Oxford University Press",
    month = "3",
    year = "2018"
}

@article{Bardeen:1985tr,
    author = "Bardeen, James M. and Bond, J. R. and Kaiser, Nick and Szalay, A. S.",
    title = "{The Statistics of Peaks of Gaussian Random Fields}",
    reportNumber = "FERMILAB-PUB-85-148-A, NSF-ITP-85-93",
    doi = "10.1086/164143",
    journal = "Astrophys. J.",
    volume = "304",
    pages = "15--61",
    year = "1986"
}

@article{eBOSS:2018qyj,
    author = "Chabanier, Sol{\`e}ne and others",
    collaboration = "eBOSS",
    title = "{The one-dimensional power spectrum from the SDSS DR14 Ly$\alpha$ forests}",
    eprint = "1812.03554",
    archivePrefix = "arXiv",
    primaryClass = "astro-ph.CO",
    doi = "10.1088/1475-7516/2019/07/017",
    journal = "JCAP",
    volume = "07",
    pages = "017",
    year = "2019"
}

@article{Sabti_2022,
   title={New Roads to the Small-scale Universe: Measurements of the Clustering of Matter with the High-redshift UV Galaxy Luminosity Function},
   volume={928},
   ISSN={2041-8213},
   url={http://dx.doi.org/10.3847/2041-8213/ac5e9c},
   DOI={10.3847/2041-8213/ac5e9c},
   number={2},
   journal={The Astrophysical Journal Letters},
   publisher={American Astronomical Society},
   author={Sabti, Nashwan and Muñoz, Julian B. and Blas, Diego},
   year={2022},
   month=apr, pages={L20} }

@article{Long:2024cak,
    author = "Long, Andrew J. and Venegas, Moira",
    title = "{Free streaming of warm wave dark matter in modified expansion histories}",
    eprint = "2412.14322",
    archivePrefix = "arXiv",
    primaryClass = "astro-ph.CO",
    doi = "10.1088/1475-7516/2025/06/043",
    journal = "JCAP",
    volume = "06",
    pages = "043",
    year = "2025"
}

@article{Hui:2016ltb,
    author = "Hui, Lam and Ostriker, Jeremiah P. and Tremaine, Scott and Witten, Edward",
    title = "{Ultralight scalars as cosmological dark matter}",
    eprint = "1610.08297",
    archivePrefix = "arXiv",
    primaryClass = "astro-ph.CO",
    doi = "10.1103/PhysRevD.95.043541",
    journal = "Phys. Rev. D",
    volume = "95",
    number = "4",
    pages = "043541",
    year = "2017"
}

@article{Amin:2025sla,
    author = "Amin, Mustafa A. and May, Simon and Mirbabayi, Mehrdad",
    title = "{Early growth of structure in warm wave dark matter}",
    eprint = "2506.12131",
    archivePrefix = "arXiv",
    primaryClass = "astro-ph.CO",
    doi = "10.1088/1475-7516/2025/10/040",
    journal = "JCAP",
    volume = "10",
    pages = "040",
    year = "2025"
}

@article{Hlozek:2014lca,
    author = "Hlozek, Ren{\'e}e and Grin, Daniel and Marsh, David J. E. and Ferreira, Pedro G.",
    title = "{A search for ultralight axions using precision cosmological data}",
    eprint = "1410.2896",
    archivePrefix = "arXiv",
    primaryClass = "astro-ph.CO",
    doi = "10.1103/PhysRevD.91.103512",
    journal = "Phys. Rev. D",
    volume = "91",
    number = "10",
    pages = "103512",
    year = "2015"
}

@article{Gorghetto:2021fsn,
    author = "Gorghetto, Marco and Hardy, Edward and Nicolaescu, Horia",
    title = "{Observing invisible axions with gravitational waves}",
    eprint = "2101.11007",
    archivePrefix = "arXiv",
    primaryClass = "hep-ph",
    doi = "10.1088/1475-7516/2021/06/034",
    journal = "JCAP",
    volume = "06",
    pages = "034",
    year = "2021"
}

@article{Gorghetto:2020qws,
    author = "Gorghetto, Marco and Hardy, Edward and Villadoro, Giovanni",
    title = "{More axions from strings}",
    eprint = "2007.04990",
    archivePrefix = "arXiv",
    primaryClass = "hep-ph",
    doi = "10.21468/SciPostPhys.10.2.050",
    journal = "SciPost Phys.",
    volume = "10",
    number = "2",
    pages = "050",
    year = "2021"
}

@article{Feix:2020txt,
    author = {Feix, Martin and Hagstotz, Steffen and Pargner, Andreas and Reischke, Robert and Sch{\"a}fer, Bjoern Malte and Schwetz, Thomas},
    title = "{Post-inflationary axion isocurvature perturbations facing CMB and large-scale structure}",
    eprint = "2004.02926",
    archivePrefix = "arXiv",
    primaryClass = "astro-ph.CO",
    doi = "10.1088/1475-7516/2020/11/046",
    journal = "JCAP",
    volume = "11",
    pages = "046",
    year = "2020"
}

@article{Feix:2019lpo,
    author = {Feix, Martin and Frank, Johann and Pargner, Andreas and Reischke, Robert and Sch{\"a}fer, Bjoern Malte and Schwetz, Thomas},
    title = "{Isocurvature bounds on axion-like particle dark matter in the post-inflationary scenario}",
    eprint = "1903.06194",
    archivePrefix = "arXiv",
    primaryClass = "astro-ph.CO",
    doi = "10.1088/1475-7516/2019/05/021",
    journal = "JCAP",
    volume = "05",
    pages = "021",
    year = "2019"
}

@article{Irsic:2017yje,
    author = "Ir{\v{s}}i{\v{c}}, Vid and Viel, Matteo and Haehnelt, Martin G. and Bolton, James S. and Becker, George D.",
    title = "{First constraints on fuzzy dark matter from Lyman-$\alpha$ forest data and hydrodynamical simulations}",
    eprint = "1703.04683",
    archivePrefix = "arXiv",
    primaryClass = "astro-ph.CO",
    doi = "10.1103/PhysRevLett.119.031302",
    journal = "Phys. Rev. Lett.",
    volume = "119",
    number = "3",
    pages = "031302",
    year = "2017"
}

@article{Kobayashi:2017jcf,
    author = "Kobayashi, Takeshi and Murgia, Riccardo and De Simone, Andrea and Ir{\v{s}}i{\v{c}}, Vid and Viel, Matteo",
    title = "{Lyman-$\alpha$ constraints on ultralight scalar dark matter: Implications for the early and late universe}",
    eprint = "1708.00015",
    archivePrefix = "arXiv",
    primaryClass = "astro-ph.CO",
    reportNumber = "SISSA-33-2017-FISI",
    doi = "10.1103/PhysRevD.96.123514",
    journal = "Phys. Rev. D",
    volume = "96",
    number = "12",
    pages = "123514",
    year = "2017"
}

@article{Irsic:2019iff,
    author = "Ir{\v{s}}i{\v{c}}, Vid and Xiao, Huangyu and McQuinn, Matthew",
    title = "{Early structure formation constraints on the ultralight axion in the postinflation scenario}",
    eprint = "1911.11150",
    archivePrefix = "arXiv",
    primaryClass = "astro-ph.CO",
    doi = "10.1103/PhysRevD.101.123518",
    journal = "Phys. Rev. D",
    volume = "101",
    number = "12",
    pages = "123518",
    year = "2020"
}

@article{Foster:2017hbq,
    author = "Foster, Joshua W. and Rodd, Nicholas L. and Safdi, Benjamin R.",
    title = "{Revealing the Dark Matter Halo with Axion Direct Detection}",
    eprint = "1711.10489",
    archivePrefix = "arXiv",
    primaryClass = "astro-ph.CO",
    reportNumber = "MIT-CTP-4964, LCTP-17-08, MIT-CTP 4964",
    doi = "10.1103/PhysRevD.97.123006",
    journal = "Phys. Rev. D",
    volume = "97",
    number = "12",
    pages = "123006",
    year = "2018"
}

@article{Maseizik:2024qly,
    author = {Maseizik, Dennis and Sigl, G{\"u}nter},
    title = "{Distributions and collision rates of ALP stars in the Milky~Way}",
    eprint = "2404.07908",
    archivePrefix = "arXiv",
    primaryClass = "astro-ph.CO",
    doi = "10.1103/PhysRevD.110.083015",
    journal = "Phys. Rev. D",
    volume = "110",
    number = "8",
    pages = "083015",
    year = "2024"
}

@ARTICLE{2018PhRvD..97h3502F,
       author = {{Fairbairn}, Malcolm and {Marsh}, David J.~E. and {Quevillon}, J{\'e}r{\'e}mie and {Rozier}, Simon},
        title = "{Structure formation and microlensing with axion miniclusters}",
      journal = {\prd},
         year = 2018,
        month = apr,
       volume = {97},
       number = {8},
          eid = {083502},
        pages = {083502},
          doi = {10.1103/PhysRevD.97.083502},
archivePrefix = {arXiv},
       eprint = {1707.03310},
 primaryClass = {astro-ph.CO},
       adsurl = {https://ui.adsabs.harvard.edu/abs/2018PhRvD..97h3502F}
}

@ARTICLE{2010MNRAS.404...60R,
       author = {{Reid}, Beth A. and {Percival}, Will J. and {Eisenstein}, Daniel J. and {Verde}, Licia and {Spergel}, David N. and {Skibba}, Ramin A. and {Bahcall}, Neta A. and {Budavari}, Tamas and {Frieman}, Joshua A. and {Fukugita}, Masataka and {Gott}, J. Richard and {Gunn}, James E. and {Ivezi{\'c}}, {\v{Z}}eljko and {Knapp}, Gillian R. and {Kron}, Richard G. and {Lupton}, Robert H. and {McKay}, Timothy A. and {Meiksin}, Avery and {Nichol}, Robert C. and {Pope}, Adrian C. and {Schlegel}, David J. and {Schneider}, Donald P. and {Stoughton}, Chris and {Strauss}, Michael A. and {Szalay}, Alexander S. and {Tegmark}, Max and {Vogeley}, Michael S. and {Weinberg}, David H. and {York}, Donald G. and {Zehavi}, Idit},
        title = "{Cosmological constraints from the clustering of the Sloan Digital Sky Survey DR7 luminous red galaxies}",
      journal = {Monthly Notices of the Royal Astronomical Society},
         year = 2010,
        month = may,
       volume = {404},
       number = {1},
        pages = {60-85},
          doi = {10.1111/j.1365-2966.2010.16276.x},
archivePrefix = {arXiv},
       eprint = {0907.1659},
 primaryClass = {astro-ph.CO},
       adsurl = {https://ui.adsabs.harvard.edu/abs/2010MNRAS.404...60R}
}

@article{Charnock:2016nzm,
    author = "Charnock, Tom and Avgoustidis, Anastasios and Copeland, Edmund J. and Moss, Adam",
    title = "{CMB constraints on cosmic strings and superstrings}",
    eprint = "1603.01275",
    archivePrefix = "arXiv",
    primaryClass = "astro-ph.CO",
    doi = "10.1103/PhysRevD.93.123503",
    journal = "Phys. Rev. D",
    volume = "93",
    number = "12",
    pages = "123503",
    year = "2016"
}

@article{Lizarraga:2016onn,
    author = "Lizarraga, Joanes and Urrestilla, Jon and Daverio, David and Hindmarsh, Mark and Kunz, Martin",
    title = "{New CMB constraints for Abelian Higgs cosmic strings}",
    eprint = "1609.03386",
    archivePrefix = "arXiv",
    primaryClass = "astro-ph.CO",
    doi = "10.1088/1475-7516/2016/10/042",
    journal = "JCAP",
    volume = "10",
    pages = "042",
    year = "2016"
}

@article{Lopez-Eiguren:2017dmc,
    author = "Lopez-Eiguren, Asier and Lizarraga, Joanes and Hindmarsh, Mark and Urrestilla, Jon",
    title = "{Cosmic Microwave Background constraints for global strings and global monopoles}",
    eprint = "1705.04154",
    archivePrefix = "arXiv",
    primaryClass = "astro-ph.CO",
    reportNumber = "HIP-2017-07-TH",
    doi = "10.1088/1475-7516/2017/07/026",
    journal = "JCAP",
    volume = "07",
    pages = "026",
    year = "2017"
}

@article{Planck:2018vyg,
    author = "Aghanim, N. and others",
    collaboration = "Planck",
    title = "{Planck 2018 results. VI. Cosmological parameters}",
    eprint = "1807.06209",
    archivePrefix = "arXiv",
    primaryClass = "astro-ph.CO",
    doi = "10.1051/0004-6361/201833910",
    journal = "Astron. Astrophys.",
    volume = "641",
    pages = "A6",
    year = "2020",
    note = "[Erratum: Astron.Astrophys. 652, C4 (2021)]"
}

@article{Khmelnitsky:2013lxt,
    author = "Khmelnitsky, Andrei and Rubakov, Valery",
    title = "{Pulsar timing signal from ultralight scalar dark matter}",
    eprint = "1309.5888",
    archivePrefix = "arXiv",
    primaryClass = "astro-ph.CO",
    doi = "10.1088/1475-7516/2014/02/019",
    journal = "JCAP",
    volume = "02",
    pages = "019",
    year = "2014"
}

@article{Kim:2023kyy,
    author = "Kim, Hyungjin and Mitridate, Andrea",
    title = "{Stochastic ultralight dark matter fluctuations in pulsar timing arrays}",
    eprint = "2312.12225",
    archivePrefix = "arXiv",
    primaryClass = "hep-ph",
    reportNumber = "DESY-23-192",
    doi = "10.1103/PhysRevD.109.055017",
    journal = "Phys. Rev. D",
    volume = "109",
    number = "5",
    pages = "055017",
    year = "2024"
}

@article{Kim:2024,
    author = "Kim, Hyungjin",
    title = "{Astrometric search for ultralight dark matter}",
    eprint = "2407.11191",
    archivePrefix = "arXiv",
    primaryClass = "astro-ph.CO",
    doi = "10.1103/PhysRevD.110.083031",
    journal = "Phys. Rev. D",
    volume = "110",
    number = "8",
    pages = "083031",
    year = "2024"
}

@article{Hu:2000ke,
    author = "Hu, Wayne and Barkana, Rennan and Gruzinov, Andrei",
    title = "{Cold and fuzzy dark matter}",
    eprint = "astro-ph/0003365",
    archivePrefix = "arXiv",
    reportNumber = "FERMILAB-PUB-00-082-A",
    doi = "10.1103/PhysRevLett.85.1158",
    journal = "Phys. Rev. Lett.",
    volume = "85",
    pages = "1158--1161",
    year = "2000"
}

@article{Consiglio_2018,
   title={PArthENoPE reloaded},
   volume={233},
   ISSN={0010-4655},
   url={http://dx.doi.org/10.1016/j.cpc.2018.06.022},
   DOI={10.1016/j.cpc.2018.06.022},
   journal={Computer Physics Communications},
   publisher={Elsevier BV},
   author={Consiglio, R. and de Salas, P.F. and Mangano, G. and Miele, G. and Pastor, S. and Pisanti, O.},
   year={2018},
   month=dec, pages={237–242} }

@misc{Dror:2025nvg,
    author = "Dror, Jeff A. and Wei, Qiushi",
    title = "{On Pulsar Timing Detection of Ultralight Vector Dark Matter}",
    eprint = "2505.22719",
    archivePrefix = "arXiv",
    primaryClass = "hep-ph",
    month = "5",
    year = "2025"
}

@article{Saikawa:2024bta,
    author = "Saikawa, Ken'ichi and Redondo, Javier and Vaquero, Alejandro and Kaltschmidt, Mathieu",
    title = "{Spectrum of global string networks and the axion dark matter mass}",
    eprint = "2401.17253",
    archivePrefix = "arXiv",
    primaryClass = "hep-ph",
    reportNumber = "KANAZAWA-24-02, MPP-2024-18",
    doi = "10.1088/1475-7516/2024/10/043",
    journal = "JCAP",
    volume = "10",
    pages = "043",
    year = "2024"
}

@article{Kaltschmidt:2025nkz,
    author = "Kaltschmidt, Mathieu and Redondo, Javier and Saikawa, Ken'ichi and Vaquero, Alejandro",
    title = "{The Spectrum of Global Axion Strings}",
    eprint = "2502.02398",
    archivePrefix = "arXiv",
    primaryClass = "hep-ph",
    doi = "10.22323/1.474.0017",
    journal = "PoS",
    volume = "COSMICWISPers2024",
    pages = "017",
    year = "2025"
}

@misc{Battye:2026whd,
    author = "Battye, Richard A. and Bunio, Lukasz P. and Cotterill, Steven J. and Manoj, Pranav B. Gangrekalve",
    title = "{Spectrum of radiation from global strings and the relic axion density}",
    eprint = "2601.19463",
    archivePrefix = "arXiv",
    primaryClass = "hep-ph",
    month = "1",
    year = "2026"
}

@article{Dine:2020pds,
    author = "Dine, Michael and Fernandez, Nicolas and Ghalsasi, Akshay and Patel, Hiren H.",
    title = "{Comments on axions, domain walls, and cosmic strings}",
    eprint = "2012.13065",
    archivePrefix = "arXiv",
    primaryClass = "hep-ph",
    reportNumber = "SCIPP 20/01",
    doi = "10.1088/1475-7516/2021/11/041",
    journal = "JCAP",
    volume = "11",
    pages = "041",
    year = "2021"
}

@article{Kirilova:2023rnl,
    author = "Kirilova, Daniela and Panayotova, Mariana and Chizhov, Emanuil",
    title = "{Big Bang Nucleosynthesis Constraints and Indications for Beyond Standard Model Neutrino Physics}",
    doi = "10.3390/sym16010053",
    journal = "Symmetry",
    volume = "16",
    number = "1",
    pages = "53",
    year = "2024"
}

@article{Benabou:2023ghl,
    author = "Benabou, Joshua N. and Buschmann, Malte and Kumar, Soubhik and Park, Yujin and Safdi, Benjamin R.",
    title = "{Signatures of primordial energy injection from axion strings}",
    eprint = "2308.01334",
    archivePrefix = "arXiv",
    primaryClass = "hep-ph",
    doi = "10.1103/PhysRevD.109.055005",
    journal = "Phys. Rev. D",
    volume = "109",
    number = "5",
    pages = "055005",
    year = "2024"
}

@ARTICLE{2020NatRP...2..245A,
       author = {{Algeri}, Sara and {Aalbers}, Jelle and {Mor{\r{a}}}, Knut Dundas and {Conrad}, Jan},
        title = "{Searching for new phenomena with profile likelihood ratio tests}",
      journal = {Nature Reviews Physics},
         year = 2020,
        month = may,
       volume = {2},
       number = {5},
        pages = {245-252},
          doi = {10.1038/s42254-020-0169-5},
archivePrefix = {arXiv},
       eprint = {1911.10237},
 primaryClass = {physics.data-an},
       adsurl = {https://ui.adsabs.harvard.edu/abs/2020NatRP...2..245A}
}

@article{Planck,
       author = {{Planck Collaboration}},
        title = "{Planck 2015 results. XIII. Cosmological parameters}",
      journal = {\aap},
         year = 2016,
        month = sep,
       volume = {594},
          eid = {A13},
        pages = {A13},
          doi = {10.1051/0004-6361/201525830},
archivePrefix = {arXiv},
       eprint = {1502.01589},
 primaryClass = {astro-ph.CO},
       adsurl = {https://ui.adsabs.harvard.edu/abs/2016A&A...594A..13P}
}

@misc{DM_review,
      title={TASI lectures on dark matter models and direct detection}, 
      author={Tongyan Lin},
      year={2019},
      eprint={1904.07915},
      archivePrefix={arXiv},
      primaryClass={hep-ph}
}

@article{Kyriazis:2025fis,
    author = "Kyriazis, Antonios and Yang, Fengwei",
    title = "{Gravitational waves from resonant transitions of tidally perturbed gravitational atoms}",
    eprint = "2503.18121",
    archivePrefix = "arXiv",
    primaryClass = "hep-ph",
    doi = "10.1007/JHEP11(2025)062",
    journal = "JHEP",
    volume = "11",
    pages = "062",
    year = "2025"
}

@article{Peccei,
  title = {$\mathrm{CP}$ Conservation in the Presence of Pseudoparticles},
  author = {Peccei, R. D. and Quinn, Helen R.},
  journal = {Phys. Rev. Lett.},
  volume = {38},
  issue = {25},
  pages = {1440--1443},
  numpages = {0},
  year = {1977},
  month = {Jun},
  publisher = {American Physical Society},
  doi = {10.1103/PhysRevLett.38.1440},
  url = {https://link.aps.org/doi/10.1103/PhysRevLett.38.1440}
}

@article{Preskill,
title = {Cosmology of the invisible axion},
journal = {Physics Letters B},
volume = {120},
number = {1},
pages = {127-132},
year = {1983},
issn = {0370-2693},
doi = {https://doi.org/10.1016/0370-2693(83)90637-8},
url = {https://www.sciencedirect.com/science/article/pii/0370269383906378},
author = {John Preskill and Mark B. Wise and Frank Wilczek}
}

@article{Weinberg,
  title = {A New Light Boson?},
  author = {Weinberg, Steven},
  journal = {Phys. Rev. Lett.},
  volume = {40},
  issue = {4},
  pages = {223--226},
  numpages = {0},
  year = {1978},
  month = {Jan},
  publisher = {American Physical Society},
  doi = {10.1103/PhysRevLett.40.223},
  url = {https://link.aps.org/doi/10.1103/PhysRevLett.40.223}
}

@article{Sikivie,
title = {A cosmological bound on the invisible axion},
journal = {Physics Letters B},
volume = {120},
number = {1},
pages = {133-136},
year = {1983},
issn = {0370-2693},
doi = {https://doi.org/10.1016/0370-2693(83)90638-X},
url = {https://www.sciencedirect.com/science/article/pii/037026938390638X},
author = {L.F. Abbott and P. Sikivie}
}

@article{Wilczek1,
  title = {Problem of Strong $P$ and $T$ Invariance in the Presence of Instantons},
  author = {Wilczek, F.},
  journal = {Phys. Rev. Lett.},
  volume = {40},
  issue = {5},
  pages = {279--282},
  numpages = {0},
  year = {1978},
  month = {Jan},
  publisher = {American Physical Society},
  doi = {10.1103/PhysRevLett.40.279},
  url = {https://link.aps.org/doi/10.1103/PhysRevLett.40.279}
}

@article{Willy,
title = {The not-so-harmless axion},
journal = {Physics Letters B},
volume = {120},
number = {1},
pages = {137-141},
year = {1983},
issn = {0370-2693},
doi = {https://doi.org/10.1016/0370-2693(83)90639-1},
url = {https://www.sciencedirect.com/science/article/pii/0370269383906391},
author = {Michael Dine and Willy Fischler}
}

@article{Rubin1,
       author = {{Rubin}, Vera C. and {Ford}, W. Kent, Jr.},
        title = "{Rotation of the Andromeda Nebula from a Spectroscopic Survey of Emission Regions}",
      journal = {\apj},
         year = 1970,
        month = feb,
       volume = {159},
        pages = {379},
          doi = {10.1086/150317},
       adsurl = {https://ui.adsabs.harvard.edu/abs/1970ApJ...159..379R}
}

@article{Rubin2,
       author = {{Rubin}, V.~C. and {Ford}, W.~K., Jr. and {Thonnard}, N.},
        title = "{Rotational properties of 21 SC galaxies with a large range of luminosities and radii, from NGC 4605 (R=4kpc) to UGC 2885 (R=122kpc).}",
      journal = {\apj},
         year = 1980,
        month = jun,
       volume = {238},
        pages = {471-487},
          doi = {10.1086/158003},
       adsurl = {https://ui.adsabs.harvard.edu/abs/1980ApJ...238..471R}
}

@PHDTHESIS{Bosma,
       author = {{Bosma}, A.},
        title = "{The distribution and kinematics of neutral hydrogen in spiral galaxies of various morphological types}",
       school = {University of Groningen, Netherlands},
         year = 1978,
        month = mar,
       adsurl = {https://ui.adsabs.harvard.edu/abs/1978PhDT.......195B}
}

@article{neeutron_edm_th,
    author = "Crewther, R. J. and Di Vecchia, P. and Veneziano, G. and Witten, Edward",
    title = "{Chiral Estimate of the Electric Dipole Moment of the Neutron in Quantum Chromodynamics}",
    reportNumber = "CERN-TH-2735",
    doi = "10.1016/0370-2693(79)90128-X",
    journal = "Phys. Lett. B",
    volume = "88",
    pages = "123",
    year = "1979",
    note = "[Erratum: Phys.Lett.B 91, 487 (1980)]"
}

@article{Pendlebury:2015lrz,
    author = "Pendlebury, J. M. and others",
    title = "{Revised experimental upper limit on the electric dipole moment of the neutron}",
    eprint = "1509.04411",
    archivePrefix = "arXiv",
    primaryClass = "hep-ex",
    doi = "10.1103/PhysRevD.92.092003",
    journal = "Phys. Rev. D",
    volume = "92",
    number = "9",
    pages = "092003",
    year = "2015"
}

@article{Arvanitaki:2009fg,
    author = "Arvanitaki, Asimina and Dimopoulos, Savas and Dubovsky, Sergei and Kaloper, Nemanja and March-Russell, John",
    title = "{String Axiverse}",
    eprint = "0905.4720",
    archivePrefix = "arXiv",
    primaryClass = "hep-th",
    doi = "10.1103/PhysRevD.81.123530",
    journal = "Phys. Rev. D",
    volume = "81",
    pages = "123530",
    year = "2010"
}

@ARTICLE{1976JPhA....9.1387K,
       author = {{Kibble}, T.~W.~B.},
        title = "{Topology of cosmic domains and strings}",
      journal = {Journal of Physics A Mathematical General},
         year = 1976,
        month = aug,
       volume = {9},
       number = {8},
        pages = {1387-1398},
          doi = {10.1088/0305-4470/9/8/029},
       adsurl = {https://ui.adsabs.harvard.edu/abs/1976JPhA....9.1387K}
}

@book{coleman1985aspects,
  title={Aspects of Symmetry: Selected Erice Lectures},
  author={Coleman, Sidney},
  year={1985},
  publisher={Cambridge University Press},
  address={Cambridge}
}

@article{Vilenkin:1984ib,
    author = "Vilenkin, Alexander",
    title = "{Cosmic Strings and Domain Walls}",
    reportNumber = "PRINT-84-0840 (TUFTS)",
    doi = "10.1016/0370-1573(85)90033-X",
    journal = "Phys. Rept.",
    volume = "121",
    pages = "263--315",
    year = "1985"
}

@article{Battye:1993jv,
    author = "Battye, R. A. and Shellard, E. P. S.",
    title = "{Global string radiation}",
    eprint = "astro-ph/9311017",
    archivePrefix = "arXiv",
    reportNumber = "DAMTP-R-93-30",
    doi = "10.1016/0550-3213(94)90573-8",
    journal = "Nucl. Phys. B",
    volume = "423",
    pages = "260--304",
    year = "1994"
}

@article{Chang:1998tb,
    author = "Chang, Sanghyeon and Hagmann, C. and Sikivie, P.",
    title = "{Studies of the motion and decay of axion walls bounded by strings}",
    eprint = "hep-ph/9807374",
    archivePrefix = "arXiv",
    reportNumber = "UFIFT-HEP-98-12",
    doi = "10.1103/PhysRevD.59.023505",
    journal = "Phys. Rev. D",
    volume = "59",
    pages = "023505",
    year = "1999"
}

@ARTICLE{1972JETP...35.1085Z,
       author = {{Zel'Dovich}, Ya. B.},
        title = "{Amplification of Cylindrical Electromagnetic Waves Reflected from a Rotating Body}",
      journal = {Soviet Journal of Experimental and Theoretical Physics},
         year = 1972,
        month = jan,
       volume = {35},
        pages = {1085},
       adsurl = {https://ui.adsabs.harvard.edu/abs/1972JETP...35.1085Z}
}

@ARTICLE{1969NCimR...1..252P,
       author = {{Penrose}, Roger},
        title = "{Gravitational Collapse: the Role of General Relativity}",
      journal = {Nuovo Cimento Rivista Serie},
         year = 1969,
        month = jan,
       volume = {1},
        pages = {252},
       adsurl = {https://ui.adsabs.harvard.edu/abs/1969NCimR...1..252P}
}
